\documentclass[preprint]{emulateapj}   
\usepackage{epsfig,natbib}
\usepackage{mathptmx}
\usepackage{amsmath,amsfonts,amssymb}

\submitted{Accepted by the Astrophysical Journal Supplements}
\shorttitle{Emission Signatures of MBHBs}
\shortauthors{Bogdanovi\'c et al.}

\newcommand{\etal}{{et al.}}

\newcommand{\Msun}{\>{\rm M_{\odot}}}

\begin{document}

\title{Modeling of Emission Signatures of Massive Black Hole Binaries: I Methods}

\author{Tamara Bogdanovi\'c\altaffilmark{1,2}, Britton D. Smith, Steinn 
Sigurdsson\altaffilmark{2}, and Michael Eracleous\altaffilmark{2}}
\affil{Department of Astronomy \&
Astrophysics, Pennsylvania State University, University Park, PA
16802} 
\altaffiltext{1}{Current Address: Department of Astronomy, University
  of Maryland, College Park, MD 20742}
\altaffiltext{2}{Center for Gravitational Wave Physics,
Pennsylvania State University, University Park, PA 16802 }
\email{tamarab@astro.umd.edu, britton, steinn, mce@astro.psu.edu}

\begin{abstract} 
We model the electromagnetic signatures of massive black hole binaries
(MBHBs) with an associated gas component. The method comprises
numerical simulations of relativistic binaries and gas coupled with
calculations of the physical properties of the emitting gas. We
calculate the UV/X-ray and the H$\alpha$ light curves and the
H$\alpha$ emission profiles. The simulations are carried out with a
modified version of the parallel tree SPH code {\it Gadget}. The
heating, cooling, and radiative processes are calculated for two
different physical scenarios, where the gas is approximated as a
black-body or a solar metallicity gas. The calculation for the solar
metallicity scenario is carried out with the photoionization code {\it
Cloudy}. We focus on sub-parsec binaries which have not yet entered
the gravitational radiation phase. The results from the first set of
calculations, carried out for a coplanar binary and gas disk, suggest
that there are pronounced outbursts in the X-ray light curve during
pericentric passages. If such outbursts persist for a large fraction
of the lifetime of the system, they can serve as an indicator of this
type of binary. The predicted H$\alpha$ emission line profiles may be
used as a criterion for selection of MBHB candidates from existing
archival data. The orbital period and mass ratio of a binary may be
inferred after carefully monitoring the evolution of the H$\alpha$
profiles of the candidates. The discovery of sub-parsec binaries is an
important step in understanding of the merger rates of MBHBs and their
evolution towards the detectable gravitational wave window.
\end{abstract}

\keywords{black hole physics---galaxies:nuclei---hydrodynamics---line:profiles---radiation mechanisms:general}

\section{Introduction}

Massive black hole binaries (hereafter MBHBs) may form as a result of
several processes: through the breakup of a supermassive protostar
\citep{br}, as a meta-stable state in the evolution of a cluster of
massive objects \citep{sva,bromm}, and as a result of galactic mergers
\citep{bbr,valtaoja89,milosavljevic01,yu02}. The last mechanism is a major 
formation route for MBHBs and it relies on hierarchical merger models
of galaxy formation \citep{haehnelt02,volonteri} and significant
dynamical evidence that the majority of galaxies harbor massive black
holes in their centers \citep[e.g.,][]{kr,richstone,peterson00}. In
this paper we investigate observational signatures of MBHBs that
result from galactic mergers.

Theoretical results imply that the evolution of a MBHB proceeds in
three stages. In the first stage, the binary separation decreases
through the process of dynamical friction when stars in the nuclear
region are scattered by the binary. In the third stage, binary orbital
angular momentum is efficiently dissipated via gravitational radiation
emission and the binary proceeds to an inevitable merger. A
possibility that in the second stage of evolution the binary merger
stalls because of depletion of stars in the nuclear region, is still a
topic of a lively scientific debate
\citep{quinlan,sigurdsson98,milosavljevic01,hemsendorf02,
merritt02,yu02,aarseth03,milosavljevic03,perets06,zier06}. If this is
indeed the case, binaries spend most of their lifetime at separations
of $0.01-1\,$pc \citep{bbr} and the majority of observed binaries
should be found in this evolutionary stage. Hereafter, we refer to
this, second evolutionary stage in the life of a binary, as to 
{\it intermediate phase}.

A number of authors have suggested that interaction of a binary with a
gaseous accretion disk may have significant effects on the inspiral
and merger rates of MBHB
\citep{ivanov99,gould00,armitage02,armitage05,escala1,escala2,
kazantzidis05,dotti06a,mayer06}.  It has been shown that massive gas
disks tend to form in centers of merging galaxies
\citep{barnes92,barnes96,mihos96}. In particular, a gas disk may
expedite the merger of a binary by dissipation of orbital angular
momentum \citep{escala1,escala2}, reducing the merger time to
$10^{6}-10^{7}$yr.

Observationally, MBHBs may be identified directly in cases when a
black hole pair is spatially resolved and when the separation and
kinematic parameters point to a close, bound pair. In practice, there
is only one such object identified so far: observations of NGC~6240,
an ultra luminous infrared galaxy, with the $Chandra$ X-ray
observatory revealed a merging pair of X-ray active nuclei with a
separation of 1.4 kpc \citep{komossa}. Recently \citet{hudson06} also
reported on the X-ray detection of a wide, proto supermassive binary
black hole at the center of cluster Abell~400. Practical obstacles in
this type of direct identification arise for several reasons. Firstly,
fairly high spatial resolution and accuracy in position measurements
are required to resolve a binary active nucleus. The spatial
resolution required to resolve an intermediate binary with orbital
separation of 1pc at the distance of $\sim$100$\,$Mpc is about
2$\,$mas. With spatial resolutions currently achievable, it is easier
to spot and resolve wide binaries, like NGC~6240, whereas it has been
suggested that MBHBs spend a major fraction of their life time as
intermediate, hard binaries \citep{bbr}. Because of their high spatial
resolution VLBA and VLBI radio observatories have a large potential
for discovery of close binaries with separations of 0.1$-$10~pc,
especially when combined with ground-based optical spectroscopic
observations \citep{rodriguez06}.

The second most appealing piece of observational evidence for a MBHB
would be a periodic manifestation of Keplerian motion. There are a
handful of galaxies which exhibit periodicities in their light curves
and thus can be considered as MBHB candidates
\citep{fan,rieger,depaolis2002,depaolis2003,sudou,xie}. A prominent
example is the blazar OJ~287 which exhibits outburst activity in the
optical light curve with a period close to 12 years. The latest
outburst of OJ~287, expected at the end of 2006, has not yet happened,
putting to the test a number of models along with the binary
hypothesis \citep{sillanpaa,lehto,valtaoja}.

Another phenomenon associated with the existence of a MBHB is the
change in the orientation of the black hole spin axis and its
precession about the total angular momentum axis, caused by its
interaction with another black hole \citep{me,zier2001,zier2002}. This
effect is one possible explanation for helical radio-jets, X-shaped,
and S-shaped radio-jets, observed in some galaxies
\citep[e.g.,][]{hunstead,leahy,parma,roos,wang,liu}, though
alternative explanations are available \citep{capetti} . In cases when
the orbit of the less massive, inspiralling black hole becomes coplanar
with an accretion disk of the primary, fueling of jets may be
interrupted and re-started \citep{liu}. This could give rise to a
second pair of jets aligned along the same axis with the initial pair,
as seen in double-double radio galaxies \citep*{schoenmakers,lwc}.

Massive binary black holes may also influence the kinematics of the
broad line region of the active galactic nucleus (AGN) which could be
detected in periodic variations of broad emission lines \citep{bbr,
gaskell83}. \citet{gaskell96} suggested that double peaked emission
lines originate in binary broad line regions. In this scenario
periodic velocity shifts of each peak in the line profile should
reflect the orbital motion of a single broad line region associated
with one of the black holes. After extensive spectroscopic monitoring
of MBHB candidates spanning two decades the predicted long term
periodic variability was not confirmed, leaving little room for a
binary broad line region scenario
\citep{halpern88,halpern92,eracleous97}. \citet{eracleous97} noted
that perturbation of a single broad-line region by a second MBH is
still an open possibility, and in fact the only good interpretation
for some objects. Another version of this hypothesis in which the
secondary black hole perturbs the broad line region of the primary was
offered by \citet*{torres} as an explanation of the variable Fe
K$\alpha$ line in 3C~273.  \citet{eh06} suggest that characteristic
spiral arm morphology can arise within the inner $1$ kpc of a gas disk
as a consequence of perturbations in a binary potential, and that in
the absence of more direct techniques, this signature may be used for
a MBHB search.

Another, soon-to-be-opened observational window should allow detection
of gravitational waves from MBHBs \citep{thorne,centrella,baker}. 
These are expected to be an important class of sources for the Laser
Interferometer Space Antenna, $LISA$
\citep{bender,danzmann,haehnelt94,hughes02,wl03,sesana04,sesana05,rw05}. 
Since it is not clear which model should be used to describe the
optical data (binary orbital motion or disk and jet precession), the
masses of current candidate binary systems are highly uncertain. The
future data from gravitational radiation observations, combined
with information on optical counterparts, should play an important role
in determining the most likely scenario \citep{komossa_rev,
armitage05}. In a broader sense, once enough information about the
statistical properties of such binaries is available, it should be
possible to constrain the merger rates and merger histories of massive
and supermassive black hole binaries
\citep{haehnelt94,menou01,hughes02}. For this reason a large
scientific effort has been recently directed towards understanding the
electromagnetic signatures in different pre-merger and post-merger
phases of MBHBs \citep{milosavljevic05,kocsis06,dotti06b}. The work
we present in this paper fits within this broader context.

In \S\ref{S_binary_simulations} we present our method for calculation
of emission signatures of MBHBs. The results are described in
\S\ref{S_results}, the discussion and future prospects in
\S\ref{S_discussion}, and in \S\ref{S_conclusions} we give our
conclusions.

\section{Numerical simulations}\label{S_binary_simulations}

We have carried out smoothed particle hydrodynamical (SPH) simulations
of the binary and a gaseous component, and we have characterized the
physical properties of the gas by calculating heating, cooling, and
radiative processes as an integral part of simulations. Based on these
results we have calculated the accretion-powered continuum and
H$\alpha$ light curves, as well as the H$\alpha$ emission line
profiles emerging from the inner parts of a gas disk on a scale of
$<0.1$ pc.

For the first time we attempt to model both gas and binary dynamics in
the nuclear region on sub-parsec scales in order to bridge the gap
between large scale hydrodynamic simulations of merging galaxies and
numerical relativity simulations of the late, pre-coalescence
evolutionary phase in the life of a binary. The binary-gas evolution
in simulations of merging galaxies has been followed down to
separations of $>0.1$ pc \citep{dotti06c}. The simulations that reach
even smaller separations are computationally expensive, mostly because
they need to be carried out at high resolution in force (or
equivalently time). Additionally, if the nuclear processes followed
include accretion, then a high resolution in mass is desired as
well. These requirements are combined with the wide spatial and
temporal dynamic range set by the nature of the problem itself. For
the above reasons, in our simulations we focus on the length scale
occupied by a nuclear accretion disk and do not consider the disk
structure on larger scales.

We have used a modified version of {\it Gadget} \citep{springel01,
springel05} to carry out the MBHB simulations. {\it Gadget} is a code
for collisionless and gas-dynamical cosmological simulations. It
evolves self-gravitating collisionless fluids with a tree $N$-body
approach, and collisional gas by SPH. {\it Gadget} was not originally
intended to carry out relativistic calculations. For this reason we
have performed several modifications to the code in order to treat the
two massive black holes relativistically. We have introduced the black
hole particles as collisionless massive particles with
pseudo-Newtonian potentials. The calculations with a pseudo-Newtonian
potential account for the decay of a black hole binary orbit through
emission of gravitational radiation (\S\ref{S_orbits}). The
gravitational drag from gas particles on the two black holes is
accounted for through a contribution from every gas particle to each
black hole's acceleration.

We set up the two Schwarzschild black holes as sink particles and
model the accretion rate of gas and resulting accretion luminosity.
The approach in which a black hole is treated as a sink particle has
been introduced before in simulations of star formation
\citep{bate03,li05} and accretion onto supermassive black holes
\citep{sdh05}. In our model, a particle is considered to contribute to
the unresolved accretion flow near a black hole once it crosses the
hole's accretion radius, R$_{acc}$. R$_{acc}$ is set to a fiducial
value of 20~$r_g$, comparable to the smallest length scale resolved in
the simulation (i.e., the smoothing length of particles in a region of
high density, $h_{sml}\sim 10 r_g$, where $r_{g}\equiv
GM_{BH}/c^{2}=1.48\times10^{13}\,M_{8}\;{\rm cm}$, and $M_{BH}=10^8
M_8\,\Msun$ is the mass of a black hole). Once the gas particles find
themselves within R$_{acc}$ their dynamics is not temporally resolved.
This choice allows to avoid a simulation hang-up when the integration
time step of such particles becomes very small. We introduce a
numerical approximation in the calculation of the accretion rate in
order to account for the fact that the unresolved accretion flow of
particles inside R$_{acc}$ radius is noisy and that their dynamics
cannot be described by the model of a steady accretion disk. Because
of the short dynamical time associated with the innermost region of
the accretion disk the particles there respond quickly to the variable
potential of the binary which causes the streams of gas to collide and
intersect with each other. From the total amount of gas that crosses
R$_{acc}$ in given time $dt$, only a fraction of the gas will be
accreted, while the rest will gain enough momentum to avoid accretion
onto the black hole. The accretion rate of a noisy flow of particles
across some radius $r$ is in general case proportional to the surface
area, $\dot{M}\propto r^2$, assuming the planar and uniform
distribution of particles within $r$.  We calculate from the
simulations the rate with which particles cross R$_{acc}$ and
extrapolate from it the accretion rate of particles that would have
crossed the Schwarzschild radius as $\dot{M}(2r_g) \approx
\dot{M}({\rm R}_{acc})\,({\rm R}_{acc}/2r_g)^{-2}$. The estimate
obtained in such way is conservative as we do not take into account
the gravitational focusing by the black holes, which would increase
the flow of particles towards the black holes and thus, their
accretion rates. In \S~\ref{S_discussion}, we discuss the effect of
our assumptions about the accretion model on the observational
signatures.

\subsection{Viscosity}\label{S_viscosity}

The viscosity is parametrized in numerical calculations in order to
account for the transfer of momentum and angular momentum without
detailed assumptions about the nature of the phenomenon. One widely
used parametrization is based on the Shakura-Sunyaev model
\citep{shakura}, where the $\alpha$-parameter takes values less than
1.0, and typically $\alpha\leq 0.1$. \citet{starling04} for example
find $0.01 \leq \alpha \leq 0.03$, for accretion disks in AGN, based
on the observed continuum variability.  Finding a suitable parametric
description for the viscosity in SPH calculations which would
faithfully describe the angular momentum transfer in different
physical regimes has been a major issue over many years
\citep{hk89,monaghan92,balsara95,steinmetz96,monaghan97,
springel05}. A notable improvement in the treatment of viscosity was
offered by parametrization that perform better in conserving the
entropy of a simulated system \citep[e.g., {\it in
Gadget-2};][]{springel05}. Although we have used the version of {\it
Gadget} released earlier \citep{springel01}, for the purposes of our
calculations we adopted the viscosity prescription given in the new
version of the code, {\it Gadget-2}. The physical properties in our
simulations that are directly affected by this choice are the
accretion rate of gas onto black holes and the internal energy of the
gas $-$ both of which play very important roles in the observational
appearance of the system.

We derive the value of the artificial viscosity parameter used in the
code, $\alpha_{Gadget}$ in such way that it corresponds to the
Shakura-Sunyaev viscosity parameter $\alpha<1$. We note that this
relation is only a convenient way of describing the viscosity
numerically and should not be interpreted as a statement about the
physical viscosity in the disk. In SPH simulations, the viscosity acts
as an excess pressure which is assigned to particles in the equation
of motion, $P\approx\,\alpha_{Gadget}\,(c_{s}\omega -
\frac{3}{2}\omega^{2})\, \rho$, where $c_{s}$ is the speed of sound
and $\omega$ is the relative velocity of the two approaching
particles. We compare this to the Shakura-Sunyaev tangential stress
component $P\approx\,-\alpha\,c_{s}^{2}\,\rho$, and obtain a local
relation between the parameters describing a single particle
interaction
\begin{equation}
\alpha\approx\alpha_{Gadget}\,\left(-\frac{\omega}{c_{s}}+\frac{3}{2}
\frac{\omega^{2}}{c_{s}^{2}}\right).  \label{eq_visc_micro}
\end{equation} 
The resulting artificial viscosity as described by
equation~(\ref{eq_visc_micro}) does not explicitly depend on the
density of gas particles. However, in simulations a major increase in
internal energy due to viscous dissipation occurs in high density
regions which are conducive to high rate of particle interactions
(i.e., collisions). The implication is that the relation connecting
the two viscosity parameters depends on the density of gas particles
in a simulation, namely,
\begin{equation}
\alpha\approx\alpha_{Gadget}\, \left(-\frac{\omega}{c_{s}}+\frac{3}{2} 
\frac{\omega^{2}}{c_{s}^{2}} \right)
\left( \frac{\rho}{\rho_{o}} \right) ^{2}, \label{eq_visc_macro}
\end{equation} 
where $\rho=\rho(r)$ is the density distribution of gas particles and
$\rho_{o}=\rho(r_{in})$ is the value of the density at the inner
radius of the accretion disk. Assuming that the largest value that
$\omega$ can take is of the order of particle orbital velocities one
can find a value of $\alpha_{Gadget}$ which corresponds to some value
of the Shakura-Sunyaev $\alpha$. From equation~(\ref{eq_visc_macro})
we find that in order to have $0.01 \leq\alpha\leq 0.1$, at radial
distances $\xi \sim 100$ from a black hole, in a disk where $c_s\sim
100~{\rm km\,s^{-1}}$, we have to choose
$\alpha_{Gadget}\approx10^{-6}$. Hereafter, we use $\xi$ as the radius
in units of gravitational radii ($\xi=r/r_{g}$).

\subsection{Elliptical Orbits in the Paczynsky-Wiita Potential and Gravitational Wave Emission} 
\label{S_orbits}

We choose a pseudo-Newtonian Paczynsky-Wiita potential
\citep{paczynsky} to describe the potentials of massive black holes 
in our calculations
\begin{equation}
\psi = \frac{-GM_{BH}}{(r-r_{s})} \; ,  \label{eq_pw_potential}
\end{equation}
where $r_{s} = 2GM_{BH}/c^2$. The two black holes can initially be
placed on circular or elliptical orbits. The properties of circular
orbits in the Paczynsky-Wiita potential are described in
\citet{paczynsky}. In addition, we derive expressions for the orbital
velocity of a test particle on an elliptical orbit in the
Paczynsky-Wiita potential in order to assign initial velocities
consistent with the potential. We adopt the reduced two-body problem
parametrization where $M=m_{1}+m_{2}$ and
$\mu=m_{1}m_{2}/(m_{1}+m_{2})$, where the energy and angular momentum
of a system are calculated with respect to its center of mass ($m_{1}$
and $m_{2}$ are masses of the two black holes). We start with the
requirement that in the Paczynsky-Wiita potential, the total energy
and orbital angular momentum at the pericenter and apocenter of the
orbit are equal. This leads to the following conditions:
\begin{equation}
\upsilon_{p}^{2}-\upsilon_{a}^{2} = 2GM \left(\frac{1}{r_{p}-r_{s}} - \frac{1}{r_{a}-r_{s}} \right) \; , \label{eq_energy}
\end{equation}
\begin{equation}
r_{p}\upsilon_{p} =  r_{a}\upsilon_{a} \; , \label{eq_momentum}
\end{equation}
where $\upsilon_{p}$, $\upsilon_{a}$, $r_{p}$, and $r_{a}$ are the
velocity and radial distance of the reduced mass body, $\mu$, at the
pericenter and apocenter, respectively. From
equations~(\ref{eq_energy}) and (\ref{eq_momentum}) we derive the
velocity at the pericenter as
\begin{equation}
\upsilon_{p} = \sqrt{\frac{2GM}{r_{a}+r_{p}}~\frac{r_{a}^{2}}{(r_{p}-r_{s})(r_{a}-r_{s})} } \; .  \label{eq_velperi}
\end{equation}
From the expression for the orbital angular momentum, $L/\mu =
r_{p}\upsilon_{p} = r^{2}\dot{\theta}$ (where $r$ is the radial
distance of the reduced mass from the center of mass of the system,
and $\theta$ is its true anomaly), we find the angular azimuthal
velocity.
\begin{equation}
\dot{\theta} = \frac{1}{r^{2}} \sqrt{\frac{2GM}{r_{a}+r_{p}}~ \frac{r_{a}^2r_{p}^{2}}{(r_{p}-r_{s})(r_{a}-r_{s})}  } \; .  \label{eq_theta_dot}
\end{equation}
 We
derive the radial velocity component from the equation of an ellipse,
$r = a(1-e^{2})/(1-e\cos \theta)$,
\begin{equation}
\dot{r} = \frac{dr}{dt} = \frac{dr}{d\theta}\dot{\theta}= - \frac{e \sin \theta\, r^{2} \dot{\theta}}{a(1-e^{2})} \; .
\label{eq_r_dot}
\end{equation}
Using the equation~(\ref{eq_velperi}) and the condition
$E(r_{p},\upsilon_{p})=E(r,\upsilon)$ we find the {\it vis viva}
expression for an elliptical orbit in the Paczynsky-Wiita potential to
be
\begin{eqnarray}
\lefteqn{\upsilon = (2GM)^{1/2}\times}\nonumber \\
& & \left( \frac{1}{r-r_{s}} +  \frac{1}{r_{p}-r_{s}}~\frac{1}{(r_{a}/r_{p})^{2}-1}
 - \frac{1}{r_{a}-r_{s}}~\frac{1}{(r_{p}/r_{a})^{2}-1} \right)^{1/2} . \label{eq_vis_viva} 
\end{eqnarray}
As the binary orbits in Paczynsky-Wiita potential it loses a fraction
of its energy and orbital angular momentum through gravitational wave
emission. We assume that as the binary contracts by emission of
gravitational waves, the two black holes remain on elliptical orbits.
This assumption is justified in our simulations where the total mass
of the gas in a simulation is much smaller than the mass of a
binary. Consequently, the perturbations of the elliptical orbits of
black holes by the underlying gas distribution are small.

Following \citet{ll75} we calculate the rate of the energy loss to
gravitational radiation using the quadrupole approximation
\begin{equation}
\frac{dE}{dt} = - \frac{32G\mu^2\Omega^6r^4}{5c^5} \;,
\label{eq_dEdt}
\end{equation}
where $\Omega(r)$ is the orbital frequency in the Paczynsky-Wiita
potential. From this equation it is possible to obtain the radiation
reaction acceleration at the center of mass frame of the binary
\citep[][and references therein]{lk99} 
\begin{equation}
{\vec{\textit a}} = - \frac{1}{\mu} \frac{dE}{dt} \frac{\vec{\upsilon}}{\upsilon^2} \;.
\end{equation}
chosen so as to preserve the symmetry of forces acting on each member
of the system. This requirement ensures that the total momentum of the
system is conserved while the individual components recoil due to the
emission of gravitational waves.

\subsection{Heating and Cooling of the Gas}\label{S_cooling}
 
We have investigated two different scenarios for radiative heating and
cooling, assuming the emitting gas can be described as a black-body
(BB) or a solar metallicity gas. The option for radiative heating and
cooling of hydrogen-helium gas is also implemented in the code,
following \citet{tw96}, \citet{black81}, \citet{cen92}, and
\citet{kwh96}, but was not used in the runs reported here. The cooling
scenarios are mutually exclusive and by comparing them one can study
the importance and impact of various cooling mechanisms in our
simulations. In our simulations, the gas is heated by shocks and
``external'' sources of illumination. We assume that the sources of
illumination are powered by accretion onto the massive black
holes. After accretion on either of the two black holes becomes
significant, UV and soft X-ray radiation emitted from the innermost
portion of the accretion flow photo-ionizes the gas. The bolometric
luminosity of an accretion-powered source of ionizing radiation can be
written as:
\begin{equation} L_{acc} = \eta\,\frac{GM_{BH}~\dot{m}}{2r_{s}} = \frac{1}{2}\,\eta \,\dot{N}
m_{p} \,c^2 
\label{eq_bol_lum} 
\end{equation} 
where $\eta = 0.01$ is
the assumed radiative efficiency, $\dot{m}$ and $\dot{N}$ are the mass
and particle accretion rates, respectively, and $m_p$ is the mass of a
gas particle.
 
\begin{enumerate}

\item{\it Black-body case.} 

In this scenario, the illuminated gas absorbs and emits the radiation
as a black-body. Each gas particle is therefore treated as an
optically thick cell of gas, in local thermal equilibrium within a
radius equal to its smoothing length, $h_{sml}$. A fraction of the
incident energy absorbed by the gas depends on a coverage factor of
the gas $\zeta$, i.e., the fraction of solid angle covered by gas, as
seen from the position of a source, and the distance of an absorbing
gas cell from the source, $r$. From our simulations we find $\zeta
\sim 10^{-4}$, corresponding to a fraction of particles exposed to
illumination from an external source at any given time. We do not take
into account radiation pressure, i.e., the momentum that is
transferred to the gas particles by the incident radiation. We write
the expression for the heating rate of the gas per unit volume as
\begin{equation} 
\Gamma_{bb} = \frac{1}{2} \,\zeta \,
L_{X}\,\frac{\rho}{m_p} \left( \frac{h_{sml}}{r} \right)^2  \;\;\;
{\rm erg\;cm^{-3}\,s^{-1}}\; .
\label{eq_bb_heating} 
\end{equation} 
Here we assume that $L_X$ is the portion of the accretion luminosity
carried by ionizing photons ($L_X/L_{acc}\approx0.1$). The black-body
cooling rate per unit volume is evaluated from the Stefan-Boltzmann
equation and it can be written as
\begin{equation}
 \Lambda_{bb} = 4\pi h_{sml}^2\,\sigma T^4 \,\frac{\rho}{m_p}  \;\;\;
{\rm erg\;cm^{-3}\,s^{-1}} \; ,
\label{eq_stefan_boltzmann}
\end{equation}
where $\sigma$ is the Stefan-Boltzmann constant.

\item{\it Solar metallicity gas.\label{S_smg}}

Spectroscopic observations of AGN show a wide range of metal
absorption and emission lines, implying that most AGN systems are
metal-rich. Additionally, metals are important coolants under the
physical conditions typically found in AGN broad line
regions. Therefore, in order to correctly model and describe these
systems it is necessary to also include metals as constituents of the
gas. However, detailed calculations of radiative transfer, coupled
with parallel hydrodynamic simulations present a large computational
challenge and are currently not possible. Including metals in heating
and cooling calculations results in a large network of equations,
especially if non-LTE processes are included. The solution of the
heating and cooling equations dominates the CPU usage in a parallel
hydrodynamics code to the extent that the calculation becomes
prohibitively expensive.

Here we describe the approximate method we used for calculation of
heating and cooling of the gas with metals. We have constructed a
number of cooling maps using the photoionization code {\it Cloudy}
\citep{ferland}. Since the maps are used in computationally expensive
numerical calculations, it is impractical to call {\it Cloudy} every
time a cooling rate is needed. Instead, we pre-calculate a grid of
cooling maps over a wide range of parameters, where the range of
parameters has been determined from preliminary calculations without
cooling. The parameter grid is read in by the simulation code, and
radiative heating and cooling rates are linearly interpolated from the
existing grid points. The cooling maps are calculated in the parameter
space of {\it density} and {\it temperature} of the gas and {\it
  intensity} of ionizing radiation. The range of parameter values for
which the maps were computed is as follows: $10^9\,{\rm cm^{-3}} < n <
10^{19}\,{\rm cm^{-3}}$, $2000\,{\rm K} < T < 10^8\,$K, and $0\; {\rm
  erg \;cm^{-2} \, s^{-1}} < J < 10^{17}\, {\rm erg \;cm^{-2} \,
  s^{-1}}$. Because one of the assumptions in {\it Cloudy} is that the
electrons are non-relativistic, the present range of its validity
extends to temperatures below roughly $10^9$ K. Gas at higher
temperatures than this is commonly encountered in our simulations; in
such cases we calculate heating and cooling rates by linearly
extrapolating the grid values. In the SPH simulations we set the lower
threshold for the gas temperature to 100~K and assign no upper
threshold. The calculation in {\it Cloudy} is explicitly one
dimensional, with results depending only on the depth coordinate.
Propagation of line photons is conducted using the escape probability
approximation. In this approach the escape probability for a photon is
calculated under the assumption that the difference between the mean
intensity averaged over the line and the source function at some
location in the emitter is due to photons leaking away from the region
\citep{ferland03}.  Although this approximation is used in many
state-of-the-art plasma codes because of its computational facility,
it is not known to what extent it affects predictions of the line
intensities.

The density and temperature of the gas are evaluated in the SPH
simulation after every time step.  The spectral energy distribution
(SED) of the ionizing radiation is taken to be a power-law with an
index $\alpha = 1$, extending from $\nu_1=1.36~{\rm eV}$ to $\nu_2=
100~{\rm keV}$, a distribution commonly assumed for the central
engines of AGN.
\begin{equation}
J_{\nu} = J_0 \; \left(\frac {\nu}{\nu_0}\right)^{-\alpha} \;\;\; {\rm erg \;cm^{-2} \, s^{-1} \, Hz^{-1}} \;,
\label{eq_sed}
\end{equation}
where $\nu_0$ is some arbitrarily chosen frequency and $J_0 =
J(\nu_0)$ is the normalization of a SED that can be directly evaluated
from the luminosity of an accretion source, eq.~(\ref{eq_bol_lum}), as
follows
\begin{equation}
F_{acc} =  \frac{\zeta L_{acc}}{4\pi r^2} = \int_{\nu_1}^{\nu_2} J_{\nu}\, d\nu \approx 11.2 \; 
J_0 \nu_0 \; ,
\label{eq_sed_normalization}
\end{equation}
where $F_{acc}$ is the total flux incident on the gas cell located at
the distance $r$ from the ionizing source. In addition to the cooling
processes included in {\it Cloudy} we also consider Compton cooling from a
thermal distribution of non-relativistic electrons, which may become a
significant coolant for the hottest and densest gas in the nuclear
region.
\begin{eqnarray}
\lefteqn{\Lambda_{Comp} = 4.48\times 10^{-5} \left(\frac{T}{10^9\,{\rm K}}\right)\left 
(\frac{n_e}{10^{10}\,{\rm cm^{-3}}}\right)} \nonumber \\
& &\times\left(\frac{F_{acc}}{10^{10}\, 
{\rm erg \;cm^{-2} \, s^{-1}}}\right) \;\;\;{\rm erg\;cm^{-3}\,s^{-1}}\;,
\label{eq_compton_cooling}
\end{eqnarray}
where $n_e$ is the electron number density. When information about the
cooling rate per unit volume, $\Lambda_k = \Lambda_{Cloudy, k} +
\Lambda_{Comp, k}$ , is obtained for every gas particle $k$ in the
simulation, their internal energy is modified accordingly. Thus, the
heating and cooling are directly coupled with the evolution of the
physical conditions in the gas. The total cooling rate from the gas
(i.e., the bolometric luminosity) is obtained by summing over all
particles, $\sum_{k=1}^{Npart}\Lambda_k \, (m_p/\rho_k)$. We do not
assume that the gas is in thermal equilibrium, and consequently
cooling is not necessarily balanced by heating. The integration scheme
used in implementation of cooling for both the black-body and solar
metallicity models is explicit. The cooling time for individual gas
particles in our simulations is typically longer than the time step
assigned based on their acceleration, $dt\propto 1/a_p$, and the
Courant condition.  Even when the cooling time is very short, $dt \leq
10^2\,{\rm s} < t_{cool}$, guarantying the numerical stability of the
integration.  In all of the calculations the default abundance pattern
is assumed to be solar. The atomic species considered by {\it Cloudy}
range from hydrogen to zinc.

Since {\it Cloudy} includes a detailed treatment of radiation
processes, we have also used it to calculate the H$\alpha$ line
intensity and optical depth, the electron scattering optical depth,
and the neutral hydrogen column density for each gas cell. As we will
describe in the next section this information allowed us to relax some
of the important assumptions commonly used to calculate the H$\alpha$
light curves and emission line profiles.

\end{enumerate}

\subsection{Calculation of H$\alpha$ light curves and emission line profiles}\label{S_attenuation}

\begin{deluxetable*}{ccccccccccccc}[t]
\tablecaption{Model Parameters\label{T_sim_params1}}
\tablewidth{0pt}
\tablecolumns{13}
\tablehead{
\colhead{$M_1$} & 
\colhead{$M_2$} & 
\colhead{$M_{disk}$} & 
\colhead{$\psi$} & 
\colhead{$P_{orb}\tablenotemark{\rm b}$} &
\colhead{$a$\tablenotemark{\rm b}} &
\colhead{$e$\tablenotemark{\rm b}} &
\colhead{$\theta$ \tablenotemark{\rm b}} &
\colhead{$\alpha_{G}$\tablenotemark{\rm c}} &
\colhead{$\eta$\tablenotemark{\rm d}} &
\colhead{$\xi_{in}$} & 
\colhead{{\it i}} &
\colhead{$\phi_{0}$} \\
\colhead{($\Msun$)} & 
\colhead{($\Msun$)} & 
\colhead{($\Msun$)} &
\colhead{} &   
\colhead{(yr)} &
\colhead{($r_g$)} &
\colhead{} &
\colhead{($\circ$)} &
\colhead{} & 
\colhead{} &
\colhead{($r_g$)} &
\colhead{($\circ$)} &
\colhead{($\circ$)}}
\startdata
$10^{8}$ & $10^{7}$ & $10^{4}$ & PW\tablenotemark{\rm a} & 15.7 & 3007 
& 0.7 & 0 (30\tablenotemark{\rm e}) & $10^{-6}$ & $10^{-2}$ & $10^2$  & 30 & 0  \\  
\enddata
\tablenotetext{a}{PW = Paczynsky-Wiita potential; equation~(\ref{eq_pw_potential}).}
\tablenotetext{b}{The initial values of binary orbital parameters.}
\tablenotetext{c}{Viscosity parameter; equations~(\ref{eq_visc_micro}) and 
(\ref{eq_visc_macro}).}
\tablenotetext{d}{Radiative efficiency; equation~(\ref{eq_bol_lum}).}
\tablenotetext{e}{in case of counter-rotating binary (SR) model only.}
\end{deluxetable*}

We chose the Balmer series H$\alpha$ line ($\lambda_{rest}=6563\,{\rm
\AA}$) to describe the emission signatures of a MBHB. The broad
H$\alpha$ line is thought to reflect the kinematics of the gas in the
broad line region of AGN \citep{sulentic00}, and it is typically found
to emerge from a wider range of radii in the broad line region than
the UV and X-ray band broad emission lines
\citep[e.g.,][]{ehl96,wandel99,sulentic00}. Also, the H$\alpha$ line
is the most prominent broad line in the optical spectrum and the least
contaminated by the neighboring narrow emission lines. It is mostly
powered by illumination of a disk by an emission source
\citep{CSD}. The emission sources in our simulations are the innermost
regions of the accretion flow around the two massive black holes that
give rise to ionizing UV and soft X-ray radiation \citep{shakura}. The
cool gas is optically thick to the incident UV/X-ray photons which can
escape the disk only after they have been converted to photons of
lower energy. One portion of the absorbed UV/X-ray radiation is
reprocessed by the gaseous disk and re-emitted as the broad H$\alpha$
line \citep[][see their discussion of broad recombination and
resonance emission lines]{shakura}.

The most important consideration in calculating the emitted H$\alpha$
light is the efficiency with which the gas reprocesses the incident
ionizing radiation and re-emits it in the H$\alpha$ band.  This
efficiency is commonly characterized by the surface emissivity of the
gas. The emissivity of the photoionized gas depends on numerous
physical parameters. Locally, it depends on the physical properties of
gas. Globally, its spatial distribution depends on the structure and
morphology of the accretion disk. The H$\alpha$ emissivity can be most
accurately assessed with photoionization calculations, thus we used
this approach in the case of a {\it solar metallicity gas}.

\begin{figure}[b] 
\epsscale{1.25}
\plotone{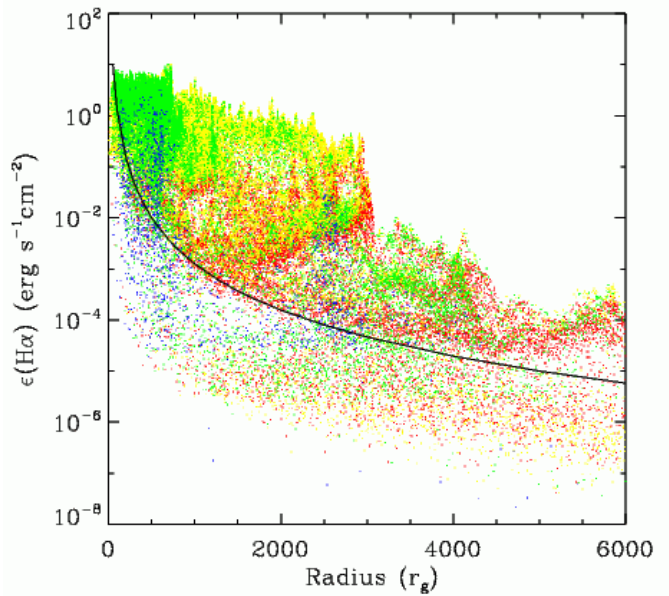}
\figcaption[b1.ps]{H$\alpha$ emissivity of the gas as a function of 
radius, at 9.4 years for a model with solar metallicity gas. The
emissivity of each gas cell plotted in the figure is weighted by the
density of the gas at that position, so that comparison with the
parametric emissivity model (plotted as a solid line with arbitrary
normalization) can be made. The temperatures of gas particles are
marked with color: red $T < 10^4$ K; yellow $10^4 {\rm K} < T< 10^6$
K; green $10^6 {\rm K}< T < 10^8$ K; blue $10^8 {\rm K} < T < 10^{10}$
K; and violet $T > 10^{10}$ K. This figure corresponds to the
morphology of the disk plotted in panel 2 of Figure~\ref{fig_xy_sr}.
\label{fig_haEmiss}}
\end{figure}

In cases when photoionization calculations are not available, the
emissivity is often parametrized as a function of radius,
$\epsilon=\epsilon_{0}\,\xi^{-q}$, where $q\approx3$. This is
justified by the photoionization calculations of \citet{CSD}, which
suggest that the emitted H$\alpha$ flux is approximately proportional
to the incident continuum flux for a wide range of values of the
density and column density. We have compared the emissivity determined
from our own photoionization calculations with the parametric model
(Figure~\ref{fig_haEmiss}) and found a qualitative agreement between
the two. We also notice a significant amount of scatter in the
emissivity values from the photoionization calculation, which is
expected due to variations in surface density of the perturbed gas
disk in our model.

\begin{figure*}[t] 
\epsscale{0.8}
\plotone{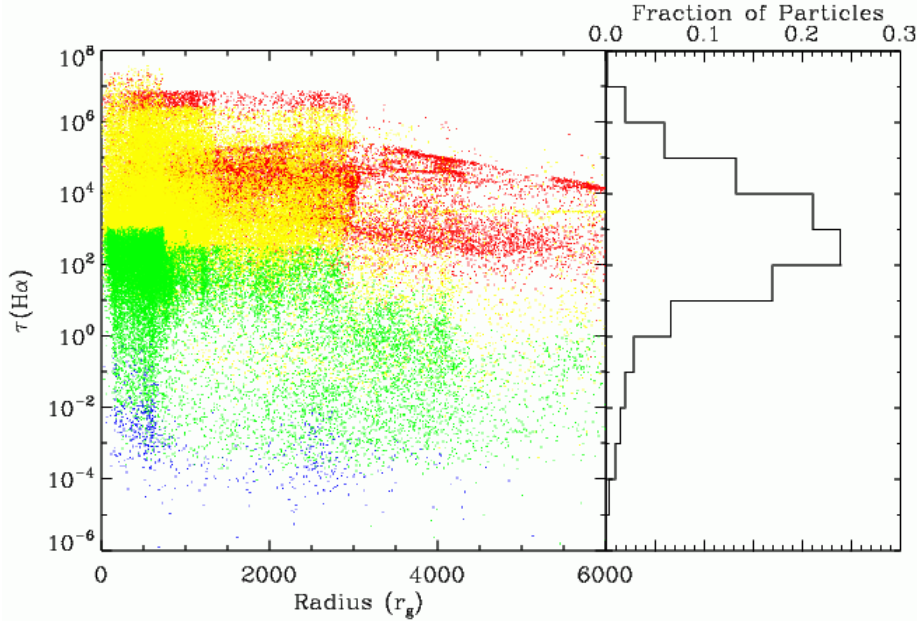}
\figcaption[b2.ps]{{\it Left:} Optical depth of gas cells to 
H$\alpha$ photons as a function of radius at 9.4 years for a model
with solar metallicity gas. The color legend is the same as in
Figure~\ref{fig_haEmiss}. This figure corresponds to the morphology of
the disk as shown in panel 2 of Figure~\ref{fig_xy_sr}. {\it Right:}
Histogram showing the fraction of particles as a
function of $\tau$(H$\alpha$).
\label{fig_tauHa}}
\end{figure*}

Once the surface emissivity of each gas cell in the simulation is
known, the H$\alpha$ luminosity can be expressed as
$L_{H\alpha}\propto\sum_{k=1}^{Npart}\epsilon_{k}$, where $k$ is a
particle index. The implicit assumption, following from the above
luminosity expression, is that the gas is optically thin to H$\alpha$
photons. We find this assumption to be justified for a cold disk since
its optical depth to H$\alpha$ light is less than $10^{-4}$. We show
the optical depth of gas cells to the H$\alpha$ photons as a function
of the disk radius in Figure~\ref{fig_tauHa}. When the gas is exposed
to an intense ionizing continuum, the optical depth to the H$\alpha$
photons increases significantly because many hydrogen atoms are in the
first excited energy state. However, in our profile calculations we do
not correct for the absorption of the H$\alpha$ photons by the
gas. This is because we are not able to correct consistently for
absorption of photons along the line of sight with the numerical code
we have used. The effects of our assumptions on the H$\alpha$
luminosity and emission line profiles are discussed in
\S~\ref{S_Halpha} of the text.

The H$\alpha$ emission line profiles have been calculated taking into
account the relativistic Doppler shift, $\nu_{obs} =
\nu_{rest}\,\sqrt{1-\beta^2}/(1+ \beta\cos \theta)$, and the
gravitational redshift in the potential well of a Schwarzschild black
hole, $\nu_{obs} = \nu_{rest} \sqrt{1-2/\xi}$, where $\nu_{obs}$ and
$\nu_{rest}$ are the observed and rest frequencies of an emitted
photon, $\beta$ is the velocity of an emitter in units of speed of
light, and $\theta$ is an angle between the direction of motion of an
emitter and the observer's line of sight. In the next section we
summarize our choice of parameters used in SPH simulations as well as
the calculation of light curves and line profiles.

\subsection{Choice of Parameters}\label{S_parameters}

In the calculations presented here, both black holes are treated as
massive particles in the pseudo-Newtonian Paczynsky-Wiita
potential. The masses of the primary and secondary black holes,
together with the other parameters are shown in
Table~\ref{T_sim_params1}. In all simulations the binary mass ratio is
chosen to be $0.1$.

Based on current theoretical and observational understanding of MBHBs
it is not obvious what kind of orbits the binaries have in the
intermediate phase of their evolution. The resulting accretion rate
and observational signature of a MBHB depend sensitively on the shape
and orientation of its orbit with respect to the gas disk. The
astrophysical significance of a predicted signature depends on the
longevity of the particular phase in the life of a binary and
consequently on the probability that binaries will be detected in that
evolutionary phase \citep{mm06}. The recent simulations of
\citet{dotti06a} imply that binaries which co-rotate with the gaseous
disk circularize and loose the memory of their initial orbital
eccentricity. On the other hand, counter-rotating binaries tend to
preserve their original eccentricity down to $\approx1$ pc, a limit
set by the resolution of their simulation. In addition, simulations of
gas disks with planets suggest that under some conditions tidally
perturbed eccentric disks can excite the eccentricity in a binary and
that the two are coupled \citep*{pnm01}. Very few constraints on the
type of orbit can be placed from the observations. An exception is the
MBHB candidate OJ~287, for which it has been suggested that the binary
orbit with and eccentricity of 0.7$-$0.8 and orbital period close to
12 yr is coplanar with the accretion disk
\citep{sillanpaa,valtaoja,liu02}.

In the calculations presented here the primary and secondary black
holes are moving on elliptical orbits.  At the beginning of a
simulation the black holes start from the apocenters of their orbits
(except in the model where the secondary starts with a true anomaly of
$\theta=30^{\circ}$, see \S~\ref{S_SR-model}). The initial values of
the semi-major axis and eccentricity are 3007$\;r_g$ and 0.7,
respectively. The period of the binary, calculated at the beginning of
the simulation is 15.67 yr.

\begin{figure}[b] 
  \epsscale{1.15} 
  \plotone{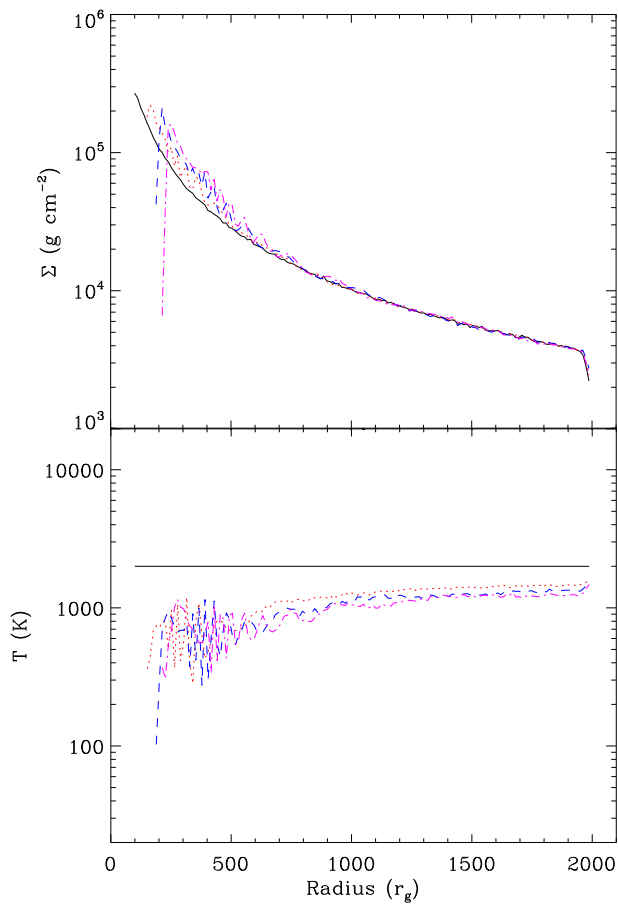}
  \figcaption[b3.ps]{Evolution of the disk surface density and
    temperature profile before the secondary black hole is introduced
    into the simulation (IC2 run). The profiles correspond to
    progressively longer times as follows: 0 years ({\it solid, black
      line}), 2.2 years ({\it dotted, red line}), 4.4 years ({\it
      dashed, blue line}), and 6.6 years ({\it dash-dot, purple line})
    after the beginning of the simulation. \label{fig_surfden_temp}}
\end{figure}

\begin{deluxetable*}{cccccccc}[t] 
\tablecaption{Results\label{T_results}}
\tablewidth{0pt}
\tablecolumns{8}
\tablehead{
\colhead{} &
\colhead{} &
\colhead{} &
\colhead{} &
\colhead{} & 
\colhead{} &
\colhead{} &
\colhead{} \\
\colhead{Model} &
\colhead{Cooling} &
\colhead{$N_{part}$} &
\colhead{Duration} &
\colhead{$t_{peri}$\tablenotemark{\rm a}} & 
\colhead{Density \tablenotemark{\rm b}} &
\colhead{$c_s$\tablenotemark{\rm b}} &
\colhead{$L_{H\alpha}$ \tablenotemark{\rm c}} \\
\colhead{} &
\colhead{} &
\colhead{} &
\colhead{(yr)} &
\colhead{(yr)} &
\colhead{(${\rm cm^{-3}}$)}&
\colhead{(${\rm km\,s^{-1}}$)} &
\colhead{(${\rm erg\,s^{-1}}$)}}
\startdata
 BB & black-body & 100k  & 37 & 8, 23 & $10^7$, $10^{12}$ & $5$, $10^3$ & $\ldots$ \\ 
 S & solar  & 100k  & 36 & 8, 23 & $10^8$, $10^{13}$ & $5$, $10^2$ & $\sim 10^{40}$ \\ 
 SR & solar & 100k & 32 & 3, 18 & $10^8$, $10^{13}$ & $5$, $10^2$ & $\sim 10^{40}$ \\ 
 S1 & solar  & 20k  & 50 & 8, 23, 38 & $10^8$, $10^{13}$ & $5$, $10^2$ &  $\sim 10^{40}$ \\ 
 S2 & solar  & 20k  & 38 & 8, 23, 38 & $10^8$, $10^{13}$ & $5$, $10^2$ & $\sim 10^{40}$ \\ 
 IC1 & solar  & 20k   & 8 & $\ldots$ & $10^{13}$ & $5$ & $\ldots$ \\ 
 IC2 & solar  & 100k  & 8 & $\ldots$ & $10^{13}$ & $5$ &  $\ldots$ \\ 
 IC3 & solar  & 500k  & 8 & $\ldots$ & $10^{13}$ & $5$ &  $\ldots$ \\ 
\enddata
\tablenotetext{a}{Approximate times of pericentric passages.}
\tablenotetext{b}{Median value for the low and high density component. IC runs: median value for the unperturbed disk.}
\tablenotetext{c}{Estimated peak luminosity.}
\end{deluxetable*}

The accretion disk is initially only associated with the primary black
hole. It is coplanar with the binary orbit and it extends from
$\xi_{in}=100$ to $\xi_{out}=2000$ ($\xi_{out}$ translates to $0.01$
pc in physical units, a size expected for AGN nuclear accretion
disks). The mass of the disk is $10^{4}\Msun$ and the number of gas
particles in the disk is $10^5$. The mass of the thin disk is
constrained by the requirement that its surface density be in the
range that is commonly found in AGN broad line regions, where the
broad H$\alpha$ emission line is expected to originate. In addition,
the disk mass is roughly consistent with the results of the larger
scale simulations \citep{escala2,dotti06a,mayer06}, where the mass of
the gas disk enclosed within central 1~pc is of the order of the mass
of the binary. Extrapolating from these results and assuming a
constant surface density for simplicity, one obtains a disk mass of
$\sim 10^4 \Msun$, for a disk size of $0.01$~pc and a $\sim
10^8\,\Msun$ binary. In \S~\ref{S_discussion}, we discuss how the
choice of the disk mass may affect the observational signatures.

\begin{figure*}[t] 
\epsscale{1.0} 
\plottwo{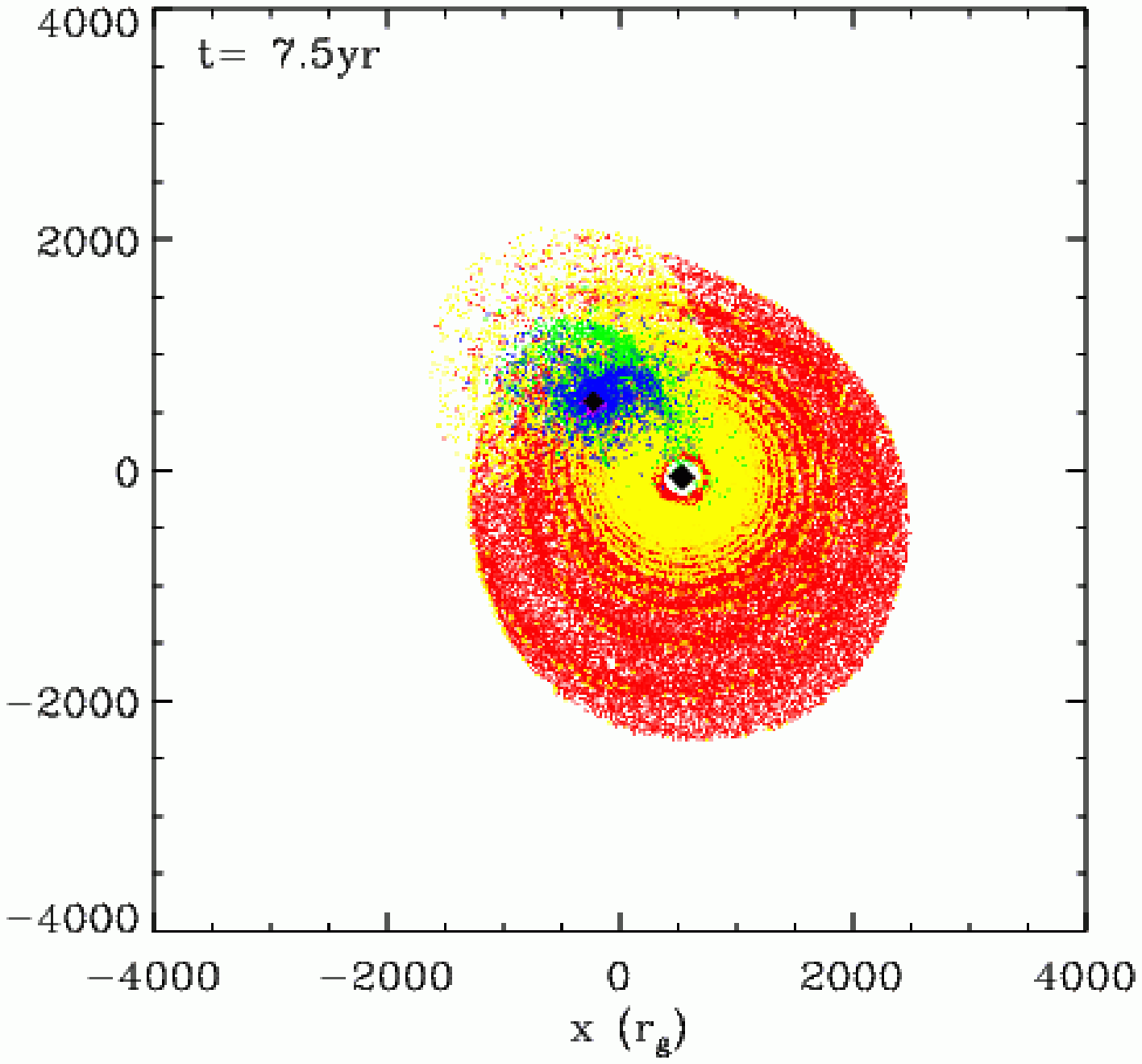}{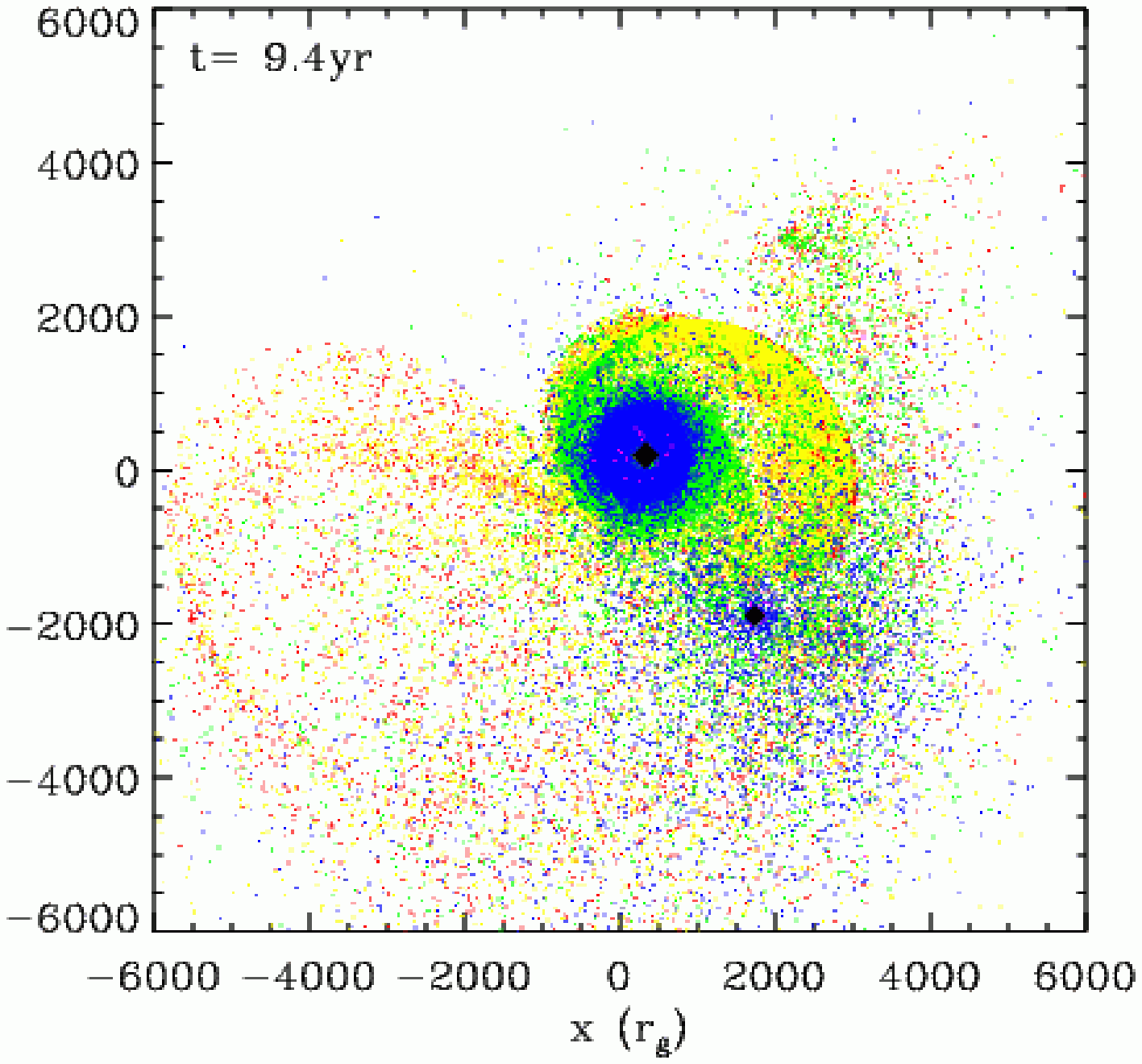}
\plottwo{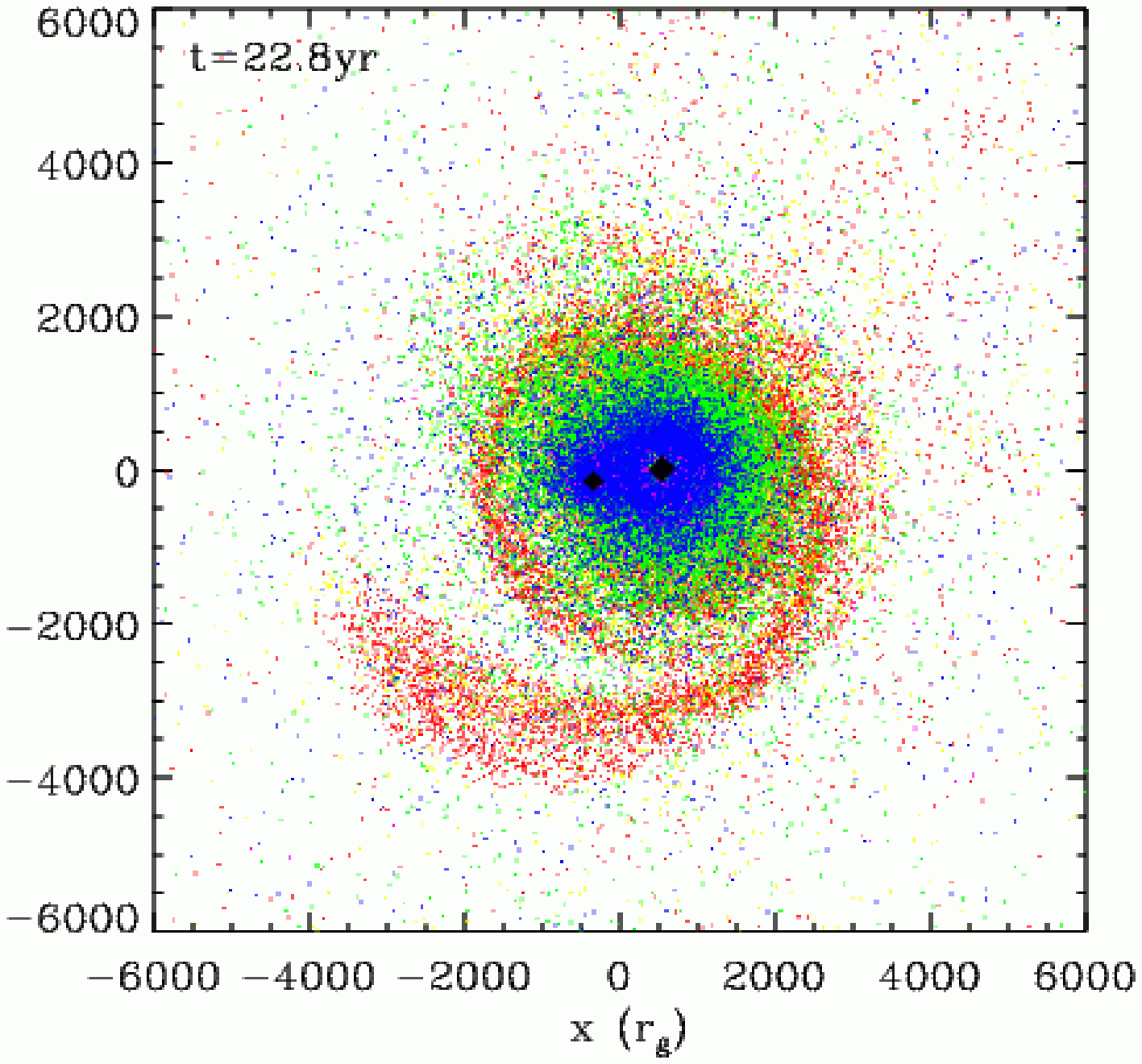}{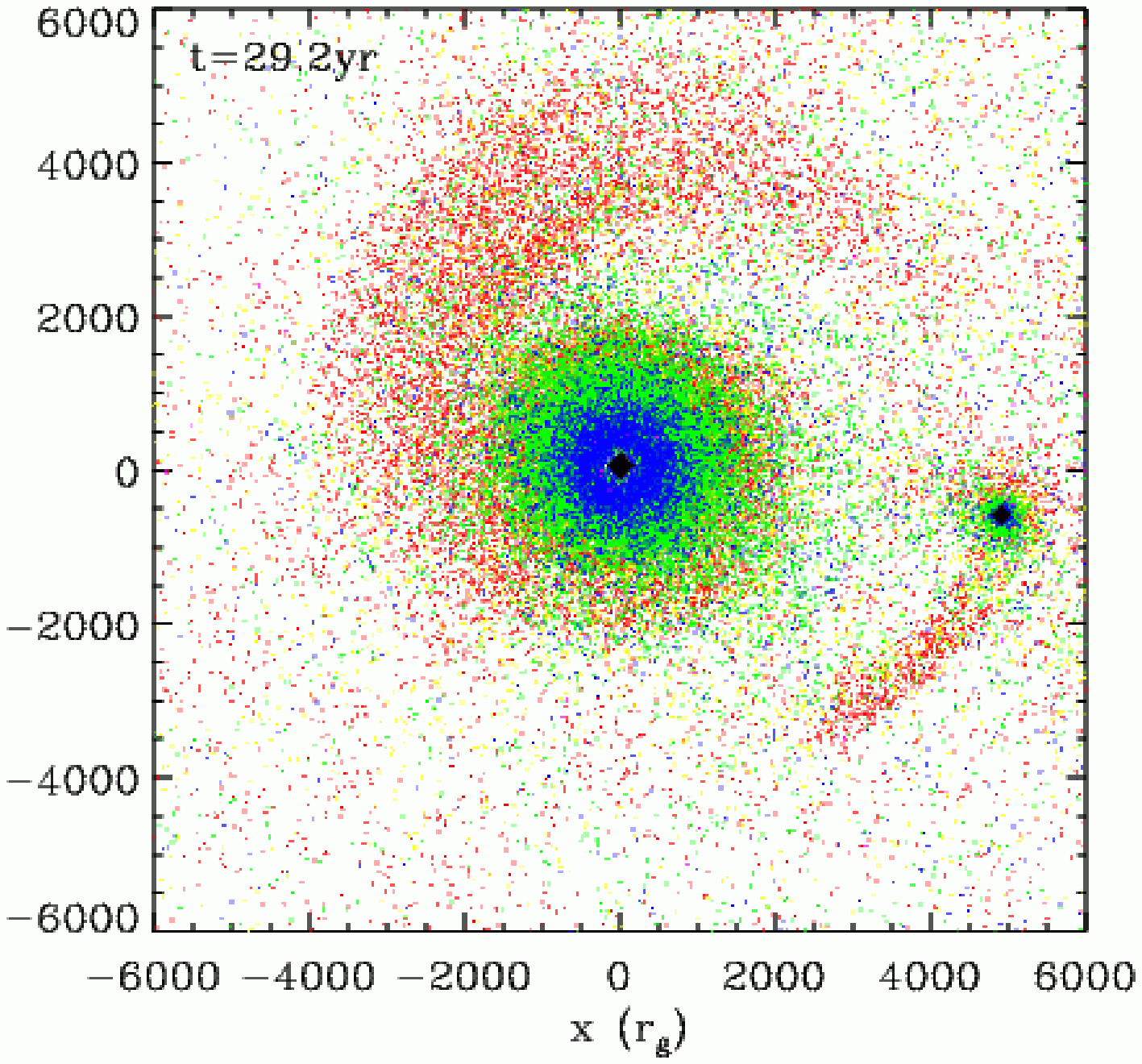}
\figcaption[b4.ps]{Sequence of snapshots from the simulation showing
the evolution of the binary and gas in the BB-model (projected into
the plane of the binary orbit). The time stamp is marked on the top of
each panel. The first snapshot shows the system shortly after the
secondary plunges into the disk for the first time. The rotation of
the binary and the disk is counter-clockwise. The color legend is the
same as in Figure~\ref{fig_haEmiss}. Note that higher temperature
particles are plotted over the lower temperature ones, with the result
that some information is hidden.  \label{fig_xy_bb}}
\end{figure*}

 The initial rotational velocities of gas particles were calculated
based on the gravitational forces only while the effect of pressure
was neglected in initially cold and unperturbed disk. We choose the
initial temperature of the disk to be $T=2000$~K everywhere. The
initial surface density distribution in the disk is
$\Sigma\propto\xi^{-1.5}$, as set by the requirement that the disk has
a uniform vortensity (defined as the ratio of vorticity to surface
density). This criterion must be satisfied in order to construct the
equilibrium model of an infinitesimally thin gaseous disk. The initial
conditions have been chosen in such way that the disk remains near
hydrostatic equilibrium before the secondary black hole plunges into
it. Figure~\ref{fig_surfden_temp} shows the evolution of the disk
surface density and temperature profile, from the initial conditions
run with 100k particles (IC2) before the secondary black hole is
introduced. The gas in the IC2 run is assumed to have a solar
metallicity. The temperature of the disk settles around $1000$~K in
this period and there are no visible transients.  No boundary
conditions were applied at the inner and outer edge of the disk in our
simulations. Figure~\ref{fig_surfden_temp} shows a gradual evolution
of the disk edges towards the regions of low density. This is a
consequence of the fact that the disk has sharp edges, where particles
are subjected to asymmetric pressure forces. We do not expect the
diffusion of particles from inner edge towards the primary black hole
to affect the results, since no particles are accreted during this
initial period. At later times, in runs where the secondary interacts
with the disk, the contribution to the accretion rate from a slow
diffusion process should be negligible in comparison with the total
accretion rate induced by the interactions.

The viscosity parameter, radiative efficiency, and metallicity of the
gas are kept constant, as described above; their values can be found
in Table~\ref{T_sim_params1}. The softening parameter, $\xi_{soft}$,
used in the expression for force in order to avoid numerical
singularity, ranges between 0.1 and 0.2 (in units of $M_8$) for gas
particles and is 4.0 for the massive black holes.

We compute line profiles emerging from the gaseous disk under the
assumption that the observer is located at a distance $d\to \infty$,
in the positive $xz$-plane, at $i=30^{\circ}$ to the
$z$-axis. Changing the inclination, changes the values of the
projected velocities (i.e., the overall width of the line profile) but
has a relatively small effect on its shape otherwise. The azimuthal
orientation of the observer is an additional parameter in the modeling
of non-axisymmetric systems. The azimuthal orientation is measured in
the orbital plane of a black hole binary, counter-clockwise with
respect to the positive $x$-axis. In the calculations we present here,
we adopt $\phi_{0}=0^{\circ}$ as an arbitrary direction towards the
observer.

\section{Results}\label{S_results}

We present and compare results of three different scenarios in which
the evolution of a massive black hole binary and coplanar gaseous
accretion disk is followed over two orbital periods of the binary
($\sim$ 30 years in total). The simulations span only two orbits
because at the temporal resolution required by the hydrodynamical
calculation, we find it computationally expensive to follow a binary
over many orbital periods. The calculation of the two orbits, with
total of $10^5$ particles, required about 20 days of CPU time on 8
2.8GHz Intel Xeon processors per simulation. We find that 50-70$\%$ of
the CPU time is consumed by the hydrodynamic calculations, including
the neighbor search, density determination, computation of hydro
forces, and heating and cooling of the gas. A fraction of 20-30$\%$ of
the CPU time is spent on computation of gravitational forces,
including the tree walks and tree construction. We find 8 CPUs to be
an optimal number of processors in terms of speed for calculations
that employ $10^5$ particles. Increasing the number of processors as
$2^n$ (i.e., to 16, 32, etc.) results in a loss of computational speed
because the CPU time consumption becomes dominated by the
communication among the processors.

In the first of three calculated scenarios the binary is co-rotating
with the accretion disk which is assumed to absorb and emit radiation
as a {\it black-body}. In the second and third scenario the binary is
co-rotating and counter-rotating with respect to the disk,
respectively, and the gas is assumed to have a {\it solar
metallicity}. In the last two cases the heating and cooling of gas are
followed with photoionization calculations.

\break

\subsection{Co-Rotating Case With Black-Body Cooling (BB-model)}\label{S_BB-model} 

The approximation in which emission from a geometrically thin
accretion disk is represented by a modified black-body spectrum has
been widely utilized since the work of \citet{shakura}. We use this
simple model as a control case for our simulations and we compare it
with the photoionization models.

Before the plunge of the secondary black hole into the disk the
density in the quasi-steady disk ranges between $10^{15}$ and
$10^{12}\;{\rm cm^{-3}}$ at the innermost and outermost edges of the
disk, and in the intermediate region it takes values according to the
assumed power-law surface density. The temperature of the quiescent
disk quickly adjusts to a two phase distribution. The higher density
gas, which occupies the inner portion of the disk cools more
efficiently (eq.~\ref{eq_stefan_boltzmann}) and reaches a temperature
of $\sim10^2\,$K (Fig~\ref{fig_xy_bb}). The lower density gas occupies
a distinct phase in temperature space at $\sim10^3\,$K. We are not
able to follow the phase of steady accretion at low rates prior to the
impact of the secondary. This is because our accretion disk is
initially set with an inner edge at $\xi_{in}=100$, where viscous
migration of gas in the radial direction is slow. For the set of
parameters we have adopted, the theory of steady thin disks predicts
an accretion rate of $10^{-4}-10^{-3}\,\Msun\,{\rm yr^{-1}}$. The
unperturbed disk is stable to gravitational collapse as determined by
the value of the Toomre parameter, $Q = c_s\,\Omega /G\Sigma$, where
$\Omega$ is the angular speed of the gas and $\Sigma$ is its surface
density. The innermost region of the gas disk has the lowest Toomre
parameter, of order of a few but consistently larger than 1. The speed
of sound, $c_{s}=(\gamma\,p/\rho)^{1/2}$, also shows a characteristic
two phase distribution with values lower than $10\; {\rm km\,s^{-1}}$
across the disk, where $\gamma=5/3$ for a mono-atomic, ideal gas. The
physical properties of the disk remain stable and do not change over a
considerable period of simulation time, until the secondary black hole
plunges into it. This is an indication that the disk has settled into
a quasi-steady state and that any further physical changes can be
attributed to the interactions with the binary.
 
The tidal perturbation by the secondary causes the disk to become
eccentric and develop a spiral pattern, even before the secondary
plunges into it. The physical properties of the disk are significantly
altered after the plunge, which occurs $\sim 6$ years after the
beginning of the simulation. The secondary forms a shock where the
temperature persists at $10^{12}$ K. While in this model the cooling
of the gas is less efficient, there is no doubt that the shock has
formed and that it trails behind the secondary as it travels through
the disk.

\begin{figure}[t]
\epsscale{1.2} 
\plotone{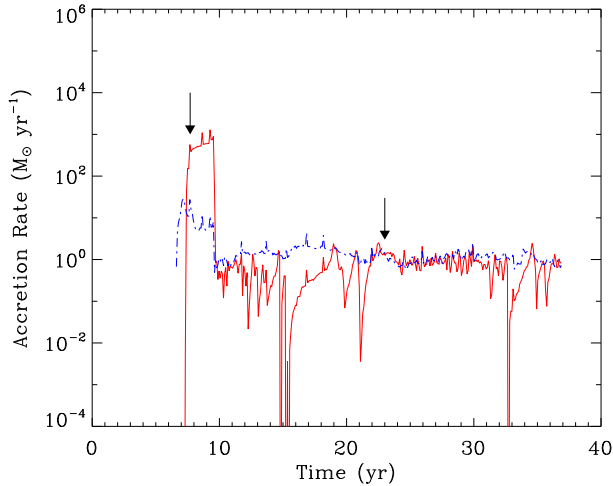} 
\figcaption[b5.ps]{Effective accretion
rate onto the primary ({\it solid, red line}) and secondary ({\it
dashed, blue line}) black holes calculated from the BB-model. The
accretion rate curves can be translated into UV/X-ray light curves, by
assuming that $1\Msun\, {\rm yr^{-1}}$ produces $\sim 10^{43}\,{\rm
erg\,s^{-1}}$ of UV/X-ray luminosity. The arrows mark the times of
pericentric passages of the binary.
\label{fig_acc_bb}}
\end{figure}

In our model, a source of illumination is assumed to turn on once the
accretion onto a black hole begins. With both black holes accreting
there are two sources illuminating and heating the gas.  As a result a
portion of the gas is blown out of the plane of the disk and forms a
hot ``corona''. The temperature of the gas is highest in the vicinity
of the black holes, where it is in the range of $10^6-10^{10}$
K. After the first passage of the secondary the disk has formed a
single trailing spiral which cools radiatively and adiabatically to
T$\sim 10^3$\,K (Fig~\ref{fig_xy_bb}). The secondary black hole
acquires a small, gravitationally bound corona of gas from the
disk. After the second passage the secondary further heats and
perturbs the disk gas which now reaches the scale of a parsec in the
plane of the disk (recall the disk initial size 0.01~pc). The
secondary black hole intersects the spiral arm of the disk and
gravitationally captures a portion of it. For a while, the secondary
moves on its orbit followed by the remainder of the spiral arm, which
resembles a cometary tail. Since a large portion of the spiral arm,
which retains a significant amount of colder gas in the disk, is
disrupted after the second orbital passage, the effective temperature
of the gas becomes more dominated by the inner hot portion of the
disk. The morphology of the system at the completion of the second
binary orbit consists of a discernible denser disk component immersed
into a spherical halo of tenuous gas. The gas density now spans the
range from $10^{10}-10^{13}\; {\rm cm^{-3}}$ in the disk component and
it extends down to $10^5\; {\rm cm^{-3}}$ in the low density regions
on the scale of $\sim 1$ pc.  The speed of sound is $\sim 5\; {\rm
km\,s^{-1}}$ and $\sim 10^3\; {\rm km\,s^{-1}}$ for the cold and hot
gas phases, respectively (Table~\ref{T_results}).

\begin{figure} 
\epsscale{1.2}
\plotone{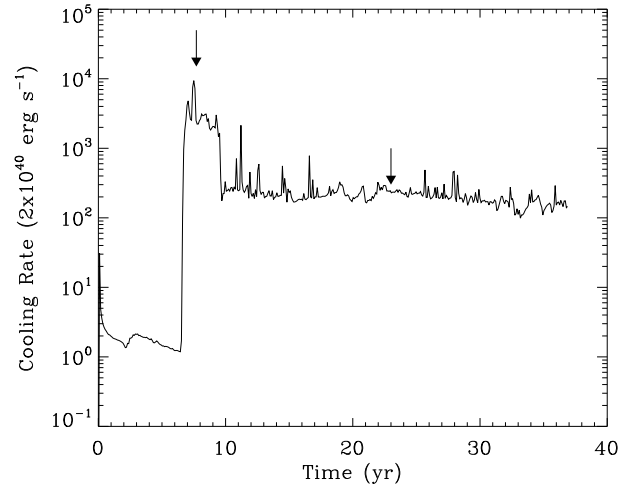}
\figcaption[b6.ps]{Cooling rate of the gas as a function of time 
calculated from the BB-model. The cooling rate shown here traces the
bolometric light curve of the gas disk (the power generated locally in the disk). 
The arrows mark the times of pericentric passages of the binary.
\label{fig_cool_bb}}
\end{figure}

During the two orbits the accretion rate onto the two black holes is
highly variable. The secondary black hole starts accreting soon after
the plunge and is followed by the primary which starts accreting a
year later at a much higher, super-Eddington rate $\dot{M}\sim 10^3
\Msun\; {\rm yr^{-1}}$ (Fig~\ref{fig_acc_bb}). The jump in accretion
rate is caused by a sudden increase in outward angular momentum
transport in the disk triggered by a gravitational perturbation of the
disk by the secondary. We observe such a high accretion rate in this
simulation only. This happens because the high density gas in the
central portion of the disk in the BB cooling model initially
equilibrates at lower temperatures than in the photoionization
models. The cold gas with a very low velocity dispersion is more
easily driven towards the primary black hole when the secondary
reaches the pericenter, at the end of year 7 of the simulation. The
second pericentric passage of the secondary occurs 23 years after the
beginning of the simulation. During the second passage there is no
large increase in the accretion rate because the gas has acquired a
significant amount of internal energy and additional kinetic energy
and it is not easily swept up by the secondary. The accretion rates
onto the primary and secondary black holes become comparable later in
the simulation. Unlike the secondary, the primary black hole exhibits
large variations in the accretion rate ($\sim$2 orders of
magnitude). This occurs because the primary black hole accretes colder
gas, confined to the plane of the disk, while the secondary black hole
accretes from the hotter, more uniform halo as it sweeps through it,
in a process resembling Bondi-Hoyle accretion. If the accretion
continues at the mean rate derived from the first two orbits the disk
mass would be exhausted in only $\sim$300 years. In order for
accretion to persist the accretion disk needs to be continually
replenished by an external supply of gas.

\begin{figure*}[t]
\epsscale{1.0} 
\plottwo{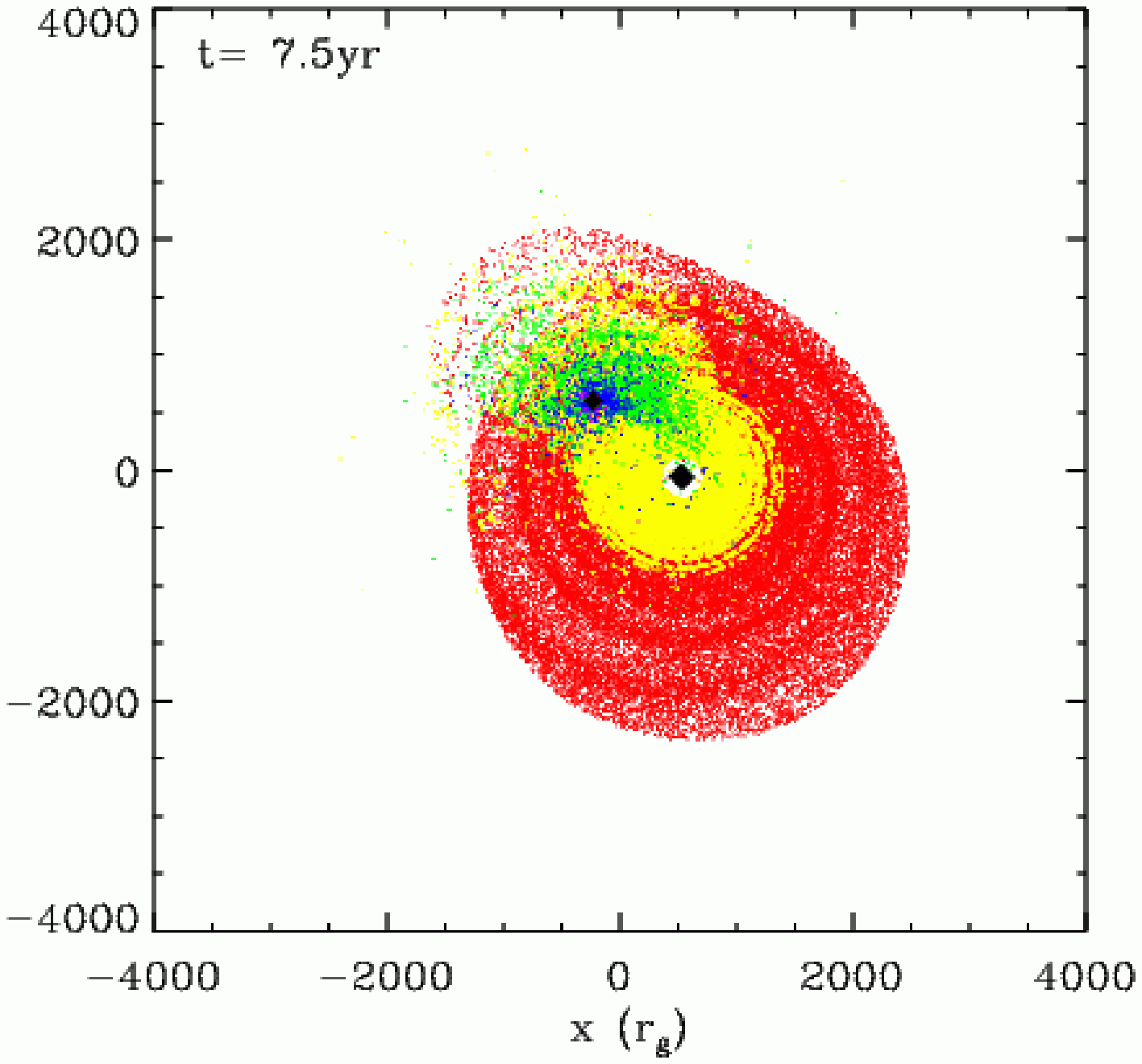}{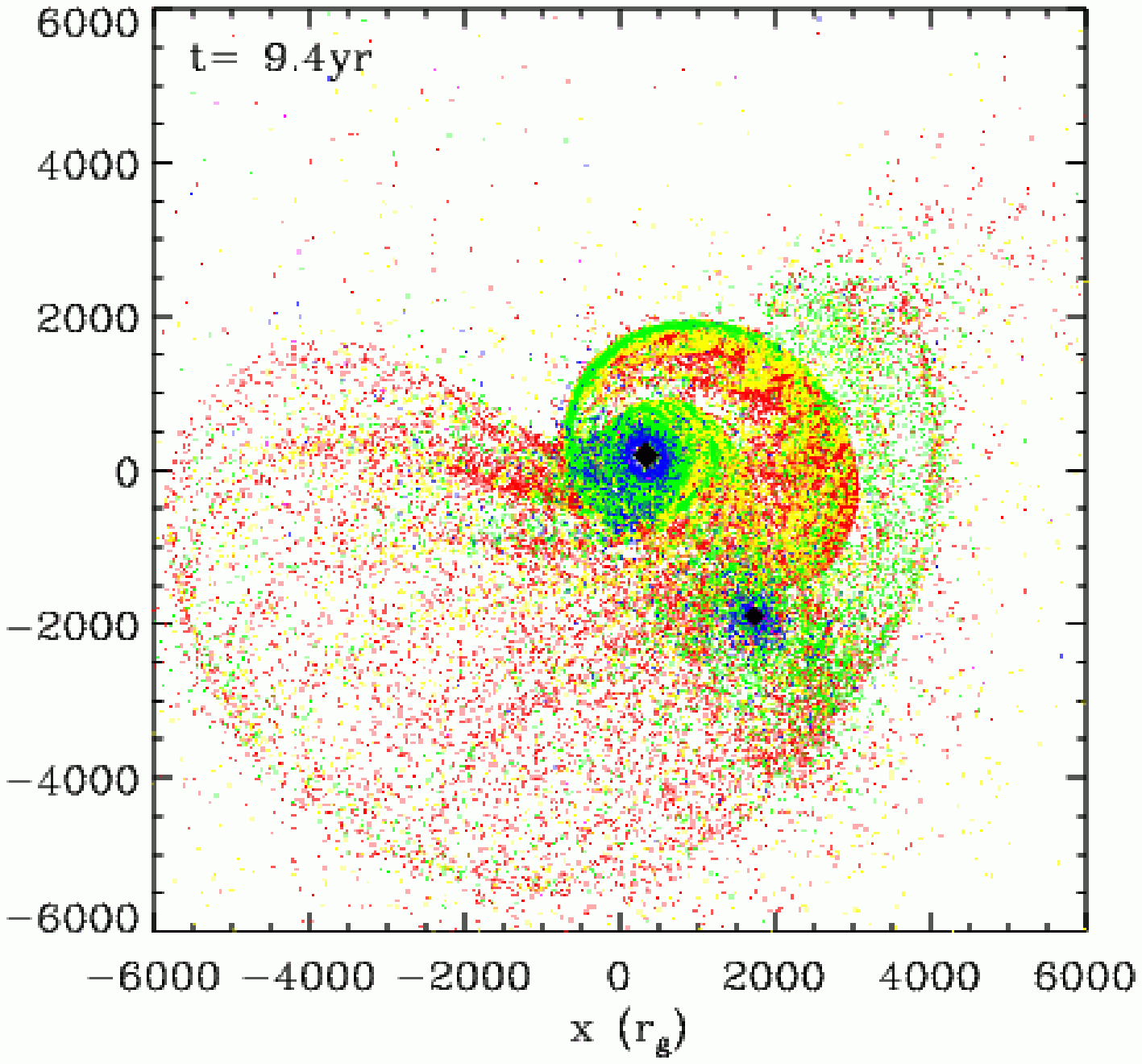}
\plottwo{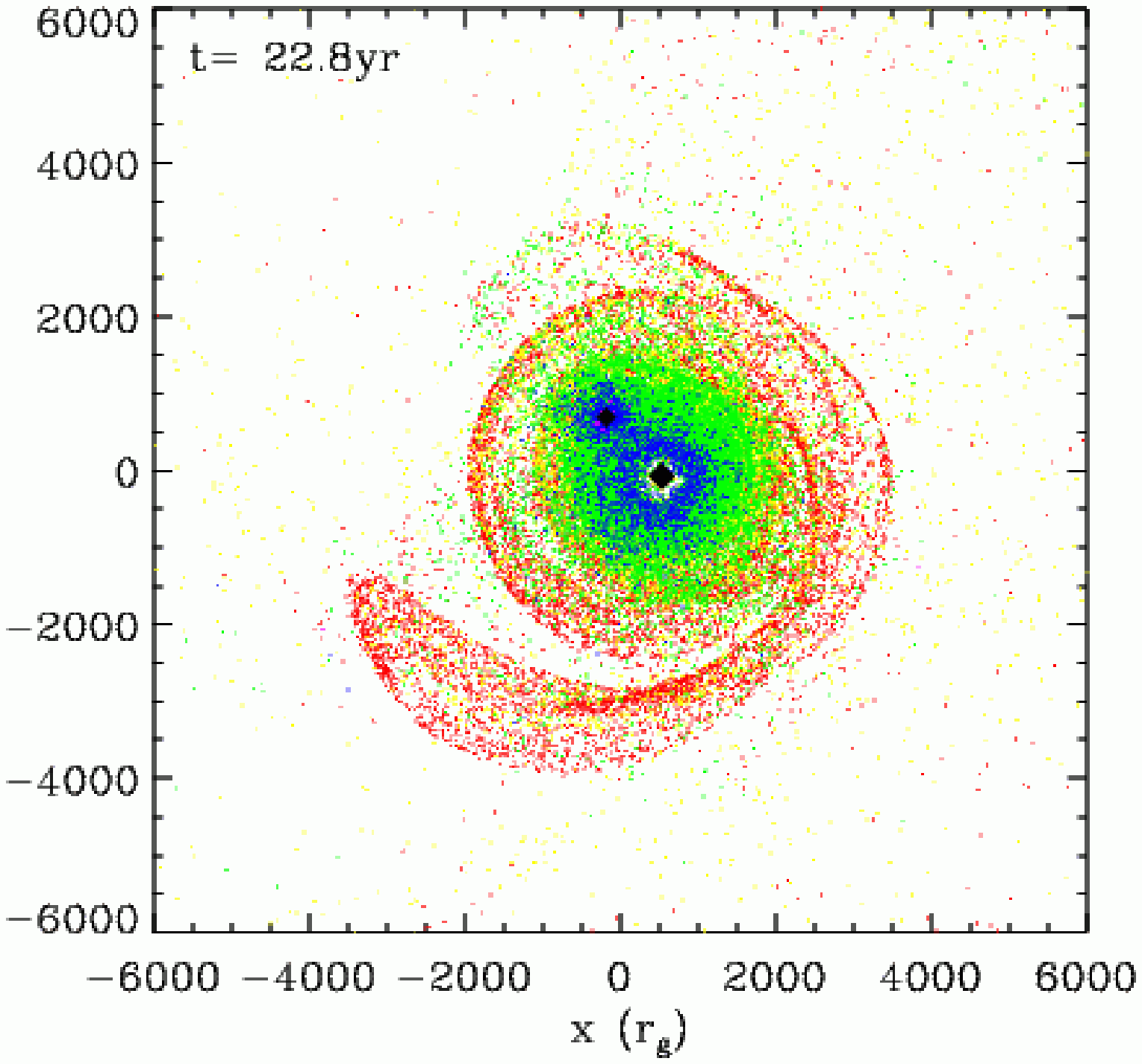}{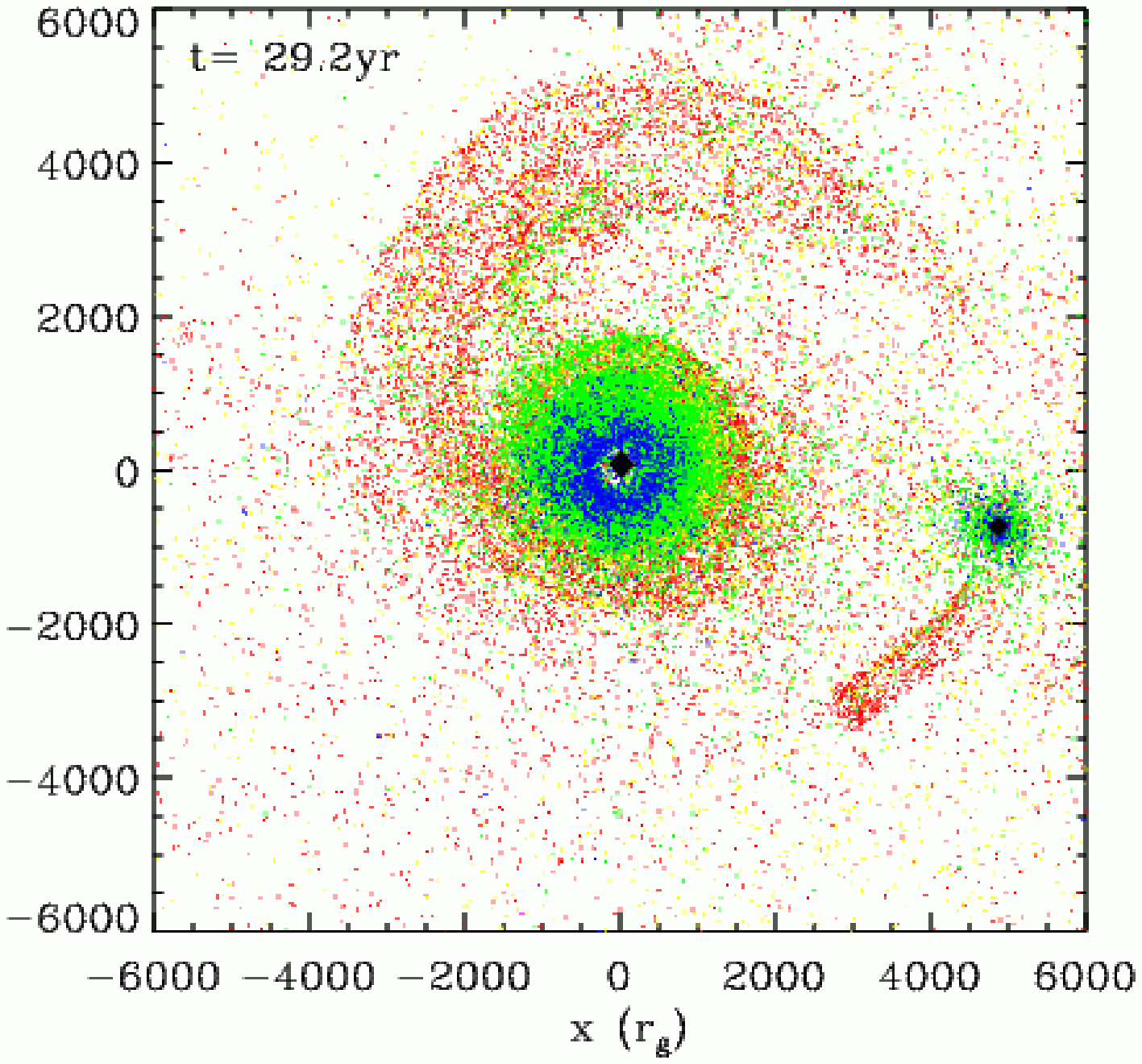}
\figcaption[b7.ps]{Sequence of snapshots from the simulation showing
the evolution of the binary and gas in the S-model (projected into the
plane of the binary orbit). The time stamps are marked on the top of
each panel and are the same as in Figure~\ref{fig_xy_bb}. The first
snapshot shows the system shortly after the secondary plunges into the
disk for the first time. Note that the second pericentric passage in
the S-model occurs about 100 days later than in the BB-model. The
rotation of the binary and the disk is counter-clockwise.  The color
legend is the same as in previous figures. The higher temperature
particles are plotted over the lower temperature ones, with the result
that some information is hidden (see Figure~\ref{fig_temp_solar}).
\label{fig_xy_sr}}
\end{figure*}

The cooling rate of the gas closely follows the heating rate
(Fig~\ref{fig_cool_bb}), corresponding to a bolometric luminosity of
the gas of $\sim 10^{43}\; {\rm erg\,s^{-1}}$. The bolometric
luminosity as a function of time does not exhibit large fluctuations
and remains at a nearly constant value after the first 10 years of
binary evolution. In the BB-model the luminosity of the accretion
powered sources dominates over the bolometric luminosity of the gas
disk (i.e., the power locally generated in the disk) by at least one
order of magnitude during the period of ``uniform'' accretion and by
several orders of magnitude during the first pericentric passage. This
is a consequence of inefficient radiative cooling (compared to
photoionization calculations). In the BB model, the inefficient
radiative cooling is compensated by adiabatic expansion of the gas,
resulting in a more pronounced spatial dispersion of the gas.

\begin{figure*}[t]
\epsscale{0.8}
\plotone{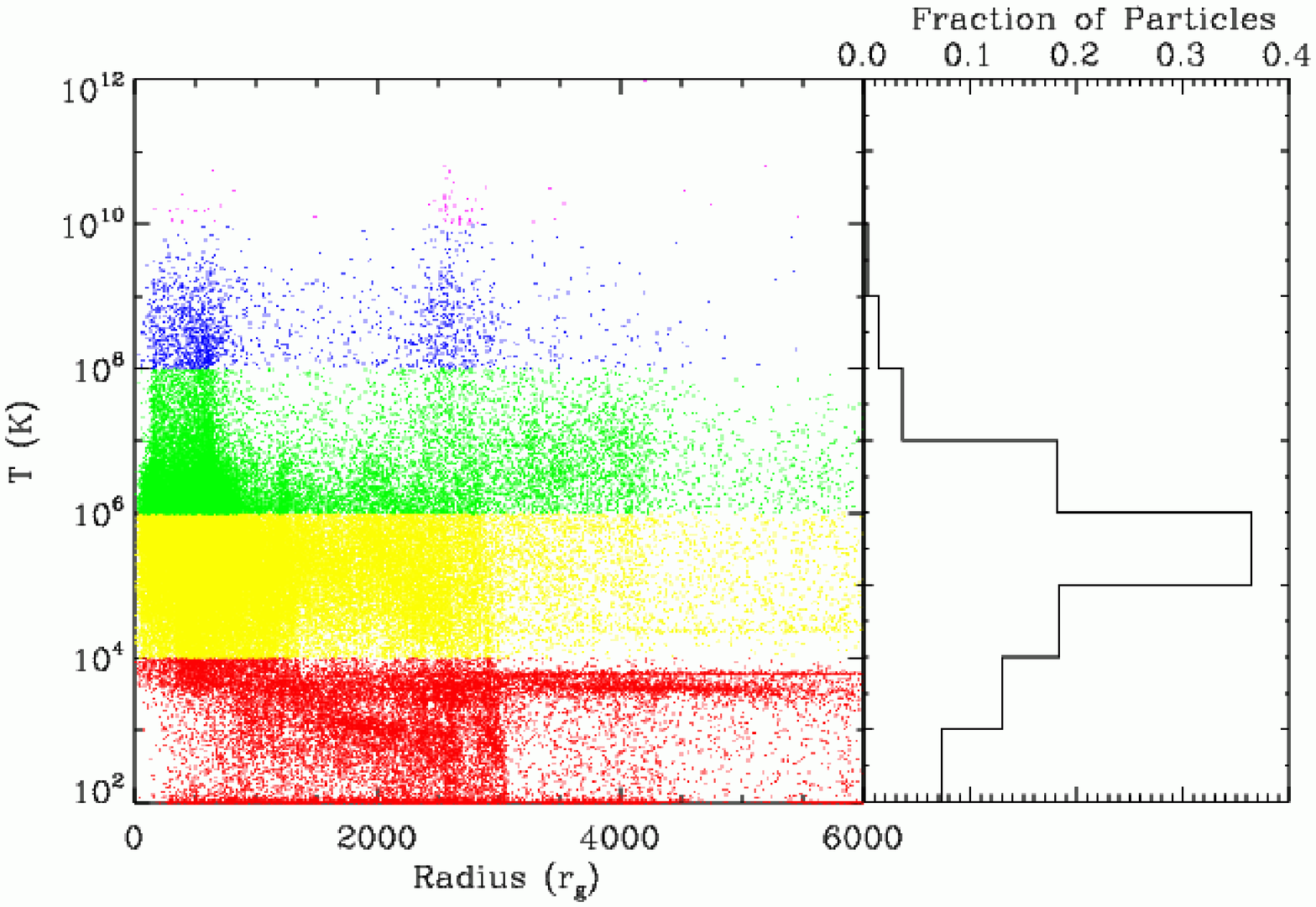}
\figcaption[b8.ps]{{\it Left:} Temperature of gas cells as a funcion
  in the disk which was partially hidden in the $xy$-projection in
  Figure~\ref{fig_xy_sr}. The colors that mark each temperature band
  are used throughout the paper. This figure corresponds to the
  morphology of the disk plotted in panel 2 of Figure~\ref{fig_xy_sr}.
  {\it Right}: Histogram showing the fraction of particles as a
  function of temperature.
\label{fig_temp_solar}}
\end{figure*}

\subsection{Co-rotating Case with Solar Metallicity Gas (S-Model)}\label{S_S-model}

\begin{figure}[b] 
\epsscale{1.2}
\plotone{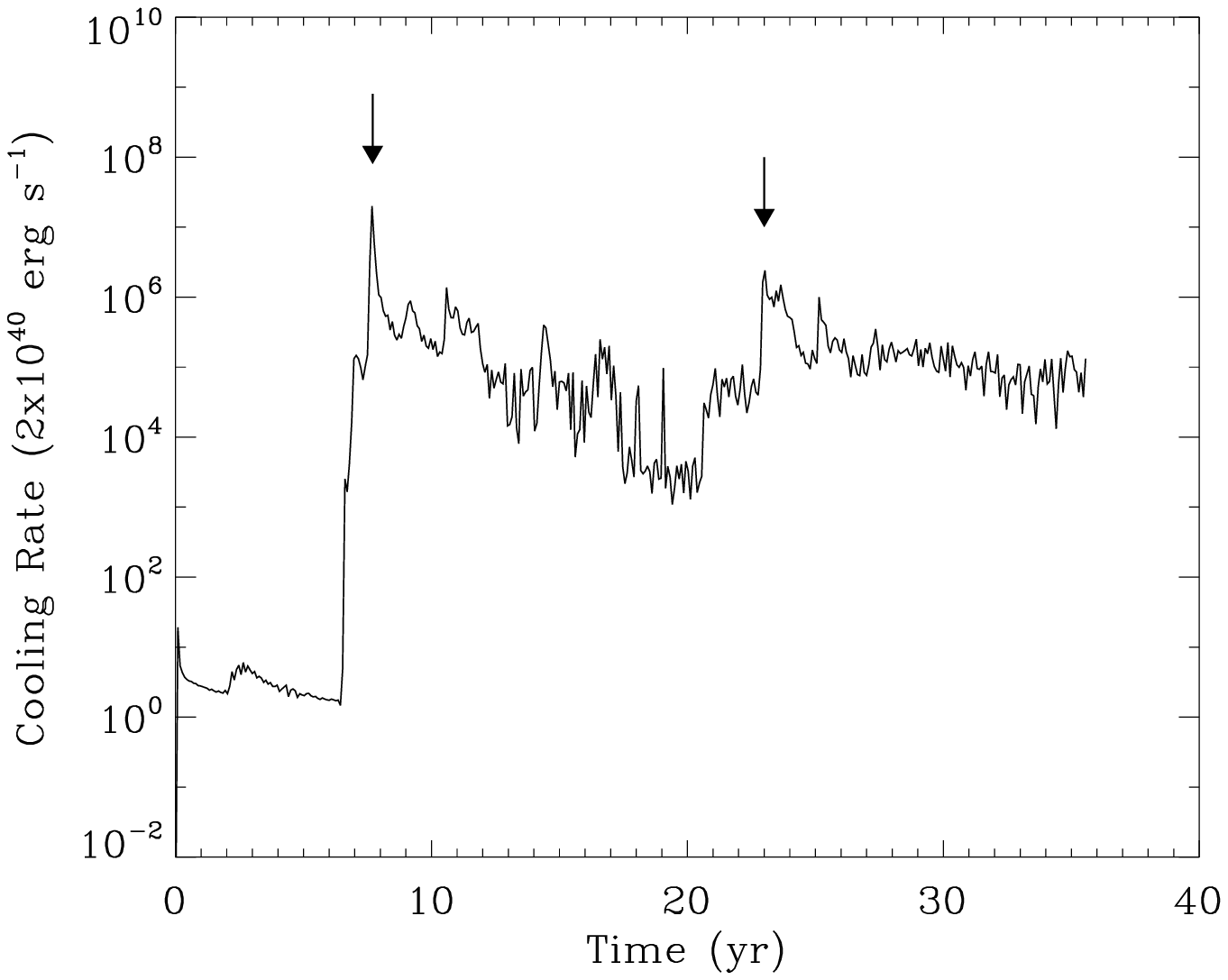}
\figcaption[b9.ps]{Cooling rate of the gas as a function of time 
calculated from the S-model. The cooling rate shown here traces the
bolometric light curve of the gas disk. The arrows mark the times of
pericentric passages of the binary.
\label{fig_cool_sr}}
\end{figure}

Figure~\ref{fig_xy_sr} shows the evolution of the morphology and
temperature of gas at the same times as Figure~\ref{fig_xy_bb}. The
initial steady state disk in the solar metallicity model (hereafter,
the S-model) has a temperature of $\sim 10^3$\,K and does not exhibit
the ``two phase'' temperature distribution, seen in the BB-model. The
speed of sound in the unperturbed disk is about $5\; {\rm
km\,s^{-1}}$. The bolometric luminosity of the gas disk during this
period is powered by collisional excitation only and it is of order
$10^{40}-10^{41}\, {\rm erg\,s^{-1}}$ (Figure~\ref{fig_cool_sr}).

The gas in the S-model shows the same main morphological features as
in the BB-model. The differences between the two are that after the
interaction with the secondary the gas in the S-model remains cooler
and shows less dispersion with the result that the spiral arm and
filaments are more pronounced. Because of the way we plot particles in
Figure~\ref{fig_xy_sr} (higher temperature particles are plotted over
the lower temperature ones), part of the information remains
hidden. For a better illustration, in Figure~\ref{fig_temp_solar} we
plot the temperature distribution in the disk as a function of radius
which shows that multiple temperature components can be present at a
single radius. The disk density at the end of the simulation is in the
range $10^{10}-10^{14}\; {\rm cm^{-3}}$, and the median density is
about one order of magnitude higher than in the BB-model (see
Table~\ref{T_results}). The disk gas in this model also remains well
above the Toomre threshold for gravitational instability. The speed of
sound is about $5\,{\rm km\,s^{-1}}$ in the cold disk and $\sim
10^2\,{\rm km\,s^{-1}}$ in the hotter gas component. The temperature
of the gas reaches the highest value (T$\sim10^{12}$ K) after the
shock is formed by the secondary. On a time scale of months, the
temperature of the gas falls below $10^{10}$ K due to the combined
effects of radiative cooling and adiabatic expansion. The dominant
radiative cooling mechanisms of the shocked gas are inverse Compton
and free-free emission. The hot component of the gas spends a
significant amount of time in the temperature range $10^4-10^8$ K. In
this regime the radiative cooling is dominated by free-free emission
and recombination radiation, while inverse Compton emission becomes
inefficient in comparison. On the other hand, the cold component of
the gas, confined to the spiral arm, has the same physical properties
as that in the BB-model and it retains a temperature of $10^3-10^4$ K.

In the period before the impact of the secondary, the radiative
cooling rate of the solar metallicity gas matches that of the BB-model
(Figs~\ref{fig_cool_bb} and \ref{fig_cool_sr}). This is an important
test of the S-model, because steady, geometrically thin disks are
indeed expected to be optically thick and behave as a black-body. The
differences in the cooling processes between the S-model and the
BB-model appear after the (accretion-powered) sources of ionizing
radiation turn on. The gas in the S-model exhibits less efficient
heating and more efficient cooling relative to the BB-model. This
happens because the ionized gas remains optically thin to ionizing
radiation and absorbs it less efficiently, while at the same time
free-free and recombination cooling processes are boosted. Once the
gas has been ionized its physical properties depart from these in the
black-body model.

\begin{figure}[b] 
\epsscale{1.2}
\plotone{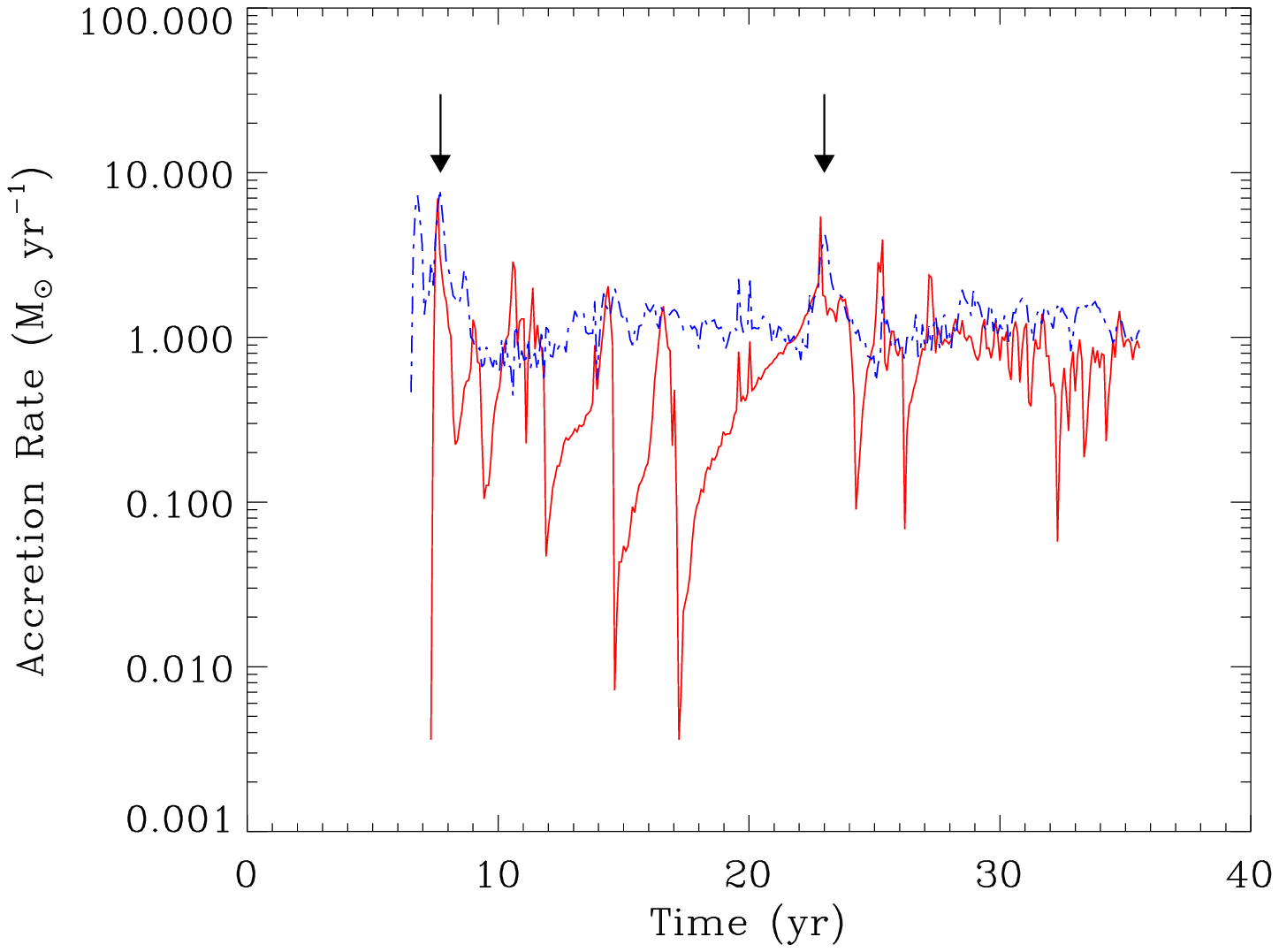}
\figcaption[b10.ps]{Effective accretion rate on the primary
({\it solid, red line}) and secondary ({\it dashed, blue line}) black
holes calculated from the S-model. The accretion rate curves
can be translated into UV/X-ray light curves by assuming that  
$1\Msun\,{\rm yr^{-1}} \sim 10^{43}\,{\rm
erg\,s^{-1}}$ of UV/X-ray luminosity. The arrows mark the times of
pericentric passages of the binary.
\label{fig_acc_sr}}
\end{figure}

\begin{figure*}[t] 
\epsscale{1.0} 
\plottwo{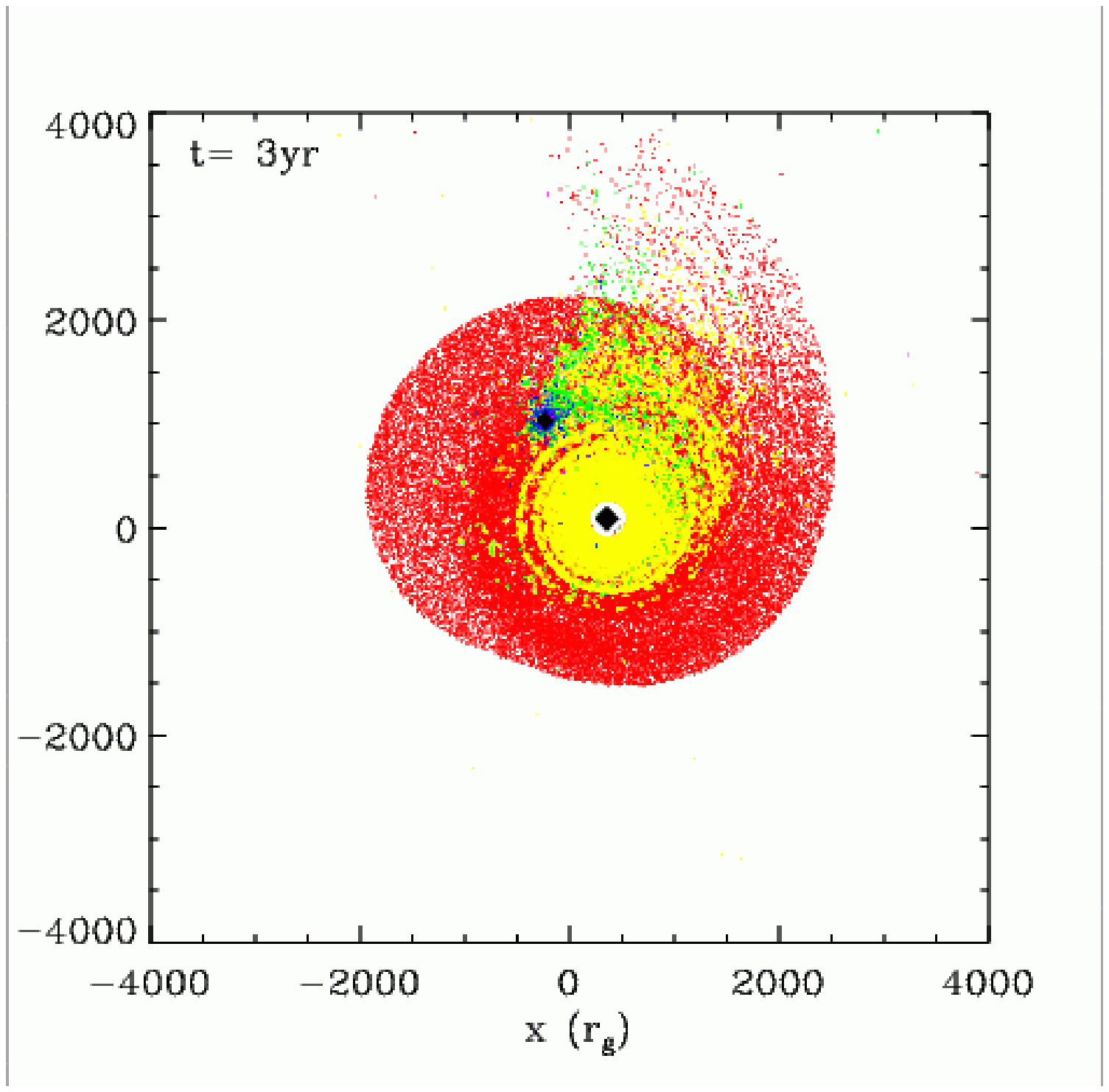}{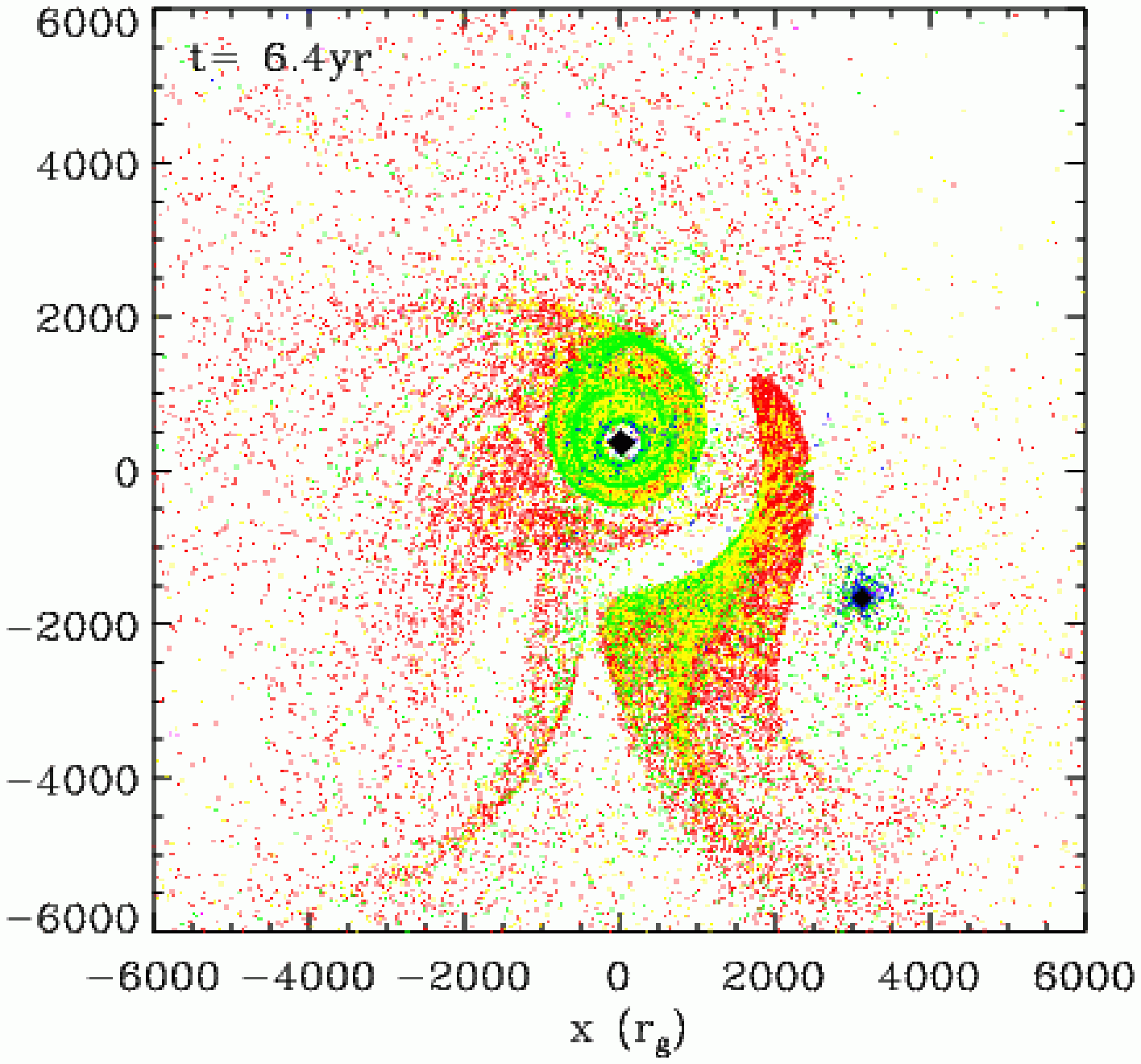}
\plottwo{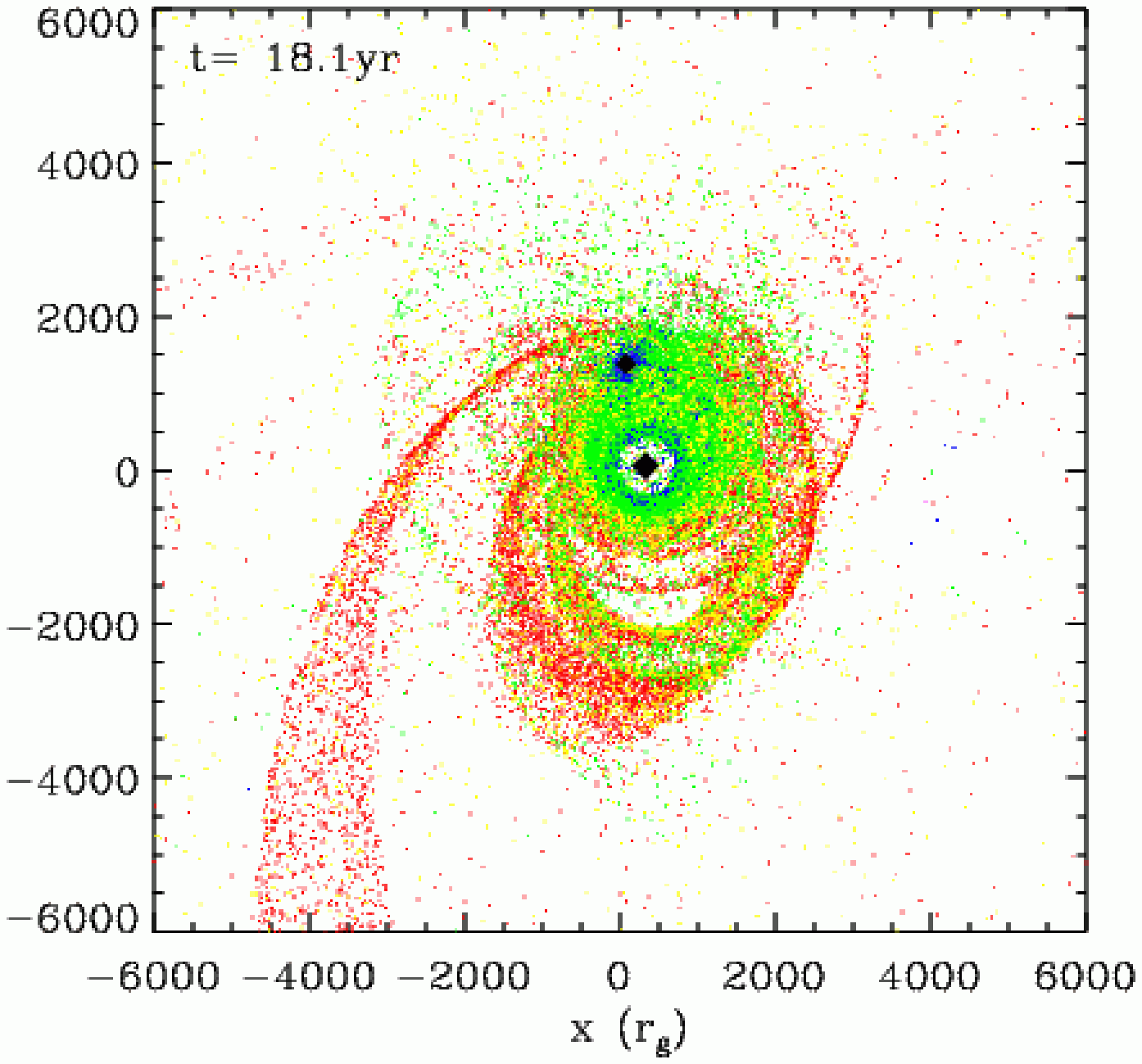}{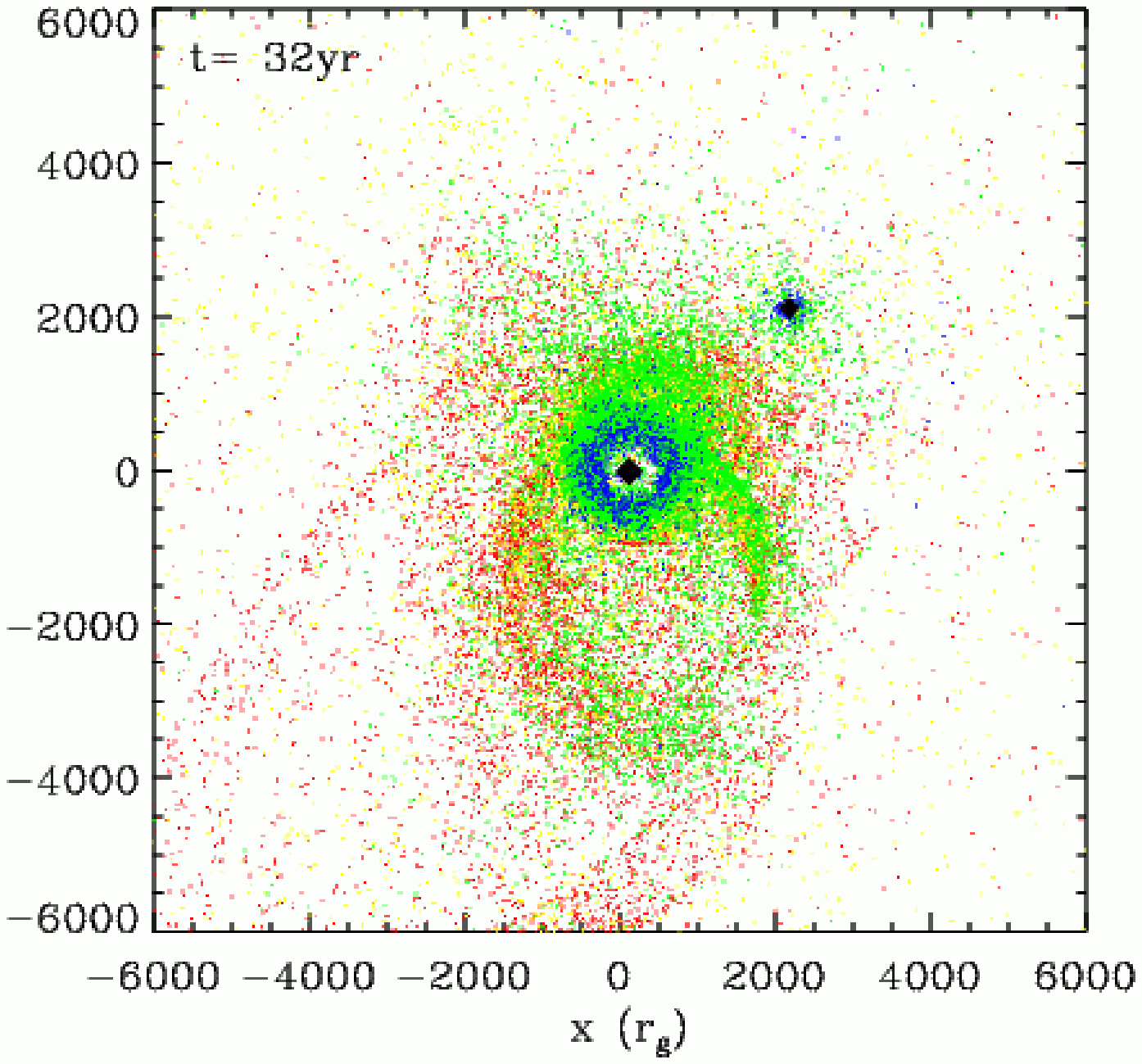}
\figcaption[b11.ps]{Sequence of snapshots from the simulation showing
the evolution of the binary and gas in the SR-model (projected into
the plane of the binary orbit). The time stamp is marked on the top of
each panel. The first snapshot shows the system shortly after the
secondary plunges into the disk for the first time. The rotation of
the binary is counter-clockwise and that of the disk is
clockwise. Note that higher temperature particles are plotted over the
lower temperature ones, with the result that some information is
hidden (also see Figure~\ref{fig_temp_retro}). \label{fig_xy_sr2}}
\end{figure*}

After the first pericentric passage of the secondary the accretion
rate reaches $10 {\rm \Msun\;yr^{-1}}$ (Figure~\ref{fig_acc_sr}),
comparable to the Eddington rate for this system, $\dot{M}_E \approx
30\,M_8\, (\eta/0.01)^{-1}\;{\rm \Msun\;yr^{-1}}$. During the
remainder of the simulation the accretion rate remains at a nearly
constant level of about $1 {\rm \Msun\;yr^{-1}}$. This implies that
the accretion luminosity is just below the Eddington luminosity
($L_{E}=1.51\times10^{46}M_{8}\,{\rm erg\,s^{-1}}$) for a few years
after the first pericentric passage of the binary and later settles at
luminosity $\sim 10^{-2}\,L_{E}$. During this period the bolometric
luminosity of the disk, powered by shocks and illumination, is found
to be $\sim 10^{45}\,{\rm erg\,s^{-1}}$ on average, with fluctuations
of up to 2 orders of magnitude. The bolometric luminosity of the gas
disk is comparable to that of the photoionization
sources. Additionally, the pericentric passages, in the 7th and 23rd
years of the simulation, can be easily discerned in the cooling curve
(Figure~\ref{fig_cool_sr}). These are not noticeable in the cooling
curve of the BB-model (Figure~\ref{fig_cool_bb}) because its
bolometric luminosity is dominated by emission from the diffuse gas
component. A common property of the BB and S-models is large
fluctuations in the accretion rate of the primary, a consequence of
the accretion of cold gas. Moreover, the largest dips in the accretion
rate of the primary coincide with the apocentric passage, when the gas
is tidally pulled away from the primary and the primary exhibits the
slowest orbital motion.

\break

\subsection{Counter-rotating Case With Solar Metallicity Gas  (SR-Model)}\label{S_SR-model}

\begin{figure*}[t] 
\epsscale{0.8}
\plotone{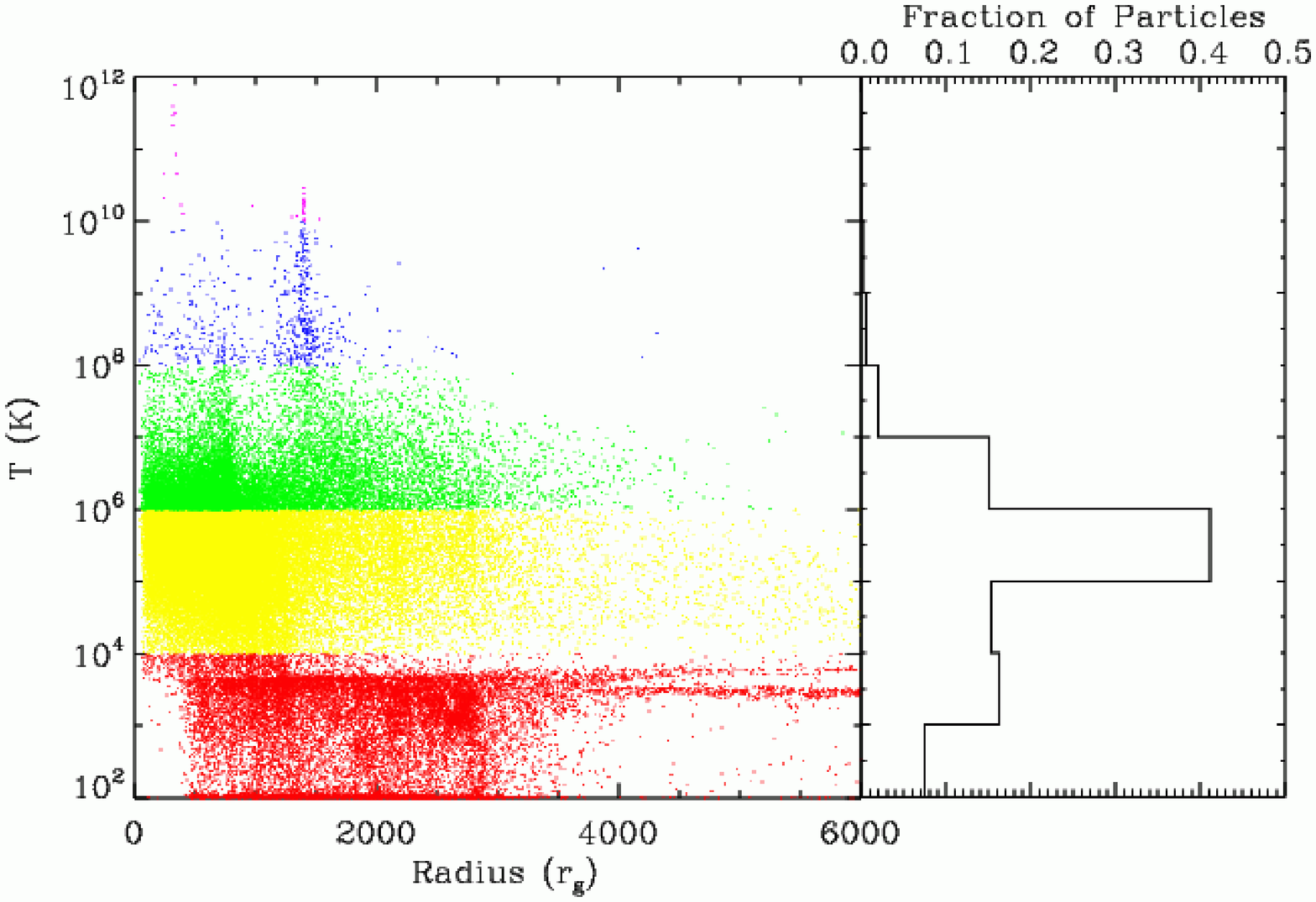}
\figcaption[b12.ps]{{\it Left:} Temperature of gas cells as a function of 
radius at 18.1 years in SR-model. This figure shows the extent of
temperature stratification in the disk which was partially hidden in
the $xy$-projection in Figure~\ref{fig_xy_sr2}. The colors that mark
each temperature band are used throughout the paper. This figure
corresponds to the morphology of the disk plotted in panel 3 of
Figure~\ref{fig_xy_sr2}.  {\it Right}: Histogram showing the fraction
of particles as a function of temperature.
\label{fig_temp_retro}}
\end{figure*}

In this scenario the binary and gas disk rotate in opposite directions
with respect to each other. In Figure~\ref{fig_xy_sr2} we show the
morphology and the temperature distribution, while in
Figure~\ref{fig_temp_retro} we plot the temperature stratification in
the disk for this model at a single time. During the pre-plunge phase
the physical properties of the disk are comparable with those in the
co-rotating S-model. The plunge of the secondary into the disk in the
counter-rotating solar metallicity model (hereafter, the SR-model) is
more energetic than in the co-rotating models, because the relative
velocity of the encounter is higher. However, we observe a lower
median gas temperature in the SR-model relative to the S-model. This
apparent discrepancy can be explained by the fact that the
counter-rotating gas disk efficiently shears the shock formed by the
impact of the secondary black hole. The hot gas from the shock front
is quickly transported away from the secondary and its thermal energy
is radiated. When compared to each other, the three models exhibit an
interesting property: while the {\it median} gas temperature is
$3\times 10^5$, $5\times 10^5$, and $10^6$\,K, the {\it mean}
temperature is $10^{11}$, $10^{10}$, and $10^9$\,K, in the SR, S, and
BB-models respectively.

\begin{figure}[t] 
\epsscale{1.2}
\plotone{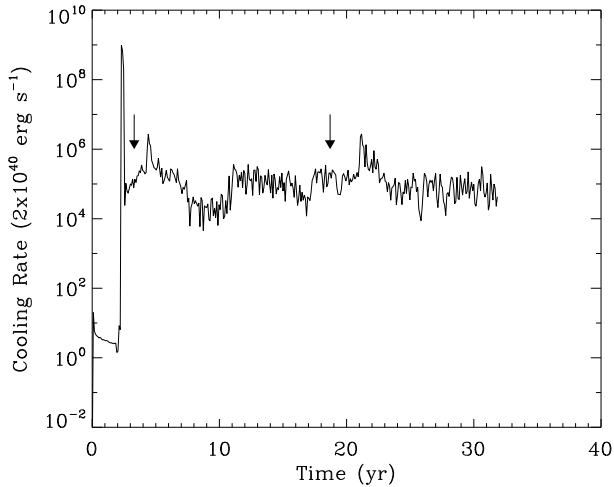}
\figcaption[b13.ps]{Cooling rate of the gas as a function of time 
calculated from the SR-model. The cooling rate shown here traces the
bolometric light curve of the gas disk. The arrows mark the times of
pericentric passages of the binary.
\label{fig_cool_sr2}}
\end{figure}

Another property of the SR-model is that its cooling curve does not
exhibit peaks at the times of the pericentric passages of the binary
(Figure~\ref{fig_cool_sr2}). The peaks noticeable in the cooling curve
occur, respectively, one and two years after the pericentric passages
and are not associated with the peaks in the accretion rate
(Figure~\ref{fig_acc_sr2}). This implies that along with accretion
there must be an additional mechanism that powers the bolometric
emission. This second heating mechanism, which competes with
photoionization, arises from shocks in the gas. The bolometric
luminosity of the gas disk settles at $\sim 10^{45}\,{\rm
erg\,s^{-1}}$ and the luminosity of the photoionization sources is
$\sim 10^{44}\,{\rm erg\,s^{-1}}$.

In the counter-rotating model, the gas is driven away from the center
of the disk, as it acquires more orbital angular momentum from the
binary, and the hollow region around the primary widens with
time. This is why during a pericentric passage there is no
characteristic jump in the accretion rate of the primary, and
accretion occurs by smooth diffusion of particles from the
disk. However, the accretion rate onto the secondary shows peaks
corresponding to pericentric passages, and smooth undulation in
between. Although over the course of the two orbits we do not notice a
decrease in the accretion rate, if the outbound migration of gas
continues and the low density gap is formed at the center, the
accretion on the primary is likely to dwindle. Observationally, this
implies that in these systems, only one of the sources (the secondary)
may remain active, until its orbit gradually becomes smaller than the
hollow region in the disk, at which point both black holes may stop
accreting. Whether both, one, or neither of the two sources are seen
in emission depends on the evolutionary stage of the binary and the
thermodynamic properties and supply of the low angular momentum gas
in the host galaxy.
  
A noticeable feature in the accretion curve of the primary is a dip
between the 8th and 12th years. This corresponds to the time when a
large portion of the gas is unbound from the disk, as shown in the
second panel of Figure~\ref{fig_xy_sr2}. The detached part of the disk
continues to revolve around the primary until it is re-captured and
wound around. After it has completed several orbits, this gas creates
a complex set of rings and filaments. A fraction of this gas will fall
into the primary black hole and boost its accretion rate back to $\sim
1\;\Msun\;{\rm yr^{-1}}$.

\subsection{X-ray Light Curves}

\begin{figure}[t]
\epsscale{1.2}
\plotone{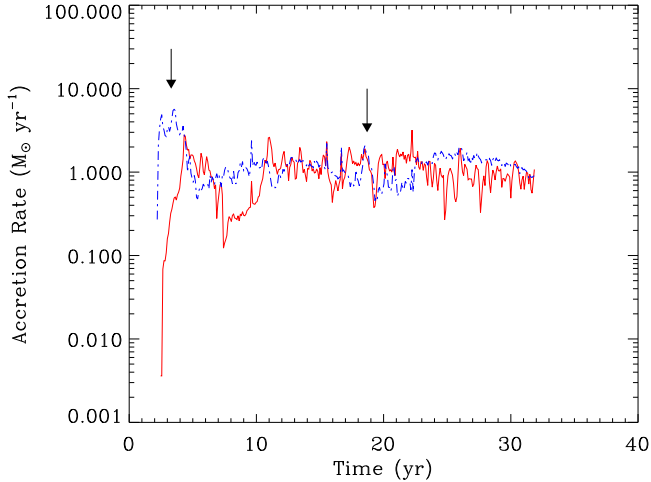}
\figcaption[b14.ps]{Effective accretion rate on the primary
({\it solid, red line}) and secondary ({\it dashed, blue line}) black
holes calculated from the SR-model. The accretion rate curves in our
calculations can be translated into UV/X-ray light curves from the
emission sources, where $1\Msun\,{\rm yr^{-1}} \sim 10^{43}\,{\rm
erg\,s^{-1}}$ of UV/X-ray luminosity. The arrows mark the times of
pericentric passages of the binary.
\label{fig_acc_sr2}}
\end{figure}

We calculate the X-ray luminosity powered by the two different
mechanisms: accretion and bremsstrahlung emission from the hot
gas. The bremsstrahlung emitting gas is heated by both, shocks and
photoionization. The accretion rates onto the primary and secondary
black holes estimated from the S and SR-models result in a UV/X-ray
luminosity of about $10^{43}~\,{\rm erg\,s^{-1}}$ and up to
$10^{44}~\,{\rm erg\,s^{-1}}$ during times of high accretion rate (the
details of the accretion curves in each model are discussed in
\S~\ref{S_S-model} and \S~\ref{S_SR-model} and shown
Figures~\ref{fig_acc_sr} and \ref{fig_acc_sr2}).

In this section we describe how the estimate of the concurrent
bremsstrahlung X-ray luminosity was obtained. We calculate the
thermal bremsstrahlung power per unit volume from the population of
relativistic electrons hotter than $10^7$ K as
\begin{eqnarray}
\Lambda_{bremss} & = &  1.4\times 10^{-27}\, T^{1/2}\, Z^2\, n_e\,
n_i\, \nonumber \\
& & \times\; \overline{g_b}\,(1+4.4\times 10^{-10}T) 
\;\;\;{\rm erg\;cm^{-3}\,s^{-1}}\;,
\label{eq_bremss}
\end{eqnarray}
where $Z$ is the atomic number, $n_e$ and $n_i$ are the number
densities of electrons and ions, respectively, $\overline{g_b}$ is a
frequency average of the velocity averaged Gaunt factor and is of
order 1. The electrons are heated in the process of photoionization of
the gas by the accretion powered sources and in shocks taking place in
the gas perturbed by the secondary black hole. In order to calculate
the resulting bremsstrahlung X-ray luminosity, we also need to know
the volume occupied by the gas giving rise to this emission. The
calculation of the size of the emitting volume from a SPH simulation
is complicated by the finite spatial resolution, as set by the
characteristic size assigned to each gas particle (i.e., the smoothing
length, $h_{sml}$). In our calculations $h_{sml} \ge 10\,r_g \approx
10^{14}\,{\rm cm}$. As a result of ``smoothing'', the bremsstrahlung
X-ray emission from shocks, which typically occur on scales smaller
than $h_{sml}$ in our simulations, is not resolved. Consequently, the
size of the emission region in this case is overestimated, while the
density and the temperature in the shock are underestimated. In order
to bypass this uncertainty we calculate the values of some physical
properties of the X-ray emitting gas following the analytical
treatment of shock structure in \citet{hm79}.

First, we attempt to discriminate between the gas particles in the
photoionized and shocked regions. We assume that gas particles that
have a Mach number $\mathcal{M} > 10$ can participate in shocks and
thus contribute to the {\it shock-powered} bremsstrahlung X-ray
emission. From the jump conditions we calculate the compression and
temperature ratios in the shocked gas cell and from these we estimate
the post-shock density and temperature of the gas. Using the
post-shock values of the physical parameters we calculate the cooling
rate (equation~[\ref{eq_bremss}]) and cooling time for the shocked gas
cell under the assumption that its dominant cooling mechanism is
bremsstrahlung. The corresponding cooling column of gas is then
$N_{cool} = n_0\, \upsilon_s\, t_{cool}$, where $n_0$ is the pre-shock
density of the gas, $\upsilon_s$ is the velocity of the shock, and $t_{cool}$
is the cooling time. The values of the cooling column and the
post-shock density of the hot gas allow us to estimate the
characteristic length scale of the cooling column, $\ell_{cool}$. For
typical values of the parameters in the shocked gas cells in our
simulations, where $n_{e} \approx n_{i} \approx n \sim
10^{12}-10^{13}\,{\rm cm^{-3}}$ and $T\sim 10^8\,{\rm K}$, we obtain
$\ell_{cool}\sim 10^9\,{\rm cm}$. We use this information to constrain
the X-ray emitting volume within a shocked gas cell. Given that
$60-70\%$ of gas particles fulfill the condition ${\mathcal M} > 10$
and post-shock $T >10^7 \,{\rm K}$, we find that the resulting
bremsstrahlung X-ray luminosity from the shocked gas is in the range
$10^{40}-10^{42}\,{\rm erg\,s^{-1}}$, with peaks reaching 
$10^{43}\,{\rm erg\,s^{-1}}$.

In the second step we estimate the bremsstrahlung X-ray luminosity
emitted from the photoionized portion of the gas. We identify the gas
cells responsible for most of the X-ray emission {\it powered by
photoionization} as low density cells with temperatures higher than $T
> 10^7\,{\rm K}$. The hot, optically thin, photoionized component of
the gas forms a halo with a median density of $n\sim 10^7-10^8\,{\rm
cm^{-3}}$. The photoionized regions are more spatially extended than
the shocked regions and their characteristic scale is $\ell_{cool}\sim
10^3\,r_g\approx 10^{16}\,{\rm cm}$. The number of photoionized gas
particles that contribute to the X-ray bremsstrahlung emission at any
given time is about a few percent of all gas particles. However, the
large volume of this halo makes up for its small mass and low number
density. Thus, the resulting bremsstrahlung luminosity from the halo
is $10^{41}-10^{42}\,{\rm erg\,s^{-1}}$, comparable to the estimated
contribution of the shocked regions.

We emphasize that the estimates of the bremsstrahlung X-ray luminosity 
presented here are based on astrophysically motivated
but simplified assumptions. For example, while ${\mathcal M} > 10$ may
be a necessary condition for a gas particle to participate in shocks,
not every such particle will do so. The shocks may be avoided in
ordered, laminar flows where the trajectories of particles do not
intersect. However, we expect that the gravitational perturbation 
from the secondary black hole will cause shocks, especially in
the inner portion of the disk where the dynamical time scale of
particles is much shorter than the period of the binary and the
density of the gas is higher. In the case of photoionization, not all hot,
 low density gas particles will be entirely photoionized, making the 
effective emitting volume (or equivalently the emitting mass) only a fraction 
of the total in a gas cell. There may also be particles which participate in
shocks and are at the same time photoionized by the accretion powered
sources. For all these reasons and because the structure of the
nuclear region is very complex, we note that the above estimates
should be regarded as constraints rather than exact values.

The total bremsstrahlung X-ray luminosity (shock plus photoionization
powered) of the gas in our simulations is lower than the
accretion-powered UV/X-ray luminosity, although during the times of
pericentric passage the peak bremsstrahlung X-ray luminosity is
comparable to the accretion-powered X-ray luminosity. The latter X-ray
light curve exhibits peaks during pericentric passages of the binary,
thus the peaks are good markers of such events. The estimated level of
the total X-ray emission (accretion powered plus bremsstrahlung)
should be observable to a redshift of $z \le 2$ during the outburst
phases.

Although our calculations follow the evolution of the binary over only
two orbital passages we suggest that in binary systems of this type,
X-ray outbursts should be expected to continue as long as there is gas
in the nuclear region for the binary to interact with. Our models
imply that at the mean level of accretion, the gaseous disks should be
depleted after $10^3-10^4$ yr. In order for the outburst activity to
last for longer time scales, the reservoir of gas in the nuclear
region needs to be continually replenished. It is also plausible,
however, that repeated collisions of the secondary black hole with the
disk will completely disrupt it and turn it into a spherical halo of
hot gas. In such a case, the accretion rate would be relatively smooth
and uniform, and no outburst would be evident.

A calculation following the X-ray light curve variability over a large
number of orbits is necessary in order to confirm that the periodicity
is a long lived signature of the binary. Although currently not
possible with the SPH method used here, simulations of the long term
evolution of the binary together with hydrodynamics and radiative
transfer may be achieved in the future by a hybrid approach including
complementary numerical techniques. In particular, they may be
realized by combining semi-analytical hydrodynamical calculations with
emphasis on the long term dynamical evolution of gas with methods that
focus on hydrodynamics and radiative transfer, like the one used here.

\begin{figure} 
\epsscale{1.2} 
\plotone{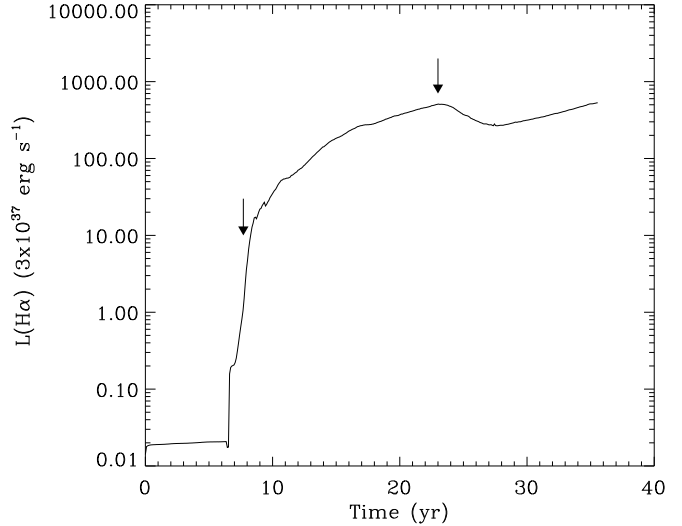} 
\figcaption[b15.ps]{H$\alpha$ light
curve for the S-model obtained with help of photoionization
calculations. The arrows mark the times of pericentric passages of the
binary. The details of the method are described in \S~\ref{S_cooling}
of the text.
\label{fig_ha_solar}}
\end{figure}

\subsection{H$\alpha$ Light Curves and Broad Emission Line Profiles}\label{S_Halpha}

\begin{figure}[t] 
\epsscale{1.2}
\plotone{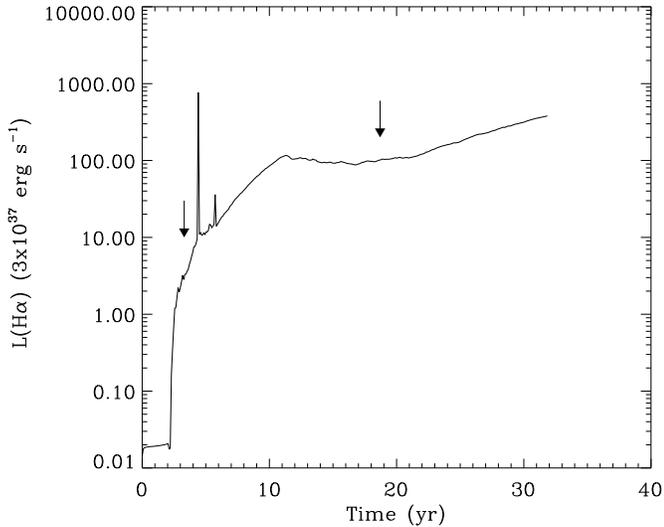}
\figcaption[b16.ps]{H$\alpha$ light curve obtained with help of 
photoionization calculations used in the SR-model. The first, large
spike in the light curve occurs after the first pericentric passage of
the binary (the arrows mark the times of pericentric passages). The
details of the method are described in
\S~\ref{S_cooling} of the text.
\label{fig_ha_retro}}
\end{figure}

By modeling the H$\alpha$ light curves for the S and SR-models we find
that after the beginning of accretion the H$\alpha$ luminosity
gradually reaches $10^{39}-10^{40}\;{\rm erg\, s^{-1}}$ in both models
(Figures~\ref{fig_ha_solar} and \ref{fig_ha_retro}). Sources with such
H$\alpha$ luminosities are observable out to the distance of the Virgo
Cluster and possibly up to a distance of 100 Mpc. During the
steady-state phase, before the plunge of the secondary into the disk,
both models are characterized by a low H$\alpha$ luminosity,
$L_{H\alpha}\sim 10^{35}\;{\rm erg\, s^{-1}}$. This is an artifact of
the calculation, a consequence of the choice of initial conditions,
because in the pre-plunge phase the emission from the gas is entirely
thermal emission from the disk and the effects of photoionization are
not taken into account. Such neutral and non-illuminated accretion
disks are unlikely to be encountered in nuclear regions of AGNs, even
if these are characterized with very low accretion rates.

Although in our model the H$\alpha$ light is closely linked with
accretion, it does not closely follow the variations in the accretion
rate. Neither the H$\alpha$ light curve exhibits easily recognizable
periodic features during the first two orbits of the binary. The only
occasion when a noticeable jump in the H$\alpha$ luminosity occurs is
in the SR-model, after the first pericentric passage
(Figure~\ref{fig_ha_retro}). The H$\alpha$ light is predominantly
contributed by a spatially extended component of the gas and is less
susceptible to shocks by the secondary. The gradual increase in the
H$\alpha$ luminosity noticeable in both light curves is caused by the
redistribution of the photoionized gas in space. As gas expands to a
scale of a parsec, its density and optical depth decrease, and the
solid angle subtended to the photoionizing source increases. As a
result, a larger fraction of ionizing photons is reprocessed to
H$\alpha$ photons and the physical properties of the gas evolve
towards those found in photoionized nebulae.

It appears difficult to infer the existence of a binary in the nuclear
region of a galaxy based on the H$\alpha$ light curve
alone. Serendipitous flares in the H$\alpha$ light curve may occur
close to pericentric passages but their duration may be fairly short
(here, $\leq$month, the upper limit set by the time resolution in the
light curve) and consequently hard to observe. The periodic outbursts
may be more pronounced in cases when the binary orbit is inclined with
respect to the disk, but such models remain to be explored.

\begin{figure*}[t] 
\epsscale{0.7}
\plottwo{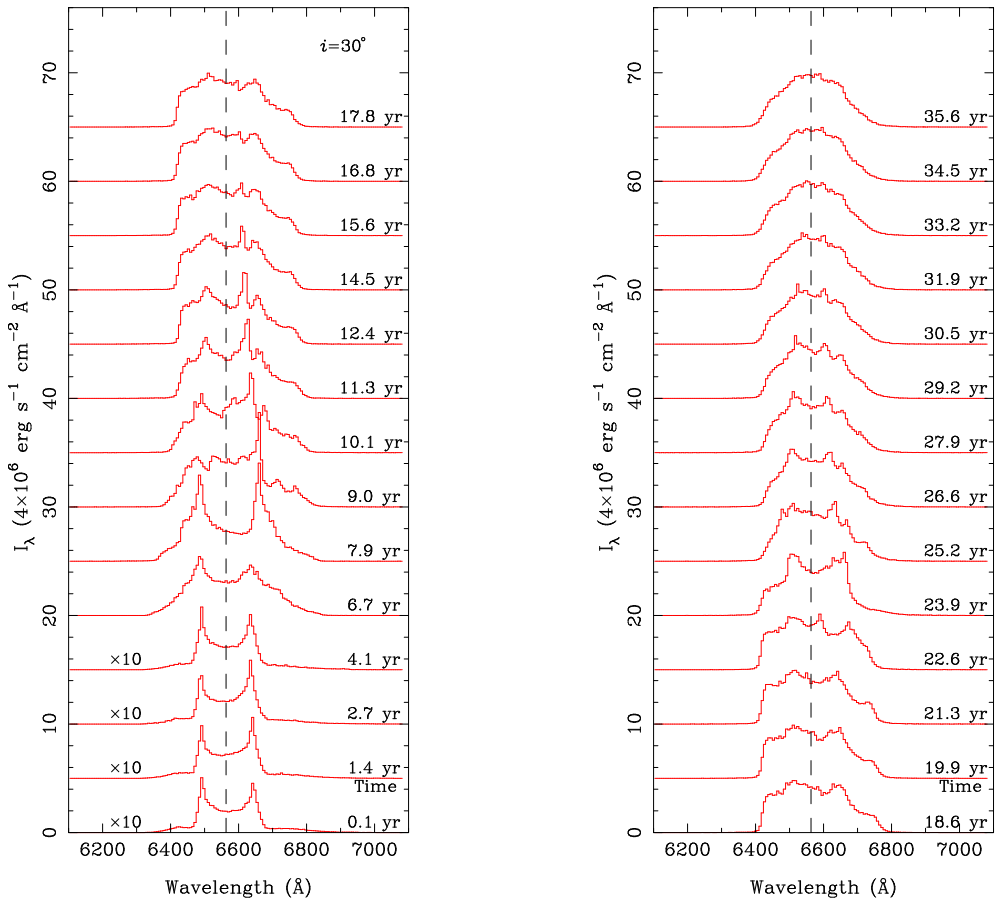}{b17b.ps}
\figcaption[b17.ps]{Sequence of the H$\alpha$ emission-line profiles 
selected from the S-model. The intrinsic intensity of profiles is
plotted against wavelength. The first 4 profiles in the sequence are
multiplied by a factor of 10, so that they can be represented on the
same intensity scale with the other profiles. The corresponding time
from the beginning of the simulation is plotted next to each
profile. The inclination of the plane of the disk with respect to the
observer is as marked on the figure. The vertical dashed line at
6563~{\rm \AA} marks the H$\alpha$ rest frame wavelength.
\label{fig_prof_solar}}
\end{figure*}

\begin{figure*}[!] 
\epsscale{0.7}
\plottwo{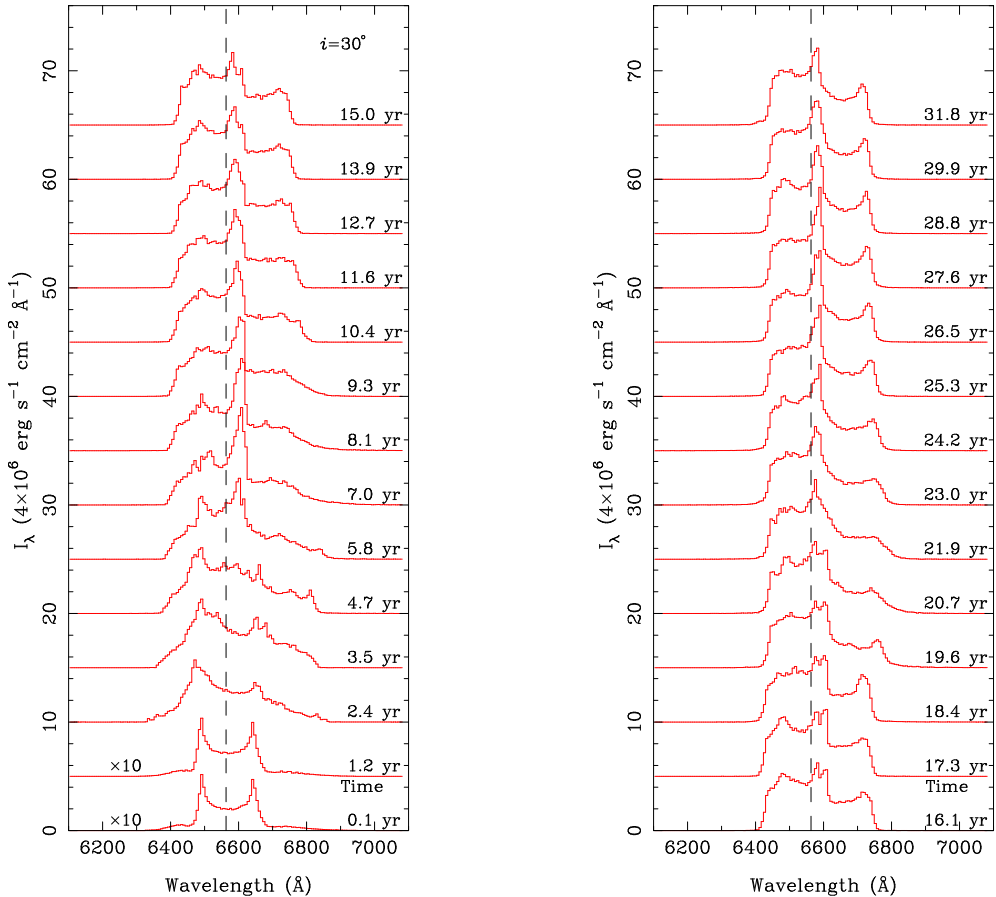}{b18b.ps}
\figcaption[b18.ps]{Same as previous figure, except for the SR-model.
The first 2 profiles in the sequence are multiplied by a factor of 10, so 
that they can be represented on the same intensity scale with the other profiles.
\label{fig_prof_retro}}
\end{figure*}

\begin{figure*}[t] 
\epsscale{0.8}
\plottwo{b19a.ps}{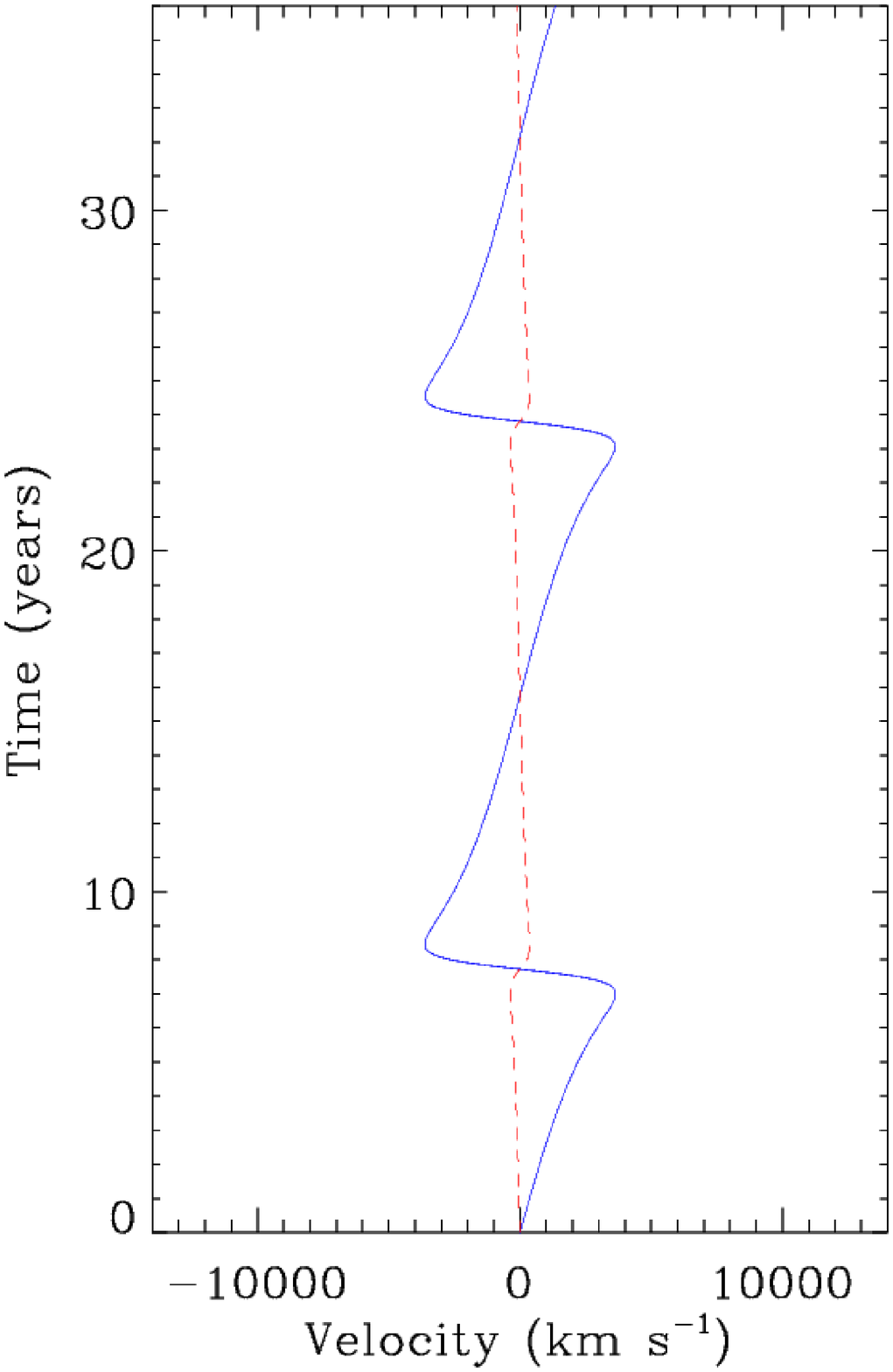}
\figcaption[b19.ps]{Trailed spectrogram (left) and black hole 
velocity curves (right) plotted for the S-model. The trailed spectrogram
is a logarithmic gray scale map of the H$\alpha$ intensity against
wavelength and observed velocity. Darker shades mark higher
intensity. Notice the low relative intensity of profiles before the
start of accretion. The velocity curve panel shows the orbital velocities
of the primary ({\it dashed, red line}) and secondary ({\it solid,
blue line}) black holes projected along the line of sight to the observer.
The velocity curves are skewed because of the non-zero eccentricity
of the orbit.
\label{fig_trail_solar}}
\end{figure*}

\begin{figure*}[!] 
\epsscale{0.8}
\plottwo{b20a.ps}{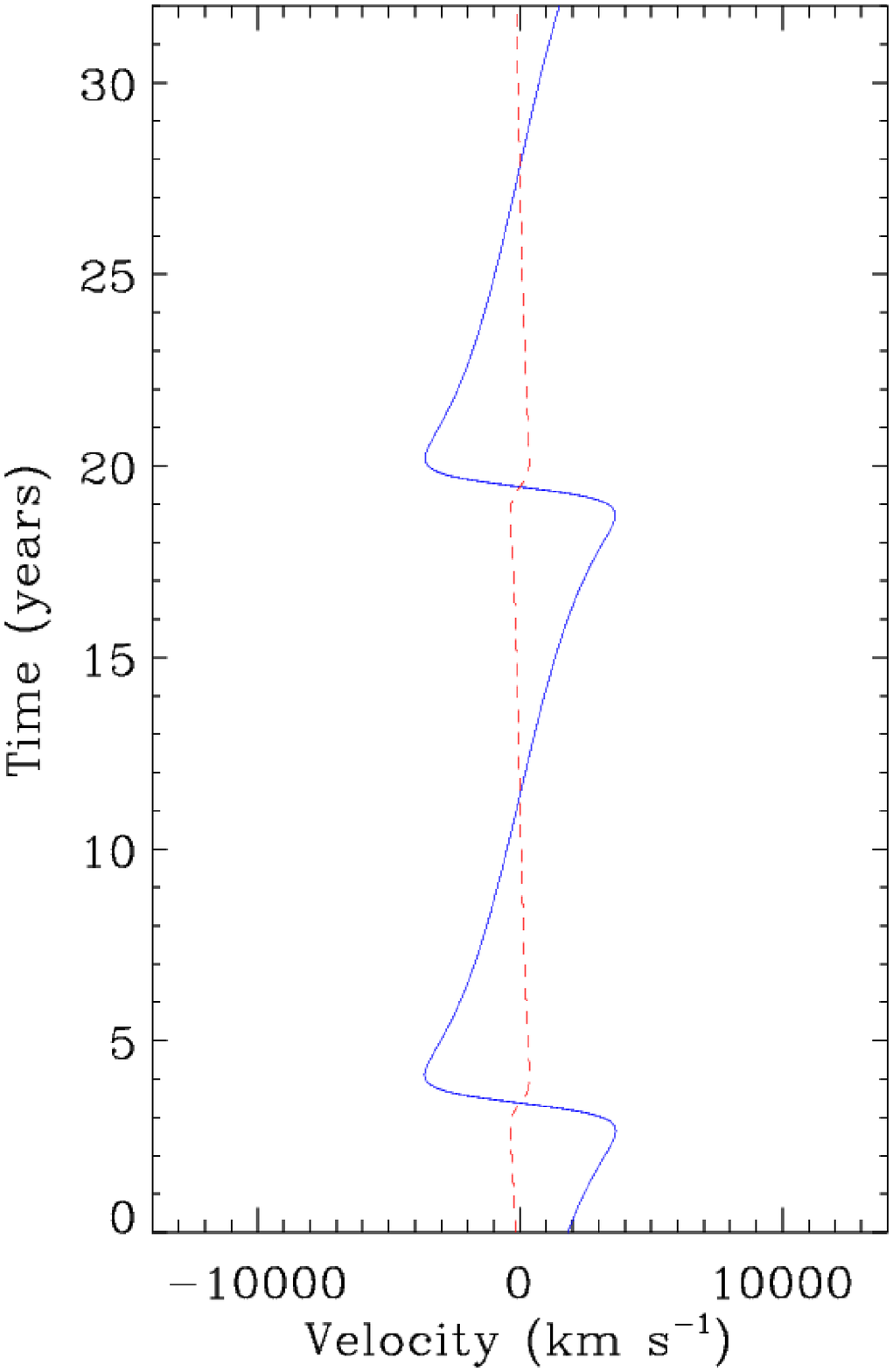}
\figcaption[b20.ps]{Trailed spectrogram (left) and black hole 
velocity curves (right) plotted for the SR-model. The trailed spectrogram
is a logarithmic gray scale map of the H$\alpha$ intensity against
wavelength and observed velocity. Darker shades mark higher
intensity. Notice the low relative intensity of profiles before the
start of accretion. The velocity curve panel shows the orbital velocities
of the primary ({\it dashed, red line}) and secondary ({\it solid,
blue line}) black holes projected along the line of sight to the
observer. The velocity curves are skewed because of the non-zero
eccentricity of the orbit.
\label{fig_trail_retro}}
\end{figure*}

We also calculate the broad H$\alpha$ emission-line profiles for both
the S and SR-models in order to better illustrate the kinematics of
the gas (see Figures~\ref{fig_prof_solar} and
\ref{fig_prof_retro}). All profiles have been calculated under the
assumption that the observer is located at a distance $d\rightarrow
\infty$ in the positive $xz$-plane, at $i=30^{\circ}$ to the $z$-axis.
Initially, from the unperturbed disk we observe double-peaked emission
line profiles but they gradually depart from this shape as the
perturbation propagates through the gas. For reasons discussed above,
the regular H$\alpha$ profiles from a quiescent, cold disk, have a low
intensity. The profiles appear variable on a time scale of months to
years, both in shape and width. The profiles in both sets show an
extended, low intensity red wing, most pronounced during the periods
of increased accretion and after the times of pericentric
passages. During such events a number of emitting gas particles find
themselves deeper in the potential well of the binary. The emitted
photons that leave the potential well are gravitationally redshifted
and Doppler boosted, thus contributing to the extended red wing and
the blue shoulder of the relativistic H$\alpha$ profiles. The changes
in the profile sequence can be better seen in the trailed
spectrograms, shown in Figures~\ref{fig_trail_solar} and
\ref{fig_trail_retro}, which are the 2D maps of the H$\alpha$
intensity against observed velocity (or wavelength). The H$\alpha$
emission in the trailed spectrograms from both S and SR-models appears
redshifted relative to the rest wavelength. In addition, the emission
in the extended red wing of the profile is easily noticeable,
especially in the SR-model (Figure~\ref{fig_trail_retro}), where this
effect is most dramatic after the pericentric passages of the
binary. Since Doppler boosting of the blue side of the line and
gravitational redshift of the red wing are pronounced during and after
the pericentric passages of the binary, they can serve as indicators
of the orbital period.  In the trailed spectrogram of the S-model it
is possible to discern the repeating behavior in the windows between 7
to 22 years and 22 to 36 years. The width of the H$\alpha$ profile
increases after the pericentric passages of the system, reflecting the
inflow of gas towards the primary. Additionally the widening of the
profile appears asymmetric and shifted towards the red with respect to
the pre-pericentric sequence of profiles. This shift is a signature of
the motion of the accretion disk which follows the primary on its
orbit. The shift is even more pronounced in the trailed spectrogram of
the SR-model which also exhibits narrower emission line profiles than
the S-model. This difference arises from the difference in kinematics
of the innermost portion of the accretion disk surrounding the
primary. In the SR-model the gas at the inner edge of the disk becomes
gradually less bound to the primary black hole over time due to the
interactions with the binary. We compare the trailed spectrograms with
the velocity curves of the two binary components projected onto the
line of sight (right panels in Figures~\ref{fig_trail_solar} and
\ref{fig_trail_retro}). Because of the nonzero eccentricity of the
orbit the velocity curves appear skewed. The variations in the profile
intensities in the S-model correspond to the features in the velocity
curve of the secondary in year 7 and 23, when it moves away from the
observer with the highest projected velocity. In the SR-model the
velocity curve of the secondary appears as a mirror image of certain
features in the trailed spectrogram because in this model the gas
rotates in the opposite direction with respect to the binary. In
general, we expect the signature of the velocity curves of the two
black holes to appear in the emission line profiles at pericentric
passages, because that is when the binary interaction with the gas is
strongest. If such features could be discerned in the observed
H$\alpha$ emission-line profiles, they would signal the presence of a
binary.

In practice, one may constrain the period and the mass ratio of the
binary from the periodicity and projected velocity components of the
two black holes, all measured from the H$\alpha$ profile sequence. The
determination of the individual black hole masses is more challenging,
because it requires knowledge of the inclination of the binary orbit
with respect to the observer. In general, the inclination of the
binary orbit may not be the same as that of the accretion disk but in
some cases it may be possible to determine it relative to the
accretion disk. If the binary orbit is {\it circular}, fewer
parameters are needed to constrain the properties of the system, and
it should be possible to determine the individual masses of black
holes, given a knowledge of the observed velocity components of the
two black holes at a single point of time, orbital period, and the
inclination of the system. To constrain the shape and the orientation
of an {\it elliptical} binary orbit one would in addition need
information about the velocity components at pericenter and apocenter,
as well as the orbital separation of the binary. Before any of these
parameters can be measured from observations of light curves and the
H$\alpha$ line sequence, it is necessary to follow at least a few
revolutions of the binary.

We note several other qualitative features of the H$\alpha$ profiles,
which can shed light on the motion of gas in the nuclear region.  In
both models, the central part of the profile, contributed by emission
from the low velocity gas, becomes more pronounced with time. This low
velocity core of the profile is contributed by filaments of gas which
have been expelled from the center at the expense of the angular
momentum of the binary. The same effect re-appears during the second
orbital passage when a new low velocity component is added to the
system. In this process filaments with a similar morphology are formed
during each passage, on a slightly different length scale, because the
older filaments have had time to expand, and with a slightly different
phase, due to the effect of rotation.

\begin{figure}[t] 
\epsscale{1.2}
\plotone{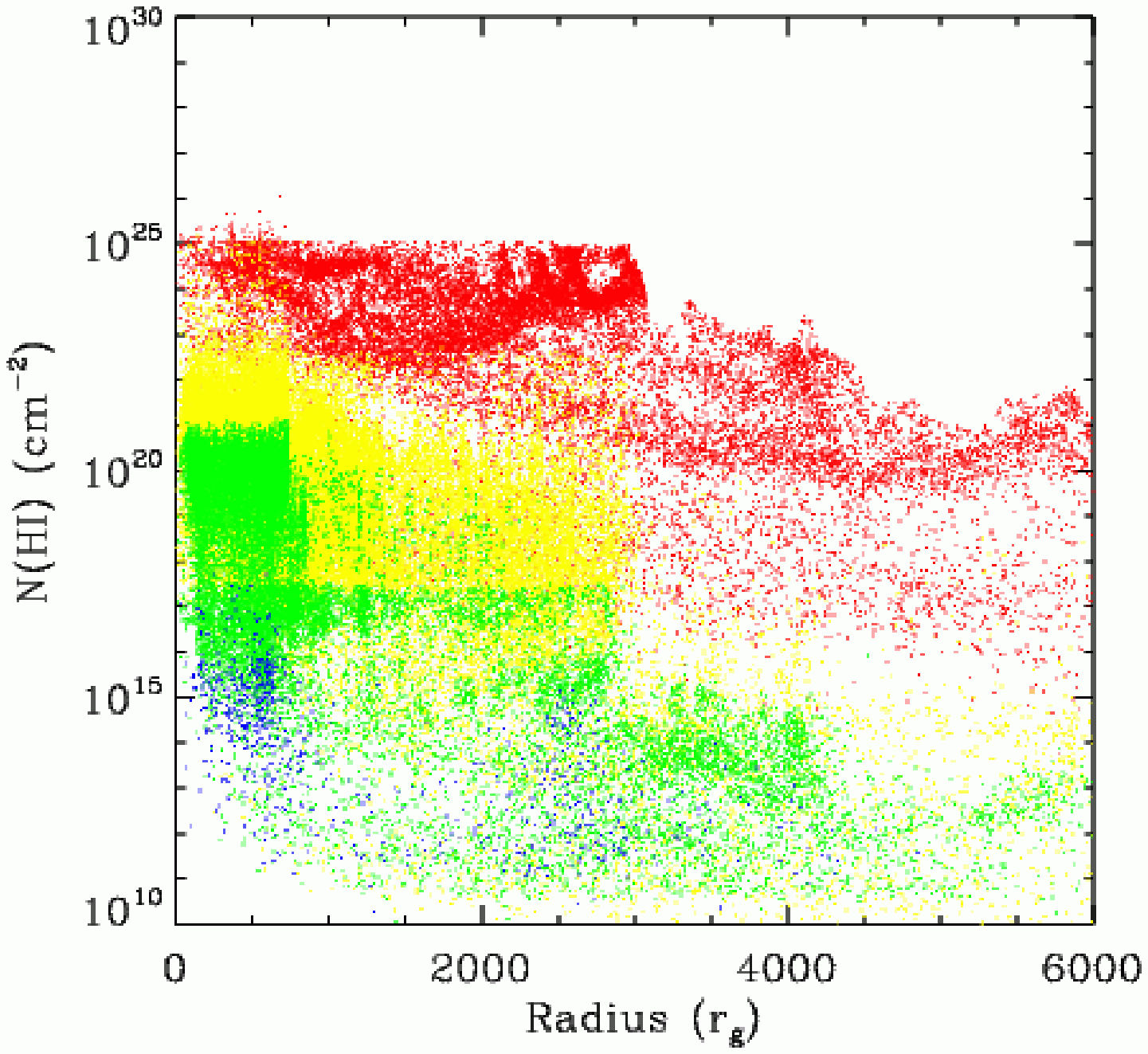}
\figcaption[b21.ps]{Neutral hydrogen column density of the gas as a 
function of radius, at a time of 9.4 years in the S-model. The colors
mark the temperatures of gas particles and are the same as in earlier
figures. This figure corresponds to the morphology of the disk as
plotted in panel 2 of Figure~\ref{fig_xy_sr}.
\label{fig_colDen}}
\end{figure}

Although it is expected that all AGNs host black holes fed by an
accretion disk, broad, double or multi-peaked emission-line profiles
are observed only in a small fraction of objects
\citep{eh03,strateva03}. The appearance of the H$\alpha$ line profiles
can be affected by physical conditions in the nuclear region, other
than the phase-space distribution of the emitting gas, such as optical
depth to H$\alpha$ photons and electron scattering.
Figure~\ref{fig_haEmiss} illustrates that a significant fraction of
the H$\alpha$ light is contributed by the hot ionized gas located in
the inner disk, via recombination. This component of the gas is
transparent to H$\alpha$ photons (Figure~\ref{fig_tauHa}).
Nevertheless, the profiles will likely be affected by the high
H$\alpha$ opacity in regions of the disk where the gas temperatures
are just below $\sim 10^6$ K and the neutral hydrogen column density
is high (Figure~\ref{fig_colDen}).

\begin{figure*}[t]
\epsscale{0.8}
\plotone{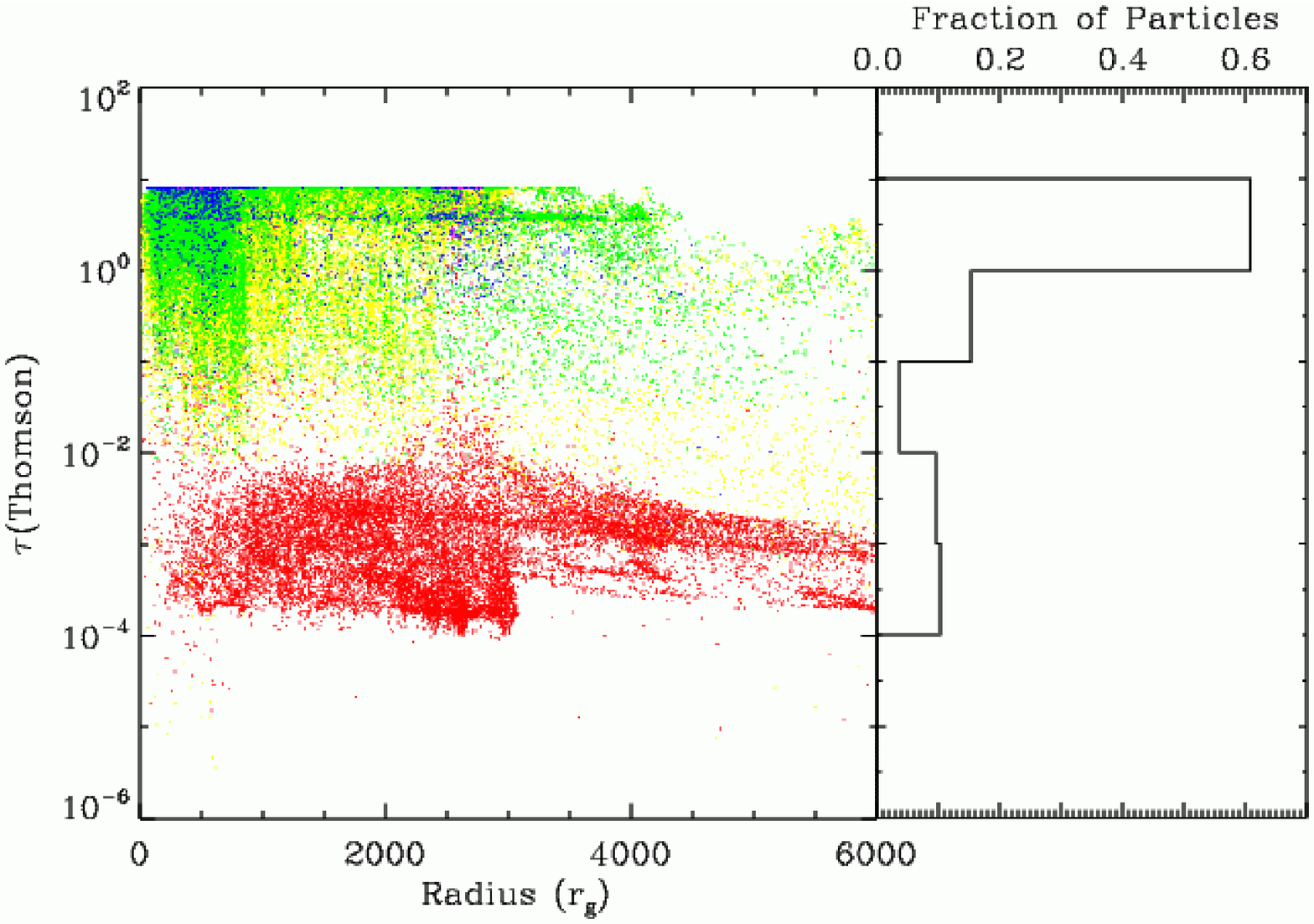}
\figcaption[b22.ps]{{\it Left:} Optical depth to Thomson scattering
of each gas cell as a function of radius, at a time of 9.4 years in
S-model. This figure corresponds to the morphology of the disk as
plotted in panel 2 of Figure~\ref{fig_xy_sr}. {\it Right}: Histogram
showing the fraction of particles as a function of Thomson optical
depth. \label{fig_tauT}}
\end{figure*}

The second factor, which impacts the shape of the H$\alpha$
emission-line profile, but to a lesser extent the H$\alpha$
luminosity, is electron scattering. As expected, the ionized gas will
have the highest optical depth to Thomson scattering
(Figure~\ref{fig_tauT}). We find that the maximum value of $\tau_T$ in
our calculation is about 10, observed for the hottest gas cells. About
60$\%$ of the mass of the gas has optical depth to Thomson scattering
larger than 1, in our simulations. When Thomson scattering is
significant, the hot corona with a high thermal velocity dispersion
($c_s\sim 10^2-10^3\, {\rm km~\,s^{-1}}$) could smooth out some
emission features in the H$\alpha$ profiles, although the emission
profiles should still show up clearly. Such conditions can be met in
nuclear regions hosting a strong photoionization source.

Although the effects of optical depth in the gas can play an important
role in estimates of H$\alpha$ luminosity and appearance of the
H$\alpha$ emission line profiles, we are not able to account for these
effects in our calculations. We used conservative values for the model
parameters, such as the radiative efficiency and covering factor,
throughout the paper so that the H$\alpha$ luminosity is not
significantly over-estimated. The problem of the propagation of
photons along the line of sight is numerically complex and is likely
to be accurately treated with the new generation of codes which
incorporate ray-tracing techniques in the radiative transfer
calculations.

\subsection{Evolution of the Binary Orbit}\label{S_orbit}

\begin{figure}[t]
\epsscale{1.2}
\plotone{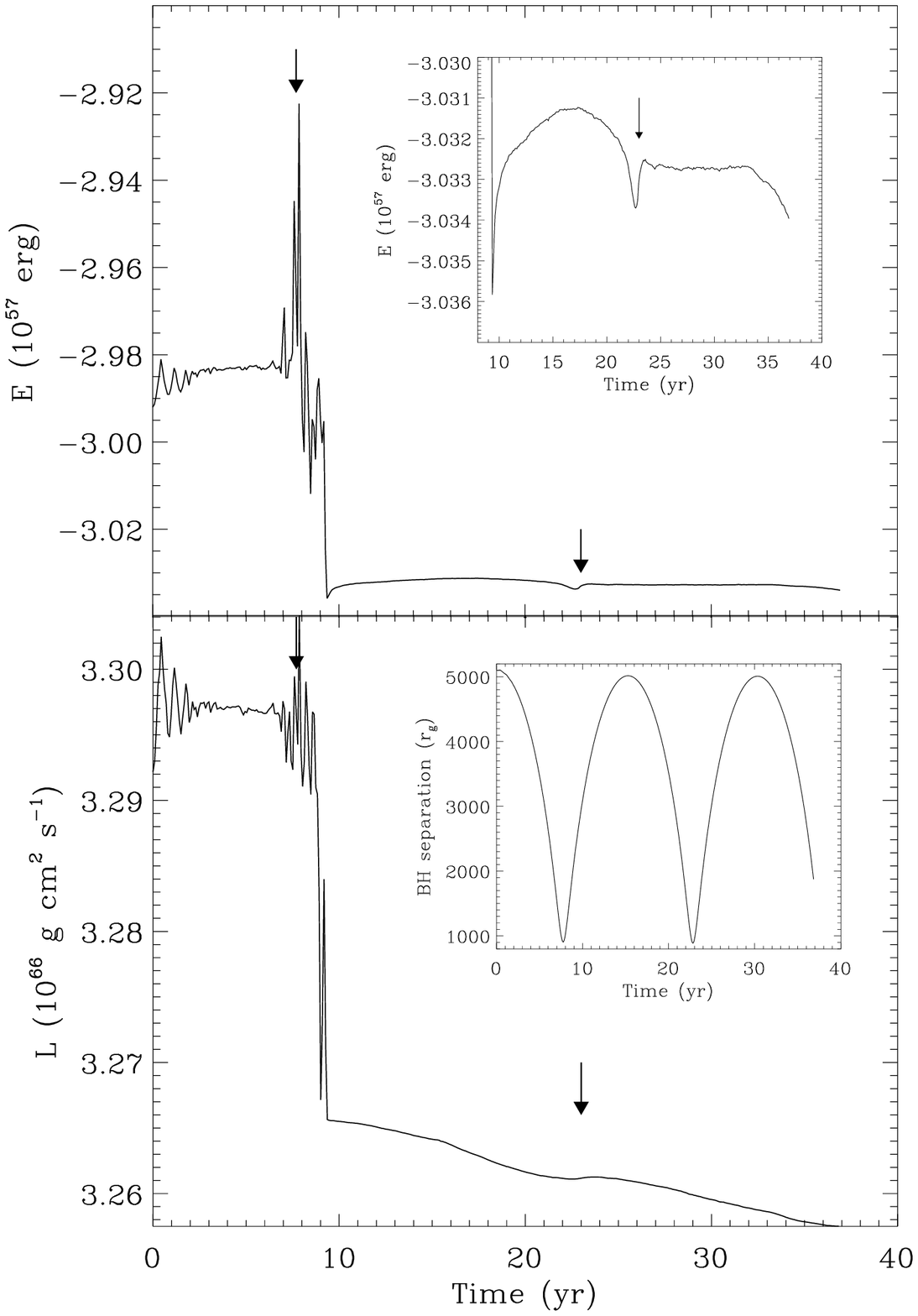}
\figcaption[b23.ps]{{\it Top:} Evolution of the binding energy of the binary
over two orbits in the BB-model. {\it Inset}: Binding energy
in the period 10 to 40 yr in more detail. {\it Bottom}: Evolution of
the orbital angular momentum of the binary. The arrows mark the times
of pericentric passages of the binary. {\it Inset}:  Evolution of the 
orbital separation of the binary. \label{fig_el_bb}}
\end{figure}

In this section we describe the changes in the binary orbit due to
interactions with the gas and compare them with analytical estimates.
The evolution of the binding energy and orbital angular momentum of
the binary in the BB, S, and SR-model is shown in
Figures~\ref{fig_el_bb}, \ref{fig_el_cs}, and \ref{fig_el_csr},
respectively. In all three models the binary system shows a net loss
of both energy and orbital angular momentum. The mean dissipation rate of
the binding energy, calculated from the simulations is $dE/dt\approx
4\times 10^{46}\,{\rm erg\,s^{-1}}$ for the BB-model and $1\times
10^{46} \,{\rm erg\,s^{-1}}$ in the S, and SR-model. As a result the
second pericentric passage in the S-model occurs about 100 days later
than in the BB-model (see Figures~\ref{fig_xy_bb} and \ref{fig_xy_sr}
at a time of 23.2 years).  The mean rate of energy dissipation due to
emission of gravitational radiation is $(dE/dt)_{gw} \approx 7 \times
10^{42}\,{\rm erg\,s^{-1}}$ and thus the evolution of the binary orbit
is dominated by interactions with the gas. Because the rate of
gravitational wave emission is very similar for all three models we
show it only once, in the inset of Figure~\ref{fig_el_cs}.

In all three models the binary is initially placed on the same orbit,
hence the differences in evolution of the orbital parameters among the
models are entirely due to the kinematic and thermodynamic properties
of the gas. Recall that in the BB-model the gas initially
redistributes into two phases with a temperature contrast of about an
order of magnitude (\S~\ref{S_BB-model}). The cold phase present in
the BB-model is missing in the S-model, where the secondary black hole
interacts with the warmer and more tenuous gas and consequently
dissipates its energy more gradually. This is consistent with the
findings of \citet{escala2} that the binary evolution depends strongly
on the ``clumpiness'' of the gas, and that the binary suffers a
stronger deceleration in gaseous media with weaker pressure
support. After the first pericentric passage, the cold gas phase in
the BB-model disappears, due to shock and photoionization heating, and
E and L evolve in a less dramatic way.

The rates of dissipation of energy and orbital angular momentum in
Figures~\ref{fig_el_bb}, \ref{fig_el_cs}, and \ref{fig_el_csr} are
functions of the orbital phase of the binary because the binary spends
a fraction of its time moving through the disk. Therefore, the
interaction of the binary and the accretion disk occur in two
different regimes studied by \citet{lp79a,lp79b}. In the first regime
the disk is confined to the Roche lobe of the primary black hole\ and
the secondary is outside the disk and interacts with it tidally.  For
a mass ratio $q$ and and a Reynolds number $\mathcal{R}$ \citet{lp79a}
show that when $q > \mathcal{R}^{-1}$, the tidal force overtakes the
viscous expansion of the disk, and the secondary effectively removes
the angular momentum at the outer edge of the disk by truncating the
disk (in our simulations $q = 0.1$ and $\mathcal{R}\sim 10^5$).  The
excess angular momentum is transported to the binary via tidal
torques.  In the second regime, the secondary is orbiting within the
circumbinary disk and truncate the inner edge of the disk
\citep{lp79b} through its tidal forces. In this regime the disk acts
as an effective energy and angular momentum sink and the binary should
evolve towards a more bound state via resonant interactions with the
disk. This is qualitatively consistent with our findings from the BB
and S-models, where the binary is co-rotating with the disk and is
found to reach the minimum binding energy when both black holes are
close to the pericenter and within the disk.

An additional torque arises from the fact that the velocity vectors of
the binary on the elliptical orbit and the gas around the primary on
circular orbits continuously change their respective orientation; thus
the torque acting on the binary changes accordingly. In the
co-rotating case, the secondary experiences a torque that pushes it
away from the primary as it approaches pericenter and a torque that
pushes it inwards as after it passes pericenter. This effect can be
seen in Figure~\ref{fig_el_cs}: the binary evolves from a more bound
to a less bound energy state as it passes pericenter and the slope of
the energy curve is negative. In the counter-rotating case, the
secondary and the disk orbit in the opposite sense and the effect of
torques is reversed. Figure~\ref{fig_el_csr} shows this reversal: the
slope of the energy curve is positive during the pericentric passage.

We first compare the rate of dissipation of orbital energy of the
binary with that predicted by \citet{ostriker99} for the case where a
massive perturber moving supersonically experiences the dynamical
friction force arising from the wake that formed in the ambient
medium. This rate can be expressed as
\begin{eqnarray}
\lefteqn{\left(\frac{dE}{dt} \right)_{df}  =    
\frac{4\pi\, (GM_p)^2 \,\rho}{\upsilon} ~
\ln\left(\frac{r_{max}}{r_{min}} \right) \approx }\nonumber \\
& &  2\times 10^{46}\, \left(\frac{M_p}{10^7\Msun}\right)^2 
\left(\frac{n}{10^{11}\,{\rm cm^{-3}}} \right) \left(\frac{\upsilon}{10^4\,{\rm km\,s^{-1}}}\right)^{-1}
{\rm erg\,s^{-1}}\; ,\nonumber \\
\end{eqnarray}
where $M_p$ and $\upsilon$ are the mass and velocity of the perturber, $\rho$
is the density of the gaseous medium, and $r_{max}$ and $r_{min}$ are
related to the linear size of the medium and the perturber. If we
assume that $r_{max}$ is comparable to the radius of the disk and
$r_{min}$ to the accretion radius of the black hole, then $\ln
(r_{max}/r_{min}) \approx 5$. In our simulations the wake of the hot
gas trailing the secondary black hole can be seen in panel 1 of
Figures~\ref{fig_xy_bb} and \ref{fig_xy_sr}.

However, as discussed by \citet{lp79a,lp79b}, tidal torques and
resonant scattering play an important role in the interaction of a
binary with a shear flow. We also note that dynamical friction
cannot be the dominant dissipation mechanism in the SR-model. In this
model, the disk rotates in the opposite sense from the binary and the
wake formed by the secondary is efficiently sheared away from the
secondary. We utilize the expression from \citet{lp79b} for the tidal
torque exerted by the binary on the disk in the process of gap
formation and obtain
$$
\left(\frac{dE}{dt} \right)_{rt} = 
\frac{8}{27}\; q^2 \Omega^3 r^4 \,\Sigma \left(\frac{r}{A} \right)^3 \approx 
\hbox to 4truein {\hfill}
$$
\begin{eqnarray}
 6\times 10^{45}\, \left(\frac{q}{0.1}\right)^2 \!
\left(\frac{M_p}{10^7\Msun}\right)^{3/2} \!
\left(\frac{n}{10^{11}\,{\rm cm^{-3}}}\right)   
\left(\frac{H}{50\, r_g}\right) \nonumber \\
\times \left(\frac{r}{10^3\, r_g}\right)^{5/2} \!
\left(\frac{A}{10^2\, r_g}\right)^{-3} \!
\,{\rm erg\,s^{-1}}
\end{eqnarray}
where $\Omega$ is the angular velocity of the gas interacting with the
black hole at radius $r$, $\Sigma$ is the surface density of the gas
disk, $A$ is the half-size of the gap, and $H$ is the half-thickness
of the disk. The results of our simulations are consistent with both
mechanisms being responsible for the dissipation of the binary orbital
energy. We also note that dynamical friction cannot be the dominant
dissipation mechanism in the case of the binary and gas disk rotating
in the opposite sense from each other. Assuming that the binary
maintains a rate of energy dissipation of $\sim 10^{46}\,{\rm
  erg\,s^{-1}}$, it would proceed to coalescence in only $\sim
10^3\,$yr. Most likely, the rate of inspiral due to interactions with
the gaseous background will not persist at this high rate over long
periods of time.  \citet{armitage02} for example discussed the
evolution of a binary on a circular orbit, in the gas disk, from
separations of 0.1 to $10^{-3}$~pc.  Extrapolating from their results
one obtains that from the separation of 0.01~pc, a combination of
disk-driven migration and inspiral due to gravitational radiation
leads to coalescence within few$\times 10^6$ years.  In order to
obtain a robust estimate of the orbital decay rate one should follow
the evolution of the binary and gas starting from much larger
separations. While this task was undertaken by a number of authors
\citep{kazantzidis05,escala1,escala2,dotti06a,mayer06} it is still a
computational challenge to extend the wide dynamic range of the
simulations to the lowest binary separations, where gravitational
radiation overtakes the dissipation caused by the gaseous background.

\section{Discussion}\label{S_discussion}

Here we discuss the impact of underlying assumptions in our models on
the observable electromagnetic signatures and the relevance of our
results in the context of the larger scale simulations.

\begin{description}

\item[\it Mass of the gas disk.]  A more massive disk would change the
gas drag, hence make the binary evolve faster and vice versa. This
would affect the number of pericentric passages of the binary and in
turn the number of peaks in the cooling rate and X-ray light curves
occurring over a given period of time. The number of shifts in
wavelength space noticeable in the H$\alpha$ trailed spectrogram would
also change, reflecting the change in the orbital period of the
binary. This may complicate the prediction and observational followup
of the outbursts, since after the first few orbital cycles the
uncertainty in the orbital period of the binary will be rather large
(weeks to months; see the example of OJ~287). If a decrease in the
orbital period of a MBHB candidate can be determined from the observed
light curves, it would imply that the binary is dissipating orbital
angular momentum via interactions with the gas disk or emission of
gravitational waves.

The MBHBs likely to be discovered by monitoring of light curves and
H$\alpha$ emission line profiles of candidate galaxies would likely
have orbital periods of order of 10~yr or less, implying binaries with
subparsec orbital separations. The dominant dissipation mechanism
could be determined if a constraint on the masses of black holes can
be put in addition to the rate of change of the orbital period. One
could then compare the observed rate of inspiral to that predicted by
the theory of general relativity for some assumed eccentricity; if the
two are discrepant this could be explained by presence of the gaseous
dynamical friction.

\item[\it Structure of the disk.] Besides the total mass, the
efficiency of dynamical friction depends on the structure of the
accretion disk. The intensity of the gas drag is stronger in systems
where the density of the gas is higher and where a massive body moves
supersonically with respect to the ambient medium
\citep{ostriker99}. The structure of nuclear accretion disks on a
subparsec scale is rather uncertain and in our calculations we assumed
that the disk is geometrically thin. The results from the larger scale
simulations by \citet{escala2} and \citet{mayer06} provide useful
insights into the effect that a different equation of state of the gas
disk can have on the orbital decay of the binary at separations of
10$-$100~pc. Both groups find that the binary inspiral is more
efficient in galactic disks with less pressure support and smaller
scale heights. Such disks resemble clumpy and dense disks in
starforming galaxies. In addition, \citet{mayer06} find that for a
galactic merger to result in a bound pair of massive black holes and
eventually a coalescence within the Hubble time, the AGN should not
play an important role as an source of heat in the disk during the
merger phase.

\item[\it Role of starburst and AGN.] A number of authors found that
when a large fraction of the disk gas is funneled to the central
1~kpc, following a galactic merger, it can lead to nuclear starburst
on timescale less than $10^8$~yr
\citep*{barnes96,dsh05,sdh05}. Because it may take $10^7$~yr or longer
for a binary to sink from kiloparsec to subparsec scales, the MBHB
inspiral and coalescence in such galaxies may coincide with an active
starburst \citep{dotti06b}. It appears that if a starburst reaches its
peak before the MBHB merges, a large fraction of the gas may be
already turned into stars while the remaining gas will be exposed to
the starburst feedback, thus diminishing the role of gas as a
dissipation agent in the MBHB inspiral. The resulting accretion rate
onto the black holes may be too low to give rise to observable
signatures in the form of periodic outbursts and variable H$\alpha$
emission line profiles; these galaxies would appear as ultraluminous
or luminous starburst galaxies \citep{fabian98, dotti06b}. However,
if the onset of a strong AGN activity in either of the parent nuclei
precedes the starburst, the mechanical and radiative feedback could
inhibit both the starburst and further accretion onto the black holes
and thus suppress the electromagnetic signatures of the binary.

\begin{figure}[t]
\epsscale{1.2}
\plotone{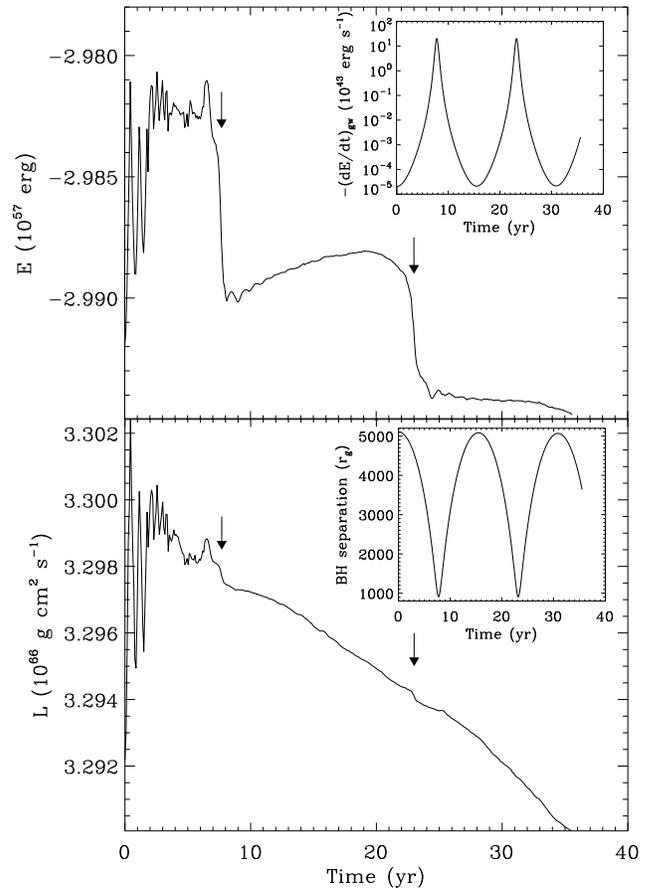}
\figcaption[b24.ps]{Similar to previous figure, except for the
  S-model.  {\it Top inset}: Energy loss rate due to emission of
  gravitational waves over time. The binary loses orbital energy via
  gravitational waves at a mean rate of about $7\times10^{42}\,{\rm
    erg\,s^{-1}}$ and emits with the highest rate close to the
  pericenter. {\it Bottom inset:} Evolution of the orbital separatio n
  of the binary. \label{fig_el_cs}}
\end{figure}

\item[\it Model of radiation physics.]  Since a fully 3D radiative
transfer calculation has not been developed yet the model of radiation
physics that is employed here relies on several assumptions. Here we
address the key assumptions that can directly affect the balance
between heating and cooling in the simulation, and consequently the
electromagnetic signatures of the binary system. The first assumption
is related to the covering factor, $\zeta$, used to parametrize the
effects of self-shielding and geometry when gas is illuminated by the
incident radiation (equation~\ref{eq_bb_heating}). In our calculation
this coefficient has a fixed value while realistically it is a
function of time and the position of a particle with respect to the
source of illumination.  Varying $\zeta$ would result in a different
temperature in the gas hence, different amounts of hot and cold
components of gas. For example, a larger value of $\zeta$ would amount
in an increase of the diffuse, hot component of gas. This would boost
the H$\alpha$ and X-ray luminosities but may not affect significantly
the conclusions about the variability of these
components. Alternatively, a smaller value of $\zeta$ would result in
more gas being in the cold phase and confined to the plane of the
disk. In this scenario both the H$\alpha$ and X-ray luminosities would
be lower, and the H$\alpha$ light could also exhibit a periodicity.
This is because the colder and denser gas is more susceptible to
shocks induced by the secondary black hole.

\begin{figure} 
\epsscale{1.2}
\plotone{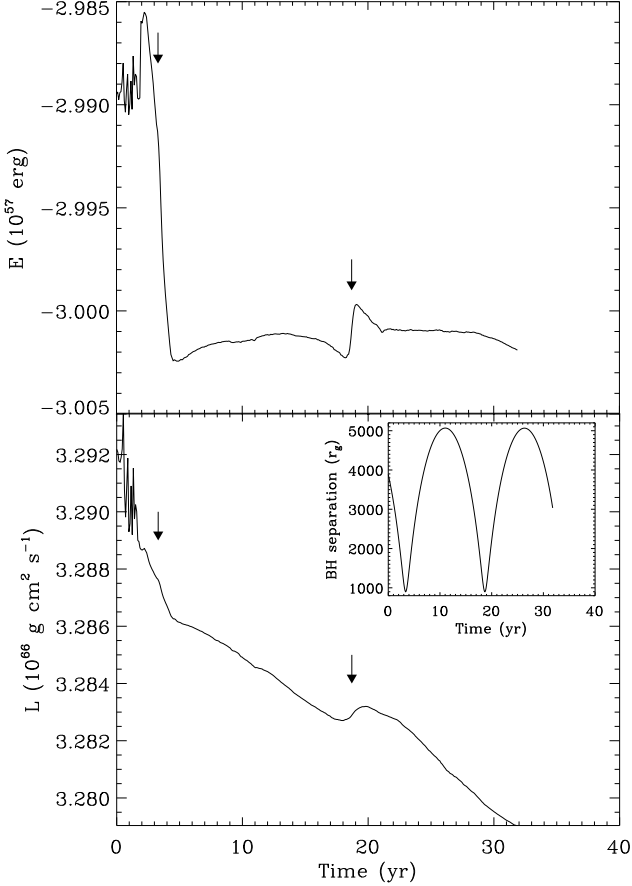}
\figcaption[b25.ps]{Similar to previous figure, except for the SR-model.
\label{fig_el_csr}}
\end{figure}

\begin{figure}[t]
\epsscale{1.2}
\plotone{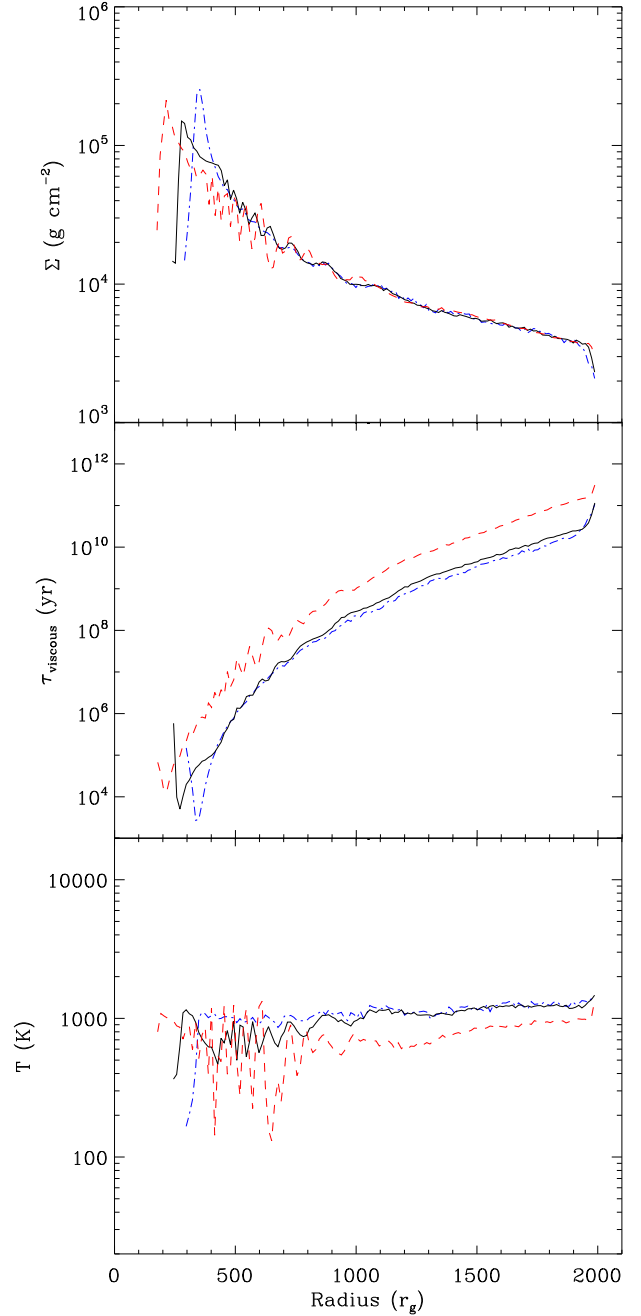}
\figcaption[b26.ps]{The effect of resolution on the disk surface
density (top), the viscous time scale (middle), and the temperature
profiles (bottom) in models, where the secondary black hole was not
introduced to the simulation. The profiles shown are for runs 20k IC1
({\it dash-dot,blue line}), 100k IC2 ({\it solid, black line}), and
500k IC3 ({\it dashed, red line}), all at a time of 8 years after the
beginning of a simulation. The gas was assumed to have a solar
metallicity. See discussion in \S~\ref{S_discussion} for more details.
\label{fig_res_sden_temp}}
\end{figure}

Another important assumption in our calculation is that radiative
cooling of the gaseous accretion disk can be approximated by a sum
over the discrete gas particles. Each particle represents a gas cell
of mass 0.1$\Msun$. The approximation should converge to an exact
solution when $N_{part}\rightarrow \infty$ in case when the gas is
optically thin to radiation. We thus conclude that for the diffuse gas
component in our simulations this approximation should be a
satisfactory description and we turn attention to the optically thick
component. This component has a higher neutral hydrogen column density
(Figure~\ref{fig_colDen}) and is confined to a disk with the thickness
corresponding to one smoothing length, $h_{sml}$. Therefore, the
vertical structure of the disk is not resolved and every gas cell
within the disk should cool by emission from the top and bottom of its
surface. This is equivalent to a geometrically thin and optically
thick accretion disk model where gas cools by emission from the two
faces of the disk. In our model the gas cells radiate isotropically,
over their entire surface area and according to the cooling rate per
unit area determined by {\it Cloudy}. Thus we overestimate the total
cooling from the optically thick component of gas by a factor of
approximately 2.  The largest error is expected for the BB-model,
where all gas particles are assumed to be optically thick, while the S
and SR models should provide a more accurate description.

\begin{figure}[t]
\epsscale{1.2}
\plotone{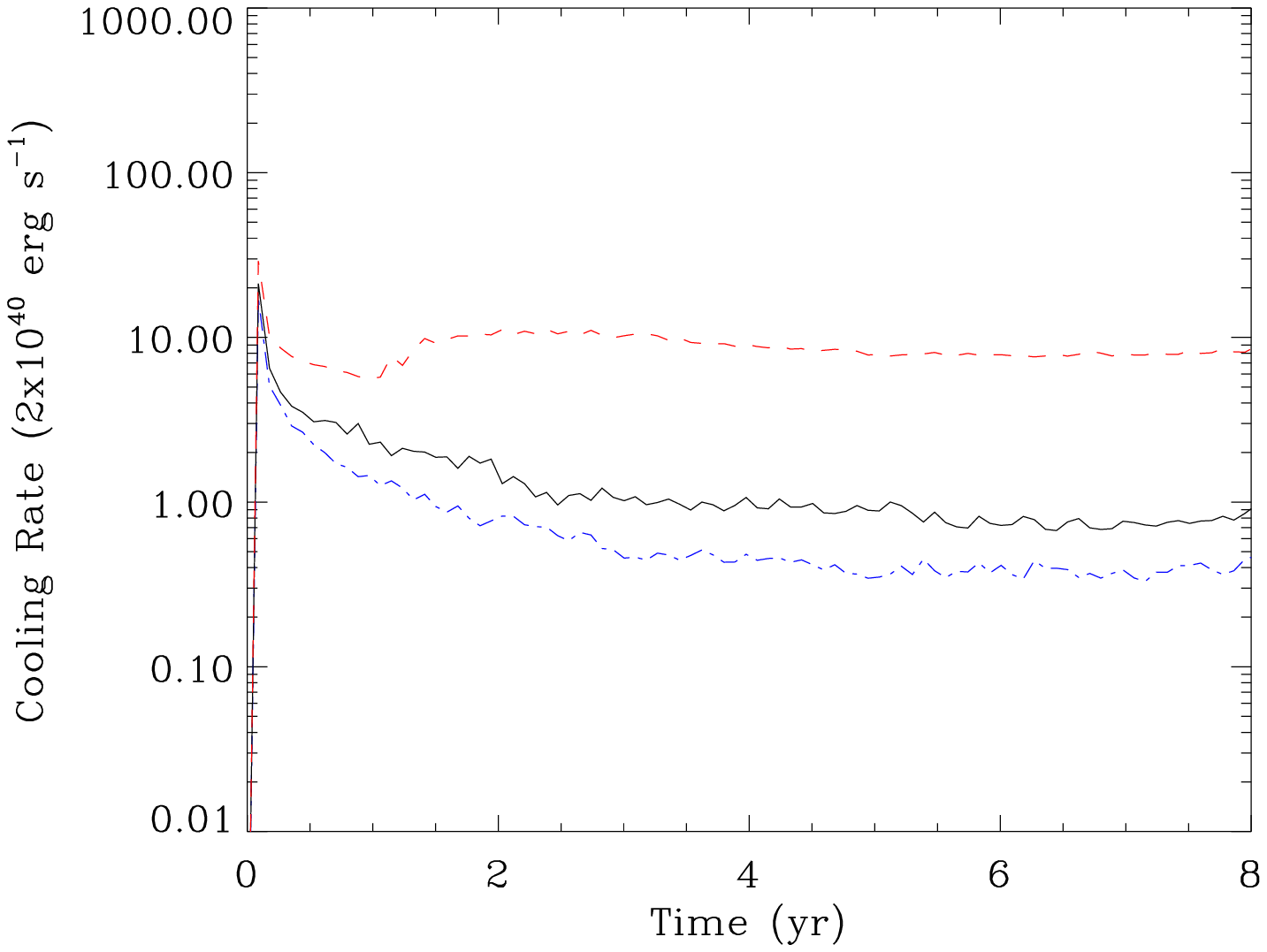}
\figcaption[b27.ps]{Dependence of the cooling rate on resolution in
a model where the secondary black hole is absent and the solar
metallicity gas disk is cooling passively. The curves are from
runs 20k IC1 ({\it dash-dot, blue line}), 100k IC2 ({\it solid, black
line}), and 500k IC3 ({\it dashed, red line}), at the time 8 years
after the beginning of each simulation.  \label{fig_res_cool}}
\end{figure}

\item[\it Role of resolution.]  Increasing the resolution yields a
better accuracy in the calculation of angular momentum transport and
at the same time it may cause larger errors in the treatment of
cooling in our models. We chose 100k particles for the BB, S, and SR
models and carried out several tests in order to assess the role of
resolution in the heating, cooling, and evolution of angular momentum
in our simulations. The first set of tests is carried out by evolving
the initial conditions (the IC-runs, see Table~\ref{T_results}), with
the primary black hole and the solar metallicity gas disk, before the
secondary black hole is introduced. Thus we can isolate the dependence
of angular momentum transport and cooling on resolution, in absence of
shocks. The results of the three runs, IC1, IC2, and IC3 with 20k,
100k, and 500k particles, respectively, are shown in
Figures~\ref{fig_res_sden_temp} and \ref{fig_res_cool}, after the
initial conditions were evolved for 8 years. The evolution of angular
momentum appears strongest in the 20k run where the pressure gradient
drives the evolution of the surface density at the inner disk radius
(see the top panel of Figure~\ref{fig_res_sden_temp}) while at larger
radii the distributions appear consistent. The dominant component of
pressure is thermal gas pressure; the viscous pressure is of lesser
importance and it operates on much longer time scales. The accretion
disks at lower resolution thus appear to have more gas pressure
support than at higher resolution. Lower resolution also results in
more viscous disks and shorter viscous time scales, as shown in middle
panel of Figure~\ref{fig_res_sden_temp} where we plot the viscous time
scale, $\tau_{viscous}\sim r^2 / (\alpha_{Gadget}\; h_{sml} \; c_s) $.

\begin{figure}[t] 
\epsscale{1.2}
\plotone{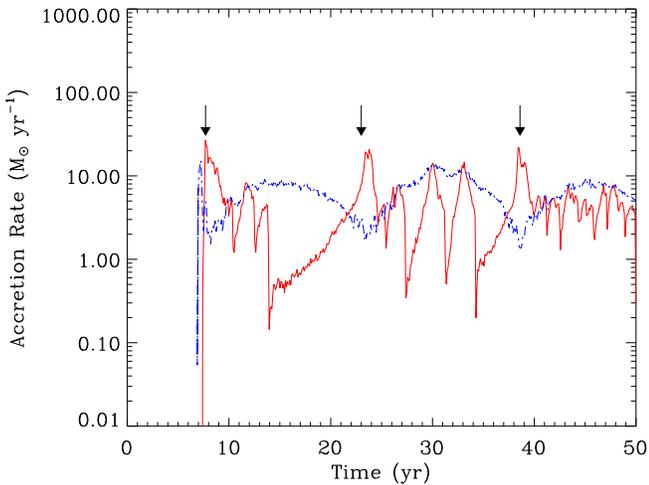}
\figcaption[b28.ps]{Effective accretion rate on the primary ({\it
solid, red line}) and secondary ({\it dashed, blue line}) black holes
calculated from the 20k S1 run, a lower resolution equivalent of the
100k S-model run (Figure~\ref{fig_acc_sr}). Note that this model has
been run for a longer time (close to 3 orbital periods) and that three
peaks are noticeable in the accretion curve. The arrows mark the times
of pericentric passages of the binary.  \label{fig_acc_20k}}
\end{figure}

The bottom panel of Figure~\ref{fig_res_sden_temp} shows the
temperature profile of the disk calculated at three different
resolutions, from the same snapshot in time. The disk in the 500k IC3
is the coldest compared to the 100k IC2 and 20k IC1 runs. This is a
consequence of the dependence of the cooling model on the resolution.
In the solar metallicity models, the disk cooling rate is expected to
scale with the total number of gas particles and their emitting volume
as $\propto N_{part}\, h_{sml}^3 = const $.  Figure~\ref{fig_res_cool}
shows the cooling curves of an unperturbed, passively cooling disk for
the three runs with the cooling being strongest in the 500k one.  The
scaling notably departs from the analytical expectation because the
adaptive smoothing length, $h_{sml}$, does not decrease with the
resolution as quickly as $\sim N_{part}^{-1/3}$. Consequently, the
higher resolution runs are characterized by a larger cooling volume
with respect to the lower resolution ones. Because the cooling rate of
a black-body emitter is characterized by its emitting surface area,
rather than volume, the dependence of the cooling on resolution is
different in the BB-model. In this case the cooling rate is nominally
expected to increase with resolution as $\propto N_{part}\, h_{sml}^2
\sim N_{part}^{1/3}$ (see S~\ref{S_cooling}).  However, the cooling
rate may have stronger dependence on $N_{part}$ due to the behavior of
$h_{sml}$ with resolution, as described previously for the solar
metallicity case.

\begin{figure*}[t] 
\epsscale{0.7}
\plottwo{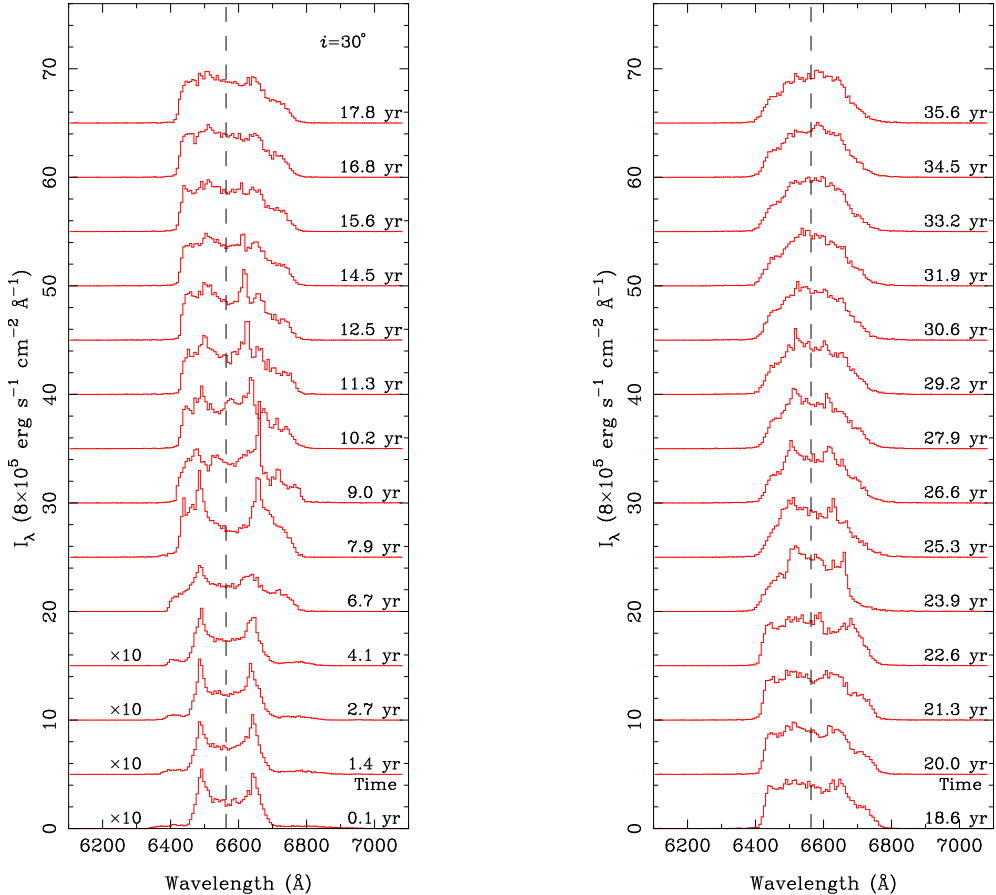}{b29b.ps}
\figcaption[b29.ps]{Sequence of H$\alpha$ emission line profiles
calculated from the 20k S1 run, a lower resolution equivalent of the
100k S-model run. The emission line profiles are plotted for the same
time slices as in Figure~\ref{fig_prof_solar}, for comparison. Note
the lower intensity scale for the profiles in 20k S1 run but otherwise
very similar profile widths and shapes to those in 100k run. The
inclination of the plane of the disk with respect to the observer is
as marked on the figure. The vertical dashed line at 6563~{\rm \AA}
marks the H$\alpha$ rest frame wavelength.  \label{fig_prof_20k}}
\end{figure*}

In order to estimate how much the cooling rates measured from 
our simulations suffer from the systematics discussed in this section, we 
compare the cooling rates for the passively cooling disk from 
the IC runs with the analytically calculated cooling rate for a black body 
with $T=1000\,$K and surface area equal to that of the disk.  The 
comparison is meaningful because the cold, nearly
isothermal, optically thick and geometrically thin disk can be
described as a black body until it is perturbed or illuminated. We
obtain a cooling rate of $L_{bb}=3\times10^{41}\,{\rm erg\,s^{-1}}$,
which is slightly higher but within an order of magnitude of the
cooling rate in the 100k IC2 run (Figure~\ref{fig_res_cool}). The same
is true for the early portions of the cooling curves in the BB, S, and
SR models, before the passage of the secondary through the disk
(Figures~\ref{fig_cool_bb}, \ref{fig_cool_sr}, and
\ref{fig_cool_sr2}). Their difference from the analytic estimate value
can be explained by slight departures of the temperature of the gas
particles from $1000\,$K. We conclude that the cooling rates
calculated in our models with 100k particles are in fair
agreement with the analytic estimate in the limit of an optically
thick gas disk. We also discuss a possibility for further improving the BB
cooling model and suppress the dependence of cooling on resolution in
\S~\ref{S_prospects}.

In the second set of tests we examine the effect of resolution on the
accretion rates and the variable H$\alpha$ emission line profiles in
the presence of the secondary black hole. These quantities are
connected to the resolution through the graininess of the gas and
through the artificial viscosity, which leads to different results for
shocks.  For the purpose of the test we carried out the S1 simulation
of the binary with a prograde disk using 20k particles (analogous to
the 100k S model). The accretion rate curves
(Figure~\ref{fig_acc_20k}) show the same prominent features as the
ones in the 100k S-model (Figure~\ref{fig_acc_sr}). The total
accretion rate in the 20k S1 run is in the range $1-10\,\Msun\,{\rm
yr^{-1}}$ and is slightly higher than in the 100k S run. In the lower
resolution run the contrast between the accretion rate of the
secondary black hole at pericenter and apocenter appears higher,
creating pronounced periodic humps in the accretion curve. This is
likely the result of the Bondi-like accretion on the secondary
coupled with a lower particle number density in the 20k S1 run.

In addition, we calculate the H$\alpha$ emission line profiles from
the 20k S1 run in the same way as in the S model; the results are
shown in Figure~\ref{fig_prof_20k}. As a consequence of the lower
cooling in the 20k run, the intrinsic intensity of the H$\alpha$ lines
is several times lower than in the 100k S-model run but the integrated
H$\alpha$ luminosity is the same, $L_{H\alpha}\sim 10^{40}\;{\rm erg\,
s^{-1}}$. This can be understood having in mind that while the
intrinsic line intensity decreases at lower resolutions, the smoothing
radius used to estimate the emitting surface of a gas particle,
increases, and the two effects cancel out in calculation of
luminosity. The shape and width of the profiles show the same dominant
features as the profiles in 100k S-run and do not appear to be very
sensitive to the change of resolution.

While employing a higher number of particles has a clear advantage we
find that a satisfactory accuracy is achieved with 100k particles
given the time scales involved in our simulations. The evolution of
angular momentum in the 100k runs is slow and the viscous evolution
even slower and they do not result in shocks and transients in the
gas. The estimated systematic error in cooling rate due to the
dependence on resolution is within one order of magnitude and the
shapes of the H$\alpha$ emission line profiles appear robust.

\item[\it Role of accretion model.] In the last paragraph of
\S~\ref{S_binary_simulations} we describe the approximation we used in
order to estimate the accretion rates of the two black holes given the
properties of their perturbed but unresolved accretion flows. We
tested whether the accretion rate onto the secondary black hole
affects the results by lowering it by a factor of 100 in one of the
models. We refer to this as the S2-model.  From this run, in which we
employed 20k particles, we calculated the observational signatures and
found that they are very similar to the 20k S1 and 100k S-models
(specifically, the cooling curve features and periodicity, H$\alpha$
profiles, and H$\alpha$ light curve). The implication is that varying
the accretion rate efficiency of the secondary black hole with respect
to that of the primary has a weak effect on the resulting
observational signatures. As long as the primary black hole maintains
equal or higher accretion rate than the secondary, the latter can be
regarded as a passive perturber. We also note that the estimates of
accretion rates in both accretion models are rather conservative, thus
it is more likely that we underestimate rather than overestimate the
accretion rate.

\end{description}

In our simulations we modeled the binary with the semi-major axis of
0.015 pc, in the regime when the dissipation due to the gaseous
background is still dominant over the gravitational radiation. This
regime corresponds to the disk-driven migration stage in the 2D grid
simulations of a binary and locally isothermal disk by
\citet[][hereafter, AN] {armitage02}.  Because there is an overlap in
the scales treated by our respective simulations we compare the
results of the two works. Note that in our simulations the binary
orbit is assumed to be elliptical and the disk surface density
initially corresponds to $\sim10^5\,{\rm g\,cm^{-2}}$ at the distance
of 100$\,r_g$ from the primary, in comparison to the circular orbit
and the surface density of $\sim10^6\,{\rm g\,cm^{-2}}$ in AN.  We
find that the low density quasi-spherical outflow in our models
reaches the scale of 1~pc in about 40 years, implying the outflow
velocities of $\sim 10^4\;{\rm km\,s^{-1}}$, in agreement with the
prediction of AN. Unlike AN, we find that in the phase of the
disk-driven migration the accretion rate of the primary is higher than
the rate expected in the absence of a binary. The rate expected in the
latter case is $\sim10^{-3}\,\Msun\,{\rm yr^{-1}}$ while the rate
measured from our simulations $\geq 10^{-2}\,\Msun\,{\rm yr^{-1}}$. We
attribute this difference to a temperature distribution of the gas and
the eccentric binary orbit. In our simulations, in general case,
different gas phases can be found to reside at the same radii in the
disk (Figures~\ref{fig_temp_solar} and \ref{fig_temp_retro}). The cold
component of gas at T$\sim10^3$~K is mostly confined to the plane of
the binary orbit where it gets swept by the secondary black hole
during every passage through the disk and accreted.  In addition, a
binary on an eccentric orbit effectively perturbs the accretion disk
and drives the spatially extended high velocity outflows. The outflows
carry away the excess angular momentum, allowing the gas at the
innermost region of the disk to be accreted.

\break
\subsection{Future Prospects}\label{S_prospects}

The physical parameter space of a binary and gas disk system is very
large and a thorough parameter study is beyond the scope of this
paper. In the future it will be useful to investigate systems with
different mass ratios in which the binary orbit and gaseous disk are
not coplanar. In these systems the secondary black hole does not
gradually migrate through the plane of the disk but crosses through
the disk at some inclination. Consequently, such encounters are
expected to have unique emission signatures. The next, even more
general set of systems to be studied are binaries with two gas
disks at different relative inclinations.

Because the model of radiation physics plays a decisive role in
determining the character of the observational signatures, we outline
a possible avenue for future improvement of the BB-model. Introducing
a flux-limited diffusion approximation \citep{cm99,wbm05,mayer06b} can
offer a more accurate treatment of the optically thick component of
gas by emulating a continuous diffuse radiation field in the
inter-particle space. In this approximation the SPH particles which
are embedded within some gas distribution diffuse heat to their
neighbor particles as determined by the temperature gradient and
coefficient of diffusion. The particles defining a surface boundary of
an emitting region in addition to diffusion can freely radiate as a
black-body or according to some other prescription. The main challenge
in the implementation is determining which gas particles are
constituents of the boundary layer in general case, when the normal of
an emitting surface is not aligned with any preferred axis in the
system and when its orientation varies in space and time
\citep[see][for the case of a disk geometry]{mayer06b}.

Because our simulations are computationally expensive, we were not
able to follow the longer-term evolution of the binary and the disk on
time scales on which a gap is expected to form in the disk. The gap is
a low density region that forms at the center of an accretion disk, on
a scale comparable to that of the binary orbit, as a consequence of
tidal torques that the binary exerts on the disk \citep{lp79a,lp79b}. 
While we observed a lower peak density in the innermost region 
of the disk at the end of simulations, whether a gap in the disk is efficiently 
cleared or continuously replenished by the gas will depend on the gas 
temperature and hence, on the treatment of heating and cooling. The simple, 
exploratory models described here can be of importance in cases when 
the gap is either replenished or its formation process is not efficient. However 
in the future we will explore models in which the gap has already formed in 
the disk and investigate its implications for the accretion rates and 
observational signatures.

\section{Summary and Conclusions}\label{S_conclusions}

We developed a method for calculation of the observational signatures
of MBHBs with an associated gas component. The method comprises SPH
calculations carried out with a modified version of the code {\it
Gadget} coupled with calculations of cooling and heating of the
gas. We implemented and tested two different schemes for heating and
cooling of gas in which the gas is described as a black-body emitter
or a photoionized gas with solar composition. In the solar metallicity
model heating and cooling of the gas are pre-calculated with the
photoionization code {\it Cloudy}, along with the emission efficiency
of the H$\alpha$ line, the optical depth to Thomson scattering, and
the neutral hydrogen column density. This information allows us to
calculate the H$\alpha$ light curves and H$\alpha$ emission line
profiles from this system.

Since MBHBs may spend the largest fraction of their lives at
sub-parsec scales \citep{bbr} it is particularly interesting to study
the observational signatures of such binaries. In this paper we
investigated the electromagnetic signatures of binaries based on a few
exploratory models; we outline the most important findings below:

\begin{itemize}

\item{} Based on the first set of models, presented here, we find that
the X-ray outbursts should occur during pericentric passages of a
coplanar binary. We suggest that periodical X-ray outbursts should
persist as long as there is a supply of gas that the MBHB
can interact with. A calculation of a much larger number of binary
orbits is needed in order to confirm that the periodic outbursts are a
long lived signature of the binary. At the estimated level of X-ray
luminosity the binary systems interacting with the gas should be
observable to a redshift of $z\leq2$.

\item{} Except for the recurrent outbursts in the X-ray light curve
the signature of a binary is most easily discernible in the H$\alpha$
emission line {\it profiles}. The variable shape of the broad
H$\alpha$ emission line profiles can be used as a first indicator when
searching for MBHBs in the nearby universe, within 100~Mpc. Once the
binary candidates are selected from a large spectroscopic survey
(e.g., the SDSS) they can be monitored over long time intervals to
look for the time-dependent signature of a binary. Based on our
simulations (for a mass ratio of 0.1), the wavelength shift in the
H$\alpha$ emission lines over the course of an orbital period should
be measurable. If one can follow the regular variations of the line
over a few cycles, one could constrain the properties of the binary,
such as orbital period and the binary mass ratio.

\item{} The observed signatures predicted here arise from the gas disk
and accretion onto the two nuclei, both of which may not be detectable
in galaxies where the nuclear region is obscured or gas exhausted by
the starburst. The resulting signatures are in addition sensitive to
the adopted model of radiation physics but are not very sensitive to
the number of particles employed in the SPH simulations.

\item{} The evolution of the intermediate phase binaries considered
here is driven by interactions of the black holes with the accretion
disk gas, while the emission of gravitational waves is secondary (3--4
orders of magnitude lower). Orbital energy is dissipated primarily via
dynamical friction against the gaseous background and tidal torques in
case of a binary co-rotating with the disk. In case of a
counter-rotating binary, the tidal torques play the dominant role in
the dissipation since dynamical friction becomes ineffective.

\item{} The current limitation of the numerical method used in this
study is that the calculation of a large number of orbits following
the hydrodynamics and radiative heating and cooling becomes
prohibitively expensive. Therefore, we are only able to model the
observational signature of the system for a brief period of its
evolution, after we have assumed a specific initial configuration.

\end{itemize}

If the signatures of binaries are found in observations, they could be
used to estimate the number of MBHBs in this evolutionary stage and
whether MBHBs indeed evolve quickly through the last parsec. Although
the intermediate phase binaries themselves emit gravitational waves at
frequencies too low to be detected with {\it LISA} their discovery is
important for understanding the evolution of the diverse supermassive
black hole binary population.

\acknowledgments

T.B. acknowledges the support of the Center for Gravitational Wave
Physics at Penn State and the Zaccheus Daniel Fellowship. T.B. also
wishes to thank Milo{\v s} Milosavljevi{\'c} and M. Coleman Miller for
insightful discussions. The simulations of binaries and gas were
carried out on computational cluster Pleiades at the Pennsylvania
State University. We wish to thank the referee for detailed and
thoughtful comments.



\begin{thebibliography}{99}

\bibitem[\protect\citeauthoryear{Aarseth}{2003}]{aarseth03} Aarseth, S.~J.\ 2003, \apss, 285, 367 

\bibitem[\protect\citeauthoryear{Armitage \& Natarajan}{2002}]{armitage02} Armitage, 
P.~J.~\& Natarajan, P.\ 2002, \apjl, 567, L9 

\bibitem[\protect\citeauthoryear{Armitage \& Natarajan}{2005}]{armitage05} 
Armitage, P.~J., \& Natarajan, P.\ 2005, \apj, 634, 921 

\bibitem[\protect\citeauthoryear{Baker}{2003}]{baker} Baker, J.\ 2003, Classical and 
Quantum Gravity, 20, 201

\bibitem[\protect\citeauthoryear{Balsara}{1995}]{balsara95} Balsara, D.~S.\ 1995, Journal 
of Computational Physics, 121, 357 

\bibitem[\protect\citeauthoryear{Barnes \& Hernquist}{1992}]{barnes92} Barnes, J.~E.~\& 
Hernquist, L.\ 1992, \araa, 30, 705 

\bibitem[\protect\citeauthoryear{Barnes \& Hernquist}{1996}]{barnes96} Barnes, J.~E.~\& 
Hernquist, L.\ 1996, \apj, 471, 115 

\bibitem[\protect\citeauthoryear{Bate et al.}{2003}]{bate03} Bate, M.~R., Bonnell, 
I.~A., \& Bromm, V.\ 2003, \mnras, 339, 577 

\bibitem[\protect\citeauthoryear{Begelman, Blandford, \& Rees}{Begelman \etal}{1980}]{bbr} 
Begelman, M.~C., Blandford, R.~D., \& Rees, M.~J.\ 1980, \nat, 287, 307 

\bibitem[\protect\citeauthoryear{Begelman \& Rees}{1978}]{br} Begelman, M.~C.~\& 
Rees, M.~J.\ 1978, \mnras, 185, 847 

\bibitem[\protect\citeauthoryear{Bender}{1998}]{bender} Bender, P.~L.\ 1998, Eighteenth 
Texas Symposium on Relativistic Astrophysics, 536 

\bibitem[\protect\citeauthoryear{Black}{1981}]{black81} Black, J.~H.\ 1981, \mnras, 197, 553 

\bibitem[\protect\citeauthoryear{Bromm \& Loeb}{2003}]{bromm} Bromm, V.~\& Loeb, A.\ 
2003, \apj, 596, 34 

\bibitem[\protect\citeauthoryear{Capetti et al.}{2002}]{capetti} Capetti, A., Zamfir, 
S., Rossi, P., Bodo, G., Zanni, C., \& Massaglia, S.\ 2002, \aap, 394, 39 

\bibitem[\protect\citeauthoryear{Cen}{1992}]{cen92} Cen, R.\ 1992, \apjs, 78, 341 

\bibitem[\protect\citeauthoryear{Centrella}{2003}]{centrella} Centrella, J.~M.\ 2003, AIP 
Conf.~Proc.~686: The Astrophysics of Gravitational Wave Sources, 686,
219 

\bibitem[\protect\citeauthoryear{Cleary \& Monaghan}{1999}]{cm99}
Cleary, P.~W. \& Monaghan J.~J. \ 1999, J. Comp. Phys., 148, 227

\bibitem[\protect\citeauthoryear{Collin-Souffrin \& Dumont}{1989}]{CSD} Collin-Souffrin, S. 
\& Dumont, A.M. 1989, A\&A, 213, 29

\bibitem[\protect\citeauthoryear{Danzmann}{1996}]{danzmann} 
Danzmann, K.~\& the LISA study team 1996, Classical and Quantum Gravity, 
13, A247 

\bibitem[\protect\citeauthoryear{De Paolis, Ingrosso, \& Nucita}{De Paolis \etal}{2002}]
{depaolis2002} De Paolis, F., Ingrosso, G., \& Nucita, A.~A.\ 2002, \aap, 388, 470 

\bibitem[\protect\citeauthoryear{De Paolis \etal}{2003}]
{depaolis2003} De Paolis, F., Ingrosso, G., Nucita, A.~A., \& Zakharov, A.~F.\ 2003, \aap, 410, 741 

\bibitem[\protect\citeauthoryear{Di Matteo, Springel, \& Hernquist}{Di Matteo et al.}{2005}]
{dsh05} Di Matteo, T., Springel, V., \& Hernquist, L.\ 2005, \nat, 433, 604

\bibitem[\protect\citeauthoryear{Dotti et al.}{2006a}]{dotti06a} Dotti, M., Colpi, M., \& 
Haardt, F.\ 2006a, \mnras, 367, 103 

\bibitem[\protect\citeauthoryear{Dotti et al.}{2006b}]{dotti06b} Dotti, M., Salvaterra, 
R., Sesana, A., Colpi, M., \& Haardt, F.\ 2006b, \mnras, 372, 869

\bibitem[\protect\citeauthoryear{Dotti et al.}{2006c}]{dotti06c}
Dotti, M., Colpi, M., Haardt, F., \& Mayer, L.\ 2006c, , \mnras,
submitted (astro-ph/0612505)

\bibitem[\protect\citeauthoryear{Eracleous \& Halpern}{2003}]{eh03} Eracleous, M., \& 
Halpern, J.~P.\ 2003, \apj, 599, 886 

\bibitem[\protect\citeauthoryear{Eracleous, Halpern, \& Livio}{Eracleous \etal}{1996}]{ehl96} 
Eracleous, M., Halpern, J.~P., \& Livio, M.\ 1996, \apj, 459, 89 

\bibitem[\protect\citeauthoryear{Eracleous et al.}{1997}]{eracleous97} Eracleous, M., 
Halpern, J.~P., Gilbert, A.~M., Newman, J.~A., \& Filippenko, A.~V.\ 1997, 
\apj, 490, 216 

\bibitem[\protect\citeauthoryear{Escala \etal}{2004}]
{escala1} Escala, A., Larson, R.~B., Coppi, P.~S., \& Mardones, D.\ 2004, \apj, 607, 765 

\bibitem[\protect\citeauthoryear{Escala \etal}{2005}]
{escala2} Escala, A., Larson, R.~B., Coppi, P.~S., \& Mardones, D.\ 2005, \apj, 630, 152

\bibitem[\protect\citeauthoryear{Etherington \& Maciejewski}{2006}]{eh06} 
Etherington, J., \& Maciejewski, W.\ 2006, \mnras, 367, 1003 

\bibitem[\protect\citeauthoryear{Fabian et al.}{1998}]{fabian98}
Fabian, A.~C., Barcons, X., Almaini, O., \& Iwasawa, K.\ 1998, \mnras,
297, L11

\bibitem[\protect\citeauthoryear{Fan et al.}{1998}]{fan} Fan, J.~H., Xie, G.~Z., 
Pecontal, E., Pecontal, A., \& Copin, Y.\ 1998, \apj, 507, 173 

\bibitem[\protect\citeauthoryear{Ferland et al.}{1998}]{ferland} Ferland, G.~J., 
Korista, K.~T., Verner, D.~A., Ferguson, J.~W., Kingdon, J.~B., \& Verner, 
E.~M.\ 1998, \pasp, 110, 761 

\bibitem[\protect\citeauthoryear{Ferland}{2003}]{ferland03} Ferland, G.~J.\ 2003, \araa, 
41, 517 
 
\bibitem[\protect\citeauthoryear{Gaskell}{1983}]{gaskell83} Gaskell, C.~M.\ 1983, Quasars 
and Gravitational Lenses, 473 

\bibitem[\protect\citeauthoryear{Gaskell}{1996}]{gaskell96} Gaskell, C.~M.\ 1996, Lecture 
Notes in Physics, Berlin Springer Verlag, 471, 165 

\bibitem[\protect\citeauthoryear{Gould \& Rix}{2000}]{gould00} Gould, A.~\& Rix, H.\ 
2000, \apjl, 532, L29 

\bibitem[\protect\citeauthoryear{Haehnelt}{1994}]{haehnelt94} Haehnelt, M.~G.\ 1994, 
\mnras, 269, 199 

\bibitem[\protect\citeauthoryear{Haehnelt \& Kauffmann}{2002}]{haehnelt02} Haehnelt, 
M.~G.~\& Kauffmann, G.\ 2002, \mnras, 336, L61 

\bibitem[\protect\citeauthoryear{Halpern \& Filippenko}{1988}]{halpern88} Halpern, 
J.~P.~\& Filippenko, A.~V.\ 1988, \nat, 331, 46 

\bibitem[\protect\citeauthoryear{Halpern \& Filippenko}{1992}]{halpern92} Halpern, J.~\& 
Filippenko, A.\ 1992, Testing the AGN Paradigm, 57 

\bibitem[\protect\citeauthoryear{Hemsendorf, Sigurdsson, \& Spurzem}{Hemsendorf \etal}{2002}]{hemsendorf02} 
Hemsendorf, M., Sigurdsson, S., \& Spurzem, R.\ 2002, \apj, 581, 1256 

\bibitem[\protect\citeauthoryear{Hernquist \& Katz}{1989}]{hk89} Hernquist, L., \& 
Katz, N.\ 1989, \apjs, 70, 419

\bibitem[\protect\citeauthoryear{Hollenbach \& McKee}{1979}]{hm79} Hollenbach, D., \& 
McKee, C.~F.\ 1979, \apjs, 41, 555 

\bibitem[\protect\citeauthoryear{Hudson et al.}{2006}]{hudson06} Hudson, D.~S., Reiprich, 
T.~H., Clarke, T.~E., \& Sarazin, C.~L.\ 2006, \aap, 453, 433 

\bibitem[\protect\citeauthoryear{Hughes}{2002}]{hughes02} Hughes, S.~A.\ 2002, \mnras, 
331, 805 

\bibitem[\protect\citeauthoryear{Hunstead \etal}{1984}] {hunstead}
Hunstead, R.~W., Murdoch, H.~S., Condon, J.~J., \& Phillips, M.~M.\
1984, \mnras, 207, 55
 
\bibitem[\protect\citeauthoryear{Ivanov, Papaloizou, \& Polnarev}{Ivanov \etal}{1999}]{ivanov99} 
Ivanov, P.~B., Papaloizou, J.~C.~B., \& Polnarev, A.~G.\ 1999, \mnras, 307, 79 

\bibitem[\protect\citeauthoryear{Katz, Weinberg, \& Hernquist}{Katz et al.}{1996}]{kwh96} 
Katz, N., Weinberg, D.~H., \& Hernquist, L.\ 1996, \apjs, 105, 19 

\bibitem[\protect\citeauthoryear{Kazantzidis et al.}{2005}]{kazantzidis05} Kazantzidis, S., et 
al.\ 2005, \apjl, 623, L67

\bibitem[\protect\citeauthoryear{Kocsis et al.}{2006}]{kocsis06} Kocsis, B., Frei, Z., 
Haiman, Z., \& Menou, K.\ 2006, \apj, 637, 27 

\bibitem[\protect\citeauthoryear{Komossa et al.}{2003a}]{komossa} Komossa, S., Burwitz, 
V., Hasinger, G., Predehl, P., Kaastra, J.~S., \& Ikebe, Y.\ 2003a, \apjl, 582, L15 

\bibitem[\protect\citeauthoryear{Komossa}{2003b}]{komossa_rev} Komossa, S.\ 2003b, AIP 
Conf.~Proc.~686: The Astrophysics of Gravitational Wave Sources, 686, 161 

\bibitem[\protect\citeauthoryear{Kormendy \& Richstone}{1995}]{kr} Kormendy, J.~\& 
Richstone, D.\ 1995, \araa, 33, 581 

\bibitem[\protect\citeauthoryear{Landau \& Lifshitz}{1975}]{ll75} Landau, L.~D., \& 
Lifshitz, E.~M.\ 1975, The classical theory of fields, Heinemann, Oxford

\bibitem[\protect\citeauthoryear{Leahy \& Williams}{1984}]{leahy} Leahy, J.~P.~\& 
Williams, A.~G.\ 1984, \mnras, 210, 929 

\bibitem[\protect\citeauthoryear{Lee \& Klu{\'z}niak}{1999}]{lk99} Lee, W.~H., \& 
Klu{\'z}niak, W.~{\L}.\ 1999, \mnras, 308, 780 

\bibitem[\protect\citeauthoryear{Lehto \& Valtonen}{1996}]{lehto} Lehto, H.~J.~\& 
Valtonen, M.~J.\ 1996, \apj, 460, 207 

\bibitem[\protect\citeauthoryear{Li et al.}{2005}]{li05} Li, Y., Mac Low, M.-M., \& Klessen,
R.~S.\ 2005, \apj, 626, 823

\bibitem[\protect\citeauthoryear{Lin \& Papaloizou}{1979a}]{lp79a} Lin, D.~N.~C., \& 
Papaloizou, J.\ 1979, \mnras, 186, 799

\bibitem[\protect\citeauthoryear{Lin \& Papaloizou}{1979b}]{lp79b} Lin, D.~N.~C., \& 
Papaloizou, J.\ 1979, \mnras, 188, 191

\bibitem[\protect\citeauthoryear{Liu}{2004}]{liu} Liu, F.~K.\ 2004, \mnras, 347, 1357 

\bibitem[\protect\citeauthoryear{Liu, Wu, \& Cao}{Liu \etal}{2003}]{lwc} Liu, F.~K., 
Wu, X., \& Cao, S.~L.\ 2003, \mnras, 340, 411 

\bibitem[\protect\citeauthoryear{Liu \& Wu}{2002}]{liu02} Liu, F.~K., \& Wu, X.-B.\ 
2002, \aap, 388, L48 

\bibitem[\protect\citeauthoryear{MacFadyen \& Milosavljevi{\'c}}{2006}]{mm06} MacFadyen, 
A.~I., \& Milosavljevi{\'c}, M.\ 2006, in prep. (astro-ph/0607467)

\bibitem[\protect\citeauthoryear{Mayer et al.}{2006a}]
{mayer06} Mayer, L., Kazantzidis, S., Madau, P., Colpi, M., Quinn, T.,
\& Wadsley, J.\ 2006a, to appear in the Proceedings of the
"Relativistic Astrophysics and Cosmology - Einstein's Legacy"
(astro-ph/0602029)

\bibitem[\protect\citeauthoryear{Mayer et al.}{2006b}]{mayer06b} Mayer, L., Lufkin, G., 
Quinn, T., \& Wadsley, J.\ 2006b, submitted to \apjl (astro-ph/0606361)

\bibitem[\protect\citeauthoryear{Menou, Haiman, \& Narayanan}{Menou \etal}{2001}]{menou01} 
Menou, K., Haiman, Z., \& Narayanan, V.~K.\ 2001, \apj, 558, 535 

\bibitem[\protect\citeauthoryear{Merritt}{2002}]{merritt02} Merritt, D.\ 2002, \apj, 568, 
998 

\bibitem[\protect\citeauthoryear{Merritt \& Ekers}{2002}]{me} Merritt, D.~\& Ekers, 
R.~D.\ 2002, Science, 297, 1310 

\bibitem[\protect\citeauthoryear{Mihos \& Hernquist}{1996}]{mihos96} Mihos, J.~C.~\& 
Hernquist, L.\ 1996, \apj, 464, 641 

\bibitem[\protect\citeauthoryear{Milosavljevi{\' c} \& Merritt}{2001}]{milosavljevic01} 
Milosavljevi{\' c}, M.~\& Merritt, D.\ 2001, \apj, 563, 34 

\bibitem[\protect\citeauthoryear{Milosavljevi{\' c} \& Merritt}{2003}]{milosavljevic03} 
Milosavljevi{\' c}, M.~\& Merritt, D.\ 2003, in AIP Conf. Proc. 686, The Astrophysics 
of Gravitational Wave Sources, ed. J. Centrella (New York: AIP), 201

\bibitem[\protect\citeauthoryear{Milosavljevi{\'c} \& Phinney}{2005}]{milosavljevic05} 
Milosavljevi{\'c}, M., \& Phinney, E.~S.\ 2005, \apjl, 622, L93 

\bibitem[\protect\citeauthoryear{Monaghan}{1992}]{monaghan92} Monaghan, J.~J.\ 1992, \araa, 
30, 543 

\bibitem[\protect\citeauthoryear{Monaghan}{1997}]{monaghan97} Monaghan, J.~J.\ 1997, 
Journal of Computational Physics, 136, 298 

\bibitem[\protect\citeauthoryear{Ostriker}{1999}]{ostriker99} Ostriker, E.~C.\ 1999, \apj, 
513, 252

\bibitem[\protect\citeauthoryear{Paczynsky \& Wiita}{1980}]{paczynsky} Paczynsky, B.~\& 
Wiita, P.~J.\ 1980, \aap, 88, 23 

\bibitem[\protect\citeauthoryear{Papaloizou, Nelson, \& Masset}{Papaloizou et al.}{2001}]
{pnm01} Papaloizou, J.~C.~B., Nelson, R.~P., \& Masset, F.\ 2001, \aap, 366, 263 

\bibitem[\protect\citeauthoryear{Parma, Ekers, \& Fanti}{Parma \etal}{1985}]{parma} Parma, P., 
Ekers, R.~D., \& Fanti, R.\ 1985, \aaps, 59, 511

\bibitem[Perets et al.(2006)]{perets06} Perets, H.~B., Hopman, 
C., \& Alexander, T.\ 2006, \apj, 656, 709

\bibitem[\protect\citeauthoryear{Peterson \& Wandel}{2000}]{peterson00} Peterson, B.~M.~\& 
Wandel, A.\ 2000, \apjl, 540, L13 

\bibitem[\protect\citeauthoryear{Quinlan \& Hernquist}{1997}]{quinlan} Quinlan, G.~D.~\& 
Hernquist, L.\ 1997, New Astronomy, 2, 533 

\bibitem[\protect\citeauthoryear{Rhook \& Wyithe}{2005}]{rw05} Rhook, K.~J., \& 
Wyithe, J.~S.~B.\ 2005, \mnras, 361, 1145 

\bibitem[\protect\citeauthoryear{Richstone et al.}{1998}]{richstone} Richstone, D., et 
al.\ 1998, \nat, 395, A14 

\bibitem[\protect\citeauthoryear{Rieger \& Mannheim}{2000}]{rieger} Rieger, F.~M.~\& 
Mannheim, K.\ 2000, \aap, 359, 948 

\bibitem[\protect\citeauthoryear{Rodriguez et al.}{2006}]{rodriguez06} Rodriguez, C., 
Taylor, G.~B., Zavala, R.~T., Peck, A.~B., Pollack, L.~K., \& Romani, 
R.~W.\ 2006, \apj, 646, 49 

\bibitem[\protect\citeauthoryear{Roos, Kaastra, \& Hummel}{Roos \etal}{1993}]{roos} 
Roos, N., Kaastra, J.~S., \& Hummel, C.~A.\ 1993, \apj, 409, 130 

\bibitem[\protect\citeauthoryear{Saslaw, Valtonen, \& Aarseth}{Saslaw \etal}{1974}]{sva} Saslaw, 
W.~C., Valtonen, M.~J., \& Aarseth, S.~J.\ 1974, \apj, 190, 253 

\bibitem[\protect\citeauthoryear{Schoenmakers et al.}{2000}]{schoenmakers} Schoenmakers, 
A.~P., de Bruyn, A.~G., R{\" o}ttgering, H.~J.~A., van der Laan, H., \& 
Kaiser, C.~R.\ 2000, \mnras, 315, 371 

\bibitem[\protect\citeauthoryear{Sesana et al.}{2004}]{sesana04} Sesana, A., Haardt, F., 
Madau, P., \& Volonteri, M.\ 2004, \apj, 611, 623 

\bibitem[\protect\citeauthoryear{Sesana et al.}{2005}]{sesana05} Sesana, A., Haardt, F., 
Madau, P., \& Volonteri, M.\ 2005, \apj, 623, 23 

\bibitem[\protect\citeauthoryear{Sigurdsson}{1998}]{sigurdsson98} Sigurdsson, S.\ 1998, AIP 
Conf.~Proc.~456: Laser Interferometer Space Antenna, Second International 
LISA Symposium on the Detection and Observation of Gravitational Waves in 
Space, 456, 53 

\bibitem[\protect\citeauthoryear{Sillanpaa et al.}{1988}]{sillanpaa} Sillanpaa, A., 
Haarala, S., Valtonen, M.~J., Sundelius, B., \& Byrd, G.~G.\ 1988, \apj, 
325, 628 

\bibitem[\protect\citeauthoryear{Shakura \& Sunyaev}{1973}]{shakura} Shakura, N.~I.~\& 
Sunyaev, R.~A.\ 1973, \aap, 24, 337 

\bibitem[\protect\citeauthoryear{Springel, Yoshida, \& White}{Springel \etal}{2001}]{springel01} Springel, 
V., Yoshida, N., \& White, S.~D.~M.\ 2001, New Astronomy, 6, 79 

\bibitem[\protect\citeauthoryear{Springel}{2005}]{springel05} Springel, V.\ 2005,\mnras, 
364, 1105  

\bibitem[\protect\citeauthoryear{Springel, Di Matteo, \& Hernquist}{Springel et al.}{2005}]
{sdh05} Springel, V., Di Matteo, T., \& Hernquist, L.\ 2005, \mnras, 361, 776

\bibitem[\protect\citeauthoryear{Starling et al.}{2004}]{starling04} Starling, R.~L.~C., 
Siemiginowska, A., Uttley, P., \& Soria, R.\ 2004, \mnras, 347, 67 

\bibitem[\protect\citeauthoryear{Steinmetz}{1996}]{steinmetz96} Steinmetz, M.\ 1996, \mnras, 
278, 1005 

\bibitem[\protect\citeauthoryear{Strateva et al.}{2003}]{strateva03} Strateva, I.~V., et 
al.\ 2003, \aj, 126, 1720 

\bibitem[\protect\citeauthoryear{Sudou \etal}{2003}]{sudou} 
Sudou, H., Iguchi, S., Murata, Y., \& Taniguchi, Y.\ 2003, Science, 300, 
1263 

\bibitem[\protect\citeauthoryear{Sulentic, Marziani, \& Dultzin-Hacyan}{Sulentic \etal}{2000}]
{sulentic00} Sulentic, J.~W., Marziani, P., \& Dultzin-Hacyan, D.\ 2000, \araa, 38, 521 

\bibitem[\protect\citeauthoryear{Thorne \& Braginskii}{1976}]{thorne} Thorne, K.~S.~\& 
Braginskii, V.~B.\ 1976, \apjl, 204, L1 

\bibitem[\protect\citeauthoryear{Thoul \& Weinberg}{1996}]{tw96} Thoul, A.~A.~\& 
Weinberg, D.~H.\ 1996, \apj, 465, 608 
 
\bibitem[\protect\citeauthoryear{Torres \etal}{2003}]{torres} 
Torres, D.~F., Romero, G.~E., Barcons, X., \& Lu, Y.\ 2003, \apj, 596, L31 

\bibitem[\protect\citeauthoryear{Valtaoja, Valtonen, \& Byrd}{Valtaoja \etal}{1989}]{valtaoja89} 
Valtaoja, L., Valtonen, M.~J., \& Byrd, G.~G.\ 1989, \apj, 343, 47 

\bibitem[\protect\citeauthoryear{Valtaoja et al.}{2000}]{valtaoja} Valtaoja, E., 
Ter{\"a}sranta, H., Tornikoski, M., Sillanp{\" a}{\" a}, A., Aller, M.~F., Aller, 
H.~D., \& Hughes, P.~A.\ 2000, \apj, 531, 744 

\bibitem[\protect\citeauthoryear{Volonteri, Haardt, \& Madau}{Volonteri \etal}{2003}]{volonteri} 
Volonteri, M., Haardt, F., \& Madau, P.\ 2003, \apj, 582, 559 

\bibitem[\protect\citeauthoryear{Wandel, Peterson, \& Malkan}{Wandel \etal}{1999}]{wandel99} Wandel, 
A., Peterson, B.~M., \& Malkan, M.~A.\ 1999, \apj, 526, 579 

\bibitem[\protect\citeauthoryear{Wang, Zhou, \& Dong}{Wang \etal}{2003}]{wang} Wang, T., Zhou, 
H., \& Dong, X.\ 2003, \aj, 126, 113 

\bibitem[\protect\citeauthoryear{Whitehouse et al.}{2005}]{wbm05} Whitehouse, S.~C., 
Bate, M.~R., \& Monaghan, J.~J.\ 2005, \mnras, 364, 1367 

\bibitem[\protect\citeauthoryear{Wyithe \& Loeb}{2003}]{wl03} Wyithe, J.~S.~B., \& 
Loeb, A.\ 2003, \apj, 590, 691

\bibitem[\protect\citeauthoryear{Xie}{2003}]{xie} Xie, G.\ 2003, Publications of the 
Yunnan Observatory, 95, 107 

\bibitem[\protect\citeauthoryear{Yu}{2002}]{yu02} Yu, Q.\ 2002, \mnras, 331, 935 

\bibitem[\protect\citeauthoryear{Zier \& Biermann}{2001}]{zier2001} Zier, C.~\& Biermann, 
P.~L.\ 2001, \aap, 377, 23 

\bibitem[\protect\citeauthoryear{Zier \& Biermann}{2002}]{zier2002} Zier, C.~\& Biermann, 
P.~L.\ 2002, \aap, 396, 91 

\bibitem[Zier(2006)]{zier06} Zier, C.\ 2006, \mnras, 371, L36

\end{thebibliography}
\end{document}